%% file: main_revised.tex
\documentclass[fleqn,usenatbib]{mnras}

\usepackage{newtxtext,newtxmath}
\usepackage{orcidlink}
\usepackage{graphicx} 
\usepackage[T1]{fontenc}
\DeclareRobustCommand{\VAN}[3]{#2}
\let\VANthebibliography\thebibliography
\def\thebibliography{\DeclareRobustCommand{\VAN}[3]{##3}\VANthebibliography}

\usepackage{graphicx}	% Including figure files
\usepackage{amsmath}	% Advanced maths commands

\defcitealias{reguitti2022}{R22}
\title[The transitioning SN 2023xgo]{SN 2023xgo: Helium-rich Type Icn or Carbon-Flash Type Ibn supernova?}

\author[Anjasha Gangopadhyay et al.]{
Anjasha Gangopadhyay\orcidlink{0000-0002-3884-5637},$^{1}$\thanks{E-mail: anjashagangopadhyay@gmail.com}
Jesper Sollerman\orcidlink{0000-0003-1546-6615}$^{1}$, Konstantinos Tsalapatas\orcidlink{0009-0004-1062-8886}$^{1}$, Keiichi Maeda\orcidlink{0000-0003-2611-7269}$^{2}$, 
\newauthor
Naveen Dukiya\orcidlink{0000-0002-0394-6745}$^{3,4}$, Steve Schulze\orcidlink{0000-0001-6797-1889}$^{5}$, Claes Fransson\orcidlink{0000-0001-8532-3594}$^{1}$, Nikhil Sarin\orcidlink{ 0000-0003-2700-1030}$^{6}$, Priscila J. Pessi\orcidlink{0000-0002-8041-8559}$^{1}$, 
\newauthor 
Mridweeka Singh\orcidlink{0000-0001-6706-2749}$^{7}$, Jacob Wise\orcidlink{0000-0003-0733-2916}$^{8}$, Tatsuya Nakaoka$^{9}$, Avinash Singh\orcidlink{0000-0003-2091-622X}$^{1}$, Raya Dastidar\orcidlink{0000-0001-6191-7160}$^{11,12}$, 
\newauthor
Miho Kawabata\orcidlink{0000-0002-4540-4928}$^{2}$, Yu-Jing Qin$^{10}$, Kaustav K. Das\orcidlink{0000-0001-8372-997X}$^{10}$, Daniel Perley\orcidlink{ 0000-0001-8472-1996}$^{8}$, Christoffer Fremling\orcidlink{0000-0002-4223-103X}$^{13}$,
\newauthor
Kenta Taguchi\orcidlink{0000-0002-8482-8993}$^{2}$, K-Ryan Hinds$^{8}$, Ragnhild Lunnan\orcidlink{0000-0001-9454-4639}$^{1}$, Rishabh Singh Teja\orcidlink{0000-0002-0525-0872}$^{7}$, Monalisa Dubey\orcidlink{0009-0002-2621-6611}$^{3}$, 
\newauthor
Bhavya Ailawadhi\orcidlink{0009-0000-1020-9711}$^{3}$, Smaranika Banerjee\orcidlink{0000-0001-6595-2238}$^{1}$, Koji S. Kawabata\orcidlink{0000-0001-6099-9539}$^{9}$, Kuntal Misra\orcidlink{0000-0003-1637-267X}$^{3}$, Devendra K. Sahu\orcidlink{0000-0002-6253-8768}$^{7}$,   
\newauthor
Se{a'}n J. Brennan\orcidlink{0000-0003-1325-6235}$^{1}$, Mansi M. Kasliwal\orcidlink{0000-0002-5619-4938}$^{16}$, Anna Y. C. Q Ho\orcidlink{0000-0002-9017-3567}$^{14}$, Aleksandra Bochenek\orcidlink{0009-0008-2714-2507}$^{8}$, 
\newauthor
Ben Rusholme\orcidlink{0000-0001-7648-4142}$^{15}$, Russ R. Laher\orcidlink{0000-0003-2451-5482}$^{15}$, Roger Smith\orcidlink{0000-0001-7062-9726}$^{13}$, Josiah Purdum$^{13}$, Niharika Sravan$^{17}$ \\
$^{1}$Oskar Klein Centre, Department of Astronomy, Stockholm University, AlbaNova, SE-106 91 Stockholm, Sweden \\
$^{2}$Department of Astronomy, Kyoto University, Kitashirakawa-Oiwake-cho, Sakyo-ku, Kyoto 606-8502, Japan \\
$^{3}$Aryabhatta Research Institute of Observational Sciences, Manora Peak 263001, India\\
$^{4}$Department of Applied Physics, Mahatma Jyotiba Phule Rohilkhand University, Bareilly, 243006, India\\
$^{5}$Center for Interdisciplinary Exploration and Research in Astrophysics (CIERA), Northwestern University, 1800 Sherman Ave, Evanston, IL 60201, USA \\
$^{6}$Oskar Klein Centre, Department of Physics, Stockholm University, AlbaNova, Stockholm SE-106 91, Sweden \\
$^{7}$Indian Institute of Astrophysics, Koramangala 2nd Block, Bangalore 560034, India \\
$^{8}$Astrophysics Research Institute, Liverpool John Moores University, IC2,  Liverpool L3 5RF, UK \\
$^{9}$Hiroshima Astrophysical Science Centre, Hiroshima University, 1-3-1 Kagamiyama, Higashi-Hiroshima, Hiroshima 739-8526, Japan \\
$^{10}$Cahill Center for Astrophysics, California Institute of Technology, MC 249-17, 
1200 E California Boulevard, Pasadena, CA, 91125, USA \\
$^{11}$Instituto de Astrofísica, Universidad Andres Bello, Fernandez Concha 700, Las Condes, Santiago RM, Chile \\
$^{12}$Millennium Institute of Astrophysics, Nuncio Monsenor Sótero Sanz 100, Providencia, Santiago, 8320000 Chile \\
$^{13}$Caltech Optical Observatories, California Institute of Technology, Pasadena, CA 91125, USA \\
$^{14}$Department of Astronomy, Cornell University, Cornell, USA \\
$^{15}$IPAC, California Institute of Technology, 1200 E. California Blvd, Pasadena, CA 91125, USA \\
$^{16}$Division of Physics, Mathematics, and Astronomy, California Institute of Technology, Pasadena, CA 91125, USA \\
$^{17}$Department of Physics, Drexel University, Philadelphia, PA 19104, USA \\
}
\date{Accepted XXX. Received YYY; in original form ZZZ}

\pubyear{2025}

\begin{document}
\label{firstpage}
\pagerange{\pageref{firstpage}--\pageref{lastpage}}
\maketitle

\begin{abstract}
We present observations of SN~2023xgo, a transitional Type Ibn/Icn SN, from $-5.6$ to 63 days relative to $r$-band peak. Early spectra show \ion{C}{iii} $\lambda$5696 emission like Type Icn SNe, shifting to Type Ibn features. The \ion{He}{i} velocities (1800-10000~km~s$^{-1}$) and pseudo-equivalent widths are among the highest in the Ibn/Icn class. The light curve declines at 0.14mag~d$^{-1}$ until 30 days, matching SNe~Ibn/Icn but slower than fast transients. SN~2023xgo is the faintest in our SN~Ibn sample ($M_r = -17.65 \pm 0.04$) but shows typical colour and host properties. Semi-analytical modelling of the light curve suggests a compact CSM shell ($\sim10^{12}-10^{13}$~cm), mass-loss rate between $10^{-4}-10^{-3}$~M$_{\odot}$~yr$^{-1}$ with CSM and ejecta masses of $\sim$0.22 and 0.12 M$_{\odot}$, respectively. Post-maximum light-curve, spectral modelling favours a $\sim$3 M$_{\odot}$ helium star progenitor with extended ($\sim10^{15}$~cm), stratified CSM (density exponent of 2.9) and mass-loss rate of $0.1-2.7$~M$_{\odot}$~yr$^{-1}$. These two mass-loss regimes imply a radially varying CSM, shaped by asymmetry or changes in the progenitor’s mass loss over time. This mass-loss behavior fits both binary and single-star evolution. Early Icn-like features stem from hot carbon ionization, fading to Ibn-like with cooling. SN~2023xgo thus offers rare insight into the connection between SNe~Icn, Ibn, and SNe~Ibn with ejecta signatures. 
\end{abstract}


%%%%%%%%%%%%%%%%%%%%%%%%%%%%%%%%%%%%%%%%%%%%%%%%%%

%%%%%%%%%%%%%%%%% BODY OF PAPER %%%%%%%%%%%%%%%%%%

\section{Introduction}
\label{intro}

Interacting stripped-envelope (SE) supernovae (SNe) is a relatively new family of SNe \cite[see for example][]{Anjasha2024review,Smith2017,Fraser2020,Dessart2024}. Whereas Type Ib SNe are classified by their spectral lack of hydrogen - and presence of helium - their interacting counterparts, SNe~Ibn, display clear narrow helium lines, but no or weak lines of narrow hydrogen. Type Icn SNe \citep{GalYamNature2022Icn, PerleyIcn} are even rarer explosions characterized by narrow carbon and nitrogen lines, but lack hydrogen and helium lines. 

The progenitors of Type Ib/c SNe must somehow have lost their outer hydrogen envelope before explosion, and the Type Ibn SNe clearly interact with helium-rich circumstellar material (CSM) previously lost by their progenitor stars. The rare subclasses of interacting SESNe thus hold great potential to explore progenitor systems and mass-loss mechanisms of massive stars \citep{Woosley1995}.  
However, this picture faces challenges. First of all, the ejected masses deduced from comparison of SN~Ib/c light curves with hydrodynamical models are only a few solar masses, which are much lower than predicted for exploding single Wolf-Rayet (WR) stars. No massive WR star has been seen in direct imaging in pre-explosion sites \citep{Kilpatrick2018,VanDyk2018} of SESNe except iPTF13bvn (\citealt{Cao2013}, \citealt{Fremling2014A&A...565A.114F}), and SNe~Ib/c are too abundant by a factor of $\sim$2 to originate from the WR population alone \citep{Smith2011}. For these reasons, binary evolution could be a likely progenitor scenario for SESNe. For SNe~Ibn, the velocities inferred from the width of the narrow helium lines attributed to the dense CSM surrounding the progenitor stars are comparable with velocities seen in the winds of local group of WR stars. However, the inferred mass-loss rates in SNe~Ibn are much higher than those seen in WR stars, and indicate that the progenitors might have undergone an enhanced phase of episodic mass loss \citep{2007Natur.447..829P}.

Type Ibn/Icn SNe often display rapid light-curve evolution, with a fast rise and often also a rapid decline \citep{Hosseinzadeh2017,GalYamNature2022Icn}, and can be quite luminous ($-$16 to $-$20~mag). It has therefore been argued that the underlying powering mechanism cannot be only due to radioactive nickel as in the noninteracting SESNe but that the CSM interaction visible in the spectral evolution contributes to the overall luminosity. The fast, blue, luminous nature of the light curves of these SNe show resemblance with Fast Blue Optical Transients (FBOTs) or Rapidly Evolving Transients (RETs) \citep{Drout2014,Ho2023}. However, there are also some slow evolving Type Ibn SN candidates like OGLE-12-SN-006 \citep{Pastorello2015-OGLE}, which shows a rise time of about 2 weeks.

The diversity in the population and the relatively low numbers of known objects mean that the boundaries between the sub classifications are still somewhat fuzzy. Spectroscopically, even if SNe~Ibn show narrow emission lines of helium, some SNe~Ibn have residual hydrogen in their spectra (e.g. ASASSN-15ed; \citealp{Pastorello2015-15ed}, SN 2019wep; \citealp{Gangopadhyay2022}). \cite{2025Gangopadhyay} discuss the continuity between SNe~IIn (with narrow hydrogen lines) and SNe~Ibn (narrow He lines around maximum light). \cite{Pursiainen2023} presented the case of SN~2023emq, which was initially classified as a SN~Icn but shows prominent \ion{He}{i} lines later on. \cite{SteveIen} suggest a new fully stripped class, SNe~Ien, with narrow silicon, sulphur and argon lines. These classes may not be distinct, but could be somehow linked by the continuity in the properties of the outer envelopes of their progenitors. A detailed investigation of these SNe would be valuable for probing how mass loss occurs during the final stages and for providing insights into the chemical structure of the progenitors.

Some SNe~Ibn may show additional emission lines early on, similar to the flash spectroscopy lines sometimes seen in early Type II SN spectra \citep{GalYam2014NatureFlash}. Some examples of these are SNe~2010al \citep{PastorelloSN2010al}, 2019uo \citep{Gangopadhyay2020}, and 2019cj \citep{Wang2024}. Also for SN~2023emq, \cite{Pursiainen2023} portrays the event as a flash ionized SN~Ibn because it shows the \ion{C}{iii} $\lambda$5696 feature only early on. SNe~Icn also show helium signatures, but very feeble and late in its evolution \citep{PerleyIcn}. In this paper, we present the analysis of SN~2023xgo, a nearby but relatively faint member of the SN~Ibn/Icn family. We discuss whether it should be regarded as a member of the Type Icn class, although it may have some residual helium, or if it is better characterized as a flash ionized Type Ibn SN with long-lived carbon emission.

The paper is organized as follows. In Section~\ref{ref:observations}, we present our comprehensive set of optical photometry and spectroscopy together with the data reduction procedures. Section~\ref{spectra} presents the spectroscopic analysis of SN~2023xgo, matching and comparing with a group of SNe~Ibn and SNe~Icn. Section~\ref{host} presents the host galaxy analysis of the SN. Section~\ref{phot} presents the light-curve evolution and comparison with another group of SNe~Ibn/Icn from the literature. In particular, we investigate if CSM interaction is required to power the light curve (Section~\ref{section:LC_fit}) by fitting analytical/semi-analytical models at early and late phases. We discuss our results and the general scenario for SN~2023xgo in Section~\ref{discussion} and summarize our conclusions in Section~\ref{summarize}.

\subsection{Discovery and classification:}

\begin{figure}
	\begin{center}
	    \includegraphics[width=\columnwidth]{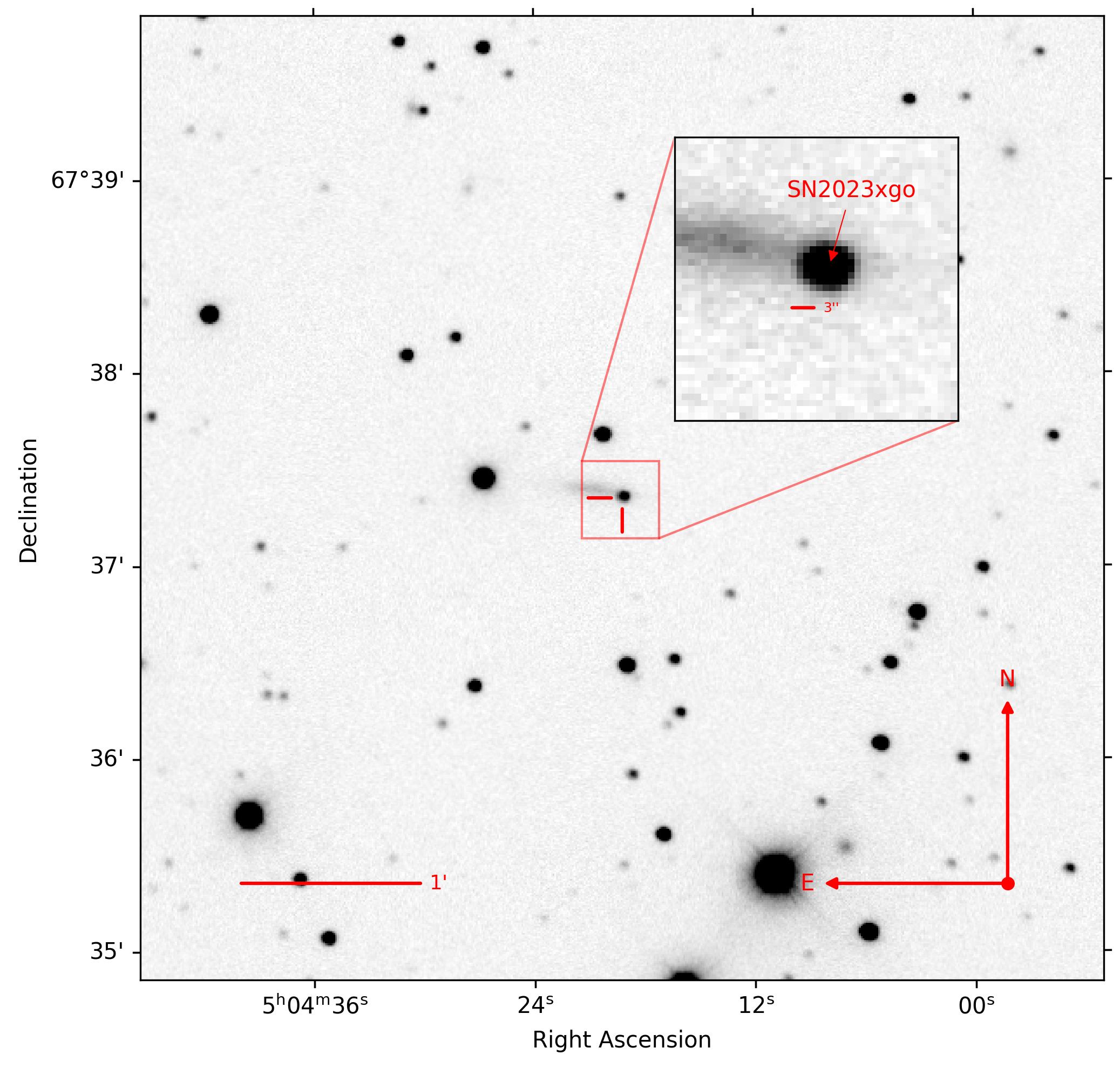}
	\end{center}
	\caption{An image of SN~2023xgo taken with the 1.3m Devasthal Fast Optical Telescope, ARIES, India in the $R$ band on 2023-11-21. The SN is 2.6 kpc away from the center of the host galaxy and is well detected in the image (highlighted in the inset).   
    }
	\label{fig:finder}
\end{figure}

SN~2023xgo was discovered by the Zwicky Transient Facility (ZTF) as ZTF23abprwou on November 9, 2023 and immediately reported to the Transient Name Server by \cite{2023TNSTR2892....1F}. The report mentions that the object at RA 05:04:19.183, DEC +67:37:21.04 (J2000.0) was first detected in the ZTF $r-$band at a magnitude of 18.56 on 2023-11-09 06:44:38.40 UT, while the last non-detection ($g > 20.41$) was from three days earlier, on 2023-11-06. 
SN~2023xgo was initially classified as a Type Icn SN \citep{2023TNSCR2931....1B} based on a low-resolution spectrum from a 0.41~m telescope obtained on November 11, 2023, in the course of the Italian Supernovae Search Project. The spectrum was reported to show a blue continuum with probable \ion{C}{iii} emission at a redshift of $z=0.014$, resembling SN~2023emq \citep{Pursiainen2023}. We later re-classified the SN on TNS as a Type Ibn SN based on the P200 spectrum (Section~\ref{spectra}) obtained on December 3, 2023 \citep{2023TNSCR3181....1S}. An image showing the precise location of the SN in its host galaxy is shown in Figure~\ref{fig:finder}.

\subsection{Explosion epoch:}
\label{expl}

To estimate the explosion epoch, we fitted a parabolic function to the rising part of the $g-$band light curve. The early light curve shape is well reproduced by a parabola. We performed the fit using 50000 iterations of Markov Chain Monte-Carlo (MCMC) simulations. Using this method, we find the explosion epoch to be MJD 60257.22 $\pm$ 0.5, which is 0.5 days prior to the first detection. This estimate is also consistent with the last non-detection of the source. We thus adopt that the explosion occurred 0.5 days before the first detection of the SN. 

\subsection{Extinction and Distance:}\label{sec:ext}

The Milky Way extinction along the line of sight to SN~2023xgo is $A_V = 0.456$~mag (from NASA Extragalactic Database\footnote{\href{https://ned.ipac.caltech.edu}{https://ned.ipac.caltech.edu}}).
We adopted this Milky Way reddening from \cite{milkyway_reddening}, applied an extinction law from \cite{Cardelli1989} assuming an R$_{\rm V}$ of 3.1. 
The \ion{Na}{iD} lines are very faint throughout the spectral evolution of SN~2023xgo.
To estimate the extinction from the host galaxy, we estimated the equivalent width of the \ion{Na}{iD} lines in the moderate resolution DBSP spectrum taken on 2023-11-15 (R $\sim$1200 - 1500), where we could detect a line at a signal-to-noise ratio of 13 (Section~\ref{ref:spectroscopy}). Using the formulation by \cite{2012Poznanski}, we estimate the extinction of the host galaxy as $A_{\rm V} = 0.043 \pm 0.011$ mag assuming R$_{\rm V}$ of 3.1. The value quoted for $A_V$ from the host is then multiplied by 0.86 to make it consistent with the recalibration of the Milky Way extinction by \cite{milkyway_reddening}. Thus, we adopt a total $A_{\rm V} = 0.49 \pm 0.01$ mag along the line of sight, considering both the contribution from the Milky Way and from the host galaxy.

There is no previously known redshift for the host galaxy. We measure the redshift of SN~2023xgo from the narrow emission lines of the host galaxy, which gives $z=0.01325$ (Section~\ref{ref:spectroscopy}).
Adopting H$_{0}$ = 73~km~s$^{-1}$ Mpc$^{-1}$ \citep{2011ApJ...730..119R}, $\Omega_{m}=0.27$ and $\Omega_{vac}=0.73$, we obtain the distance to be 54.4 $\pm$ 1.4 Mpc, and a distance modulus of 33.68 $\pm$ 0.06~mag. 
\cite{Carrick2015} found a peculiar velocity on the order of 100~km~s$^{-1}$. The peculiar velocity associated could contribute to the distance uncertainty, and we consider this as part of the error in both the distance and distance modulus. We use this value throughout the manuscript.

\section{Observations and Data Reductions}
\label{ref:observations}
\subsection{Optical and infrared photometry}
\label{phot_obs}
We observed SN~2023xgo in the \textit{uvBVRIgrizJHK} bands from day $-$5.6 to $\sim$63 d post $r-$band maximum (in the observer frame of reference) (see Section~\ref{phot}). The photometry of SN~2023xgo in the $gri-$bands was obtained using the ZTF camera \citep{Dekany2020} on the Palomar 48 inch (P48) telescope and also with the Palomar 60 inch Rainbow Camera. The Zwicky Transient Facility \citep{Graham2019, Bellm2019} surveys the entire observable northern sky with a typical 2-day cadence in $g-$ and $r-$band. The rise of the SN light curve was covered in the ZTF data, and complemented this with $r-$band data from the P60 Rainbow Camera. The P60 data were reduced with the \texttt{FPipe} pipeline described in \cite{Fleming2016}. The ZTF forced point spread function (PSF)-fit photometry was requested from the Infrared Processing and Analysis Center \citep{Masci2019} for the P48 $gri-$bands. To obtain the light curve, we follow the ZTF data processing procedure\footnote{\url{https://irsa.ipac.caltech.edu/data/ZTF/docs/ZTF_zfps_userguide.pdf}} including baseline correction, validation of flux uncertainties, combining measurements obtained on the same night, and converting the differential fluxes to the AB magnitude system. 

Optical photometric observations of SN~2023xgo in $griz-$bands were also performed using the robotic 0.7~m GROWTH-India telescope (GIT, \citealp{2022growth}) located at the Indian Astronomical Observatory (IAO) in Hanle, India. The observations were carried out in the Sloan Digital Sky Survey (SDSS) $griz-$bands beginning 2023 May 20. 
The data were processed with the standard GIT image processing pipeline described in \citet{2022growth}, and the steps followed are described in \citet{2023teja}.

SN~2023xgo was observed using the 1.3m Devasthal Fast Optical Telescope (DFOT) in $BVRI$-filters.
The bias and flat-field corrections were done following the standard procedures using \texttt{ccdproc} \citep{ccdproc}, and cosmic ray removal was done using \texttt{astrosrappy} \citep{asctroscrappy, lacosmic}. Image subtraction was performed using High Order Transform of PSF ANd Template Subtraction (HOTPANTS; \citealp{hotpants}) with the reference images taken after the SN had faded. The PSF photometry on the subtracted image was performed using a custom \texttt{PSFEx} \citep{psfex_ascl} and \texttt{photutils} \citep{photutils} based pipeline. A local star sequence was calibrated using Landolt fields observed on the same night, which was used to calibrate the \textit{BVRI} instrumental magnitudes to apparent magnitudes.

The SN was also monitored with the 2.0~m Liverpool Telescope (LT; \citealt{Steele2004}) using the IO:O imager at the Roque de los Muchachos Observatory in the $griz-$bands. The images were retrieved from the LT data archive\footnote{\url{https://telescope.livjm.ac.uk/cgi-bin/lt_search}} and processed through a PSF photometry script developed by Hinds and Taggart et al. (in preparation), and subtracting the host galaxy contribution from the images. Each measurement was calibrated using stars from the Pan-STARRS \citep{Flewelling2020} catalog.

One epoch of photometry in the $gr-$bands was obtained with the Alhambra Faint Object Spectrograph and Camera (ALFOSC) at the 2.56m Nordic Optical Telescope (NOT). For the reduction, the \texttt{PyNOT}\footnote{\url{https://github.com/jkrogager/PyNOT}} data processing pipeline was utilized. In case of multiple exposures, we computed the weighted average of the magnitudes estimated from different images with the same filter setup.

Imaging observations in $BVRIJHK$-bands were also carried out using the 1.5m Kanata telescope (KT; \citealp{2008SPIE.7014E..4LK}) of Hiroshima University in Japan. We used the standard tasks available in the data reduction software \texttt{IRAF} \citep{Tody1986} for carrying out the pre-processing. Multiple frames were taken on some nights and co-added in respective bands after the geometric alignment of the images to increase the signal-to-noise ratio, and then template subtracted photometry was performed to obtain the magnitudes. A smaller version of the photometry table in reported in Table~A\ref{tab:2023xgo_phot_obs}.

\subsection{Ultraviolet photometry from {\it Swift}}
The Ultraviolet Optical Telescope (UVOT; \citealp{2005roming}) onboard the Neil Gehrels \textit{Swift} Observatory \citep{gehrels2004} monitored SN~2023xgo in four epochs. We analysed these data with the online tools of the UK \textit{Swift} team\footnote{\href{https://www.swift.ac.uk/user_objects/}{https://www.swift.ac.uk/user\_objects}} that use the software package \texttt{HEASoft}\footnote{\href{https://heasarc.gsfc.nasa.gov/docs/software/heasoft/}{https://heasarc.gsfc.nasa.gov/docs/software/heasoft/}} version 6.26.1 and methods described in \cite{Evans2007a, Evans2009a}. Since the host galaxy was faint for SN~2023xgo, we did not use the host subtracted photometry and used the full, unsubtracted light curve for our analysis.

\subsection{X-ray observations from {\it Swift}}
While monitoring SN~2023xgo with UVOT, \textit{Swift} also observed the field with its onboard X-ray telescope XRT between 0.3 and 10~keV in photon-counting mode \citep{Burrows2005a}. The X-ray data was reduced using the same technique as the UV data. SN~2023xgo evaded detection at all epochs. The median $3\sigma$ count-rate limit of each observing block is $9\times10^{-3}~\rm s^{-1}$ (0.3--10~keV). Coadding all data pushes the $3\sigma$ count-rate limits to $1.2\times10^{-3}~\rm s^{-1}$. To convert the count-rate limits into a flux, we assume a power-law spectrum with a photon index\footnote{The photon index is defined as the power-law index of the photon flux density ($N(E)\propto E^{-\Gamma}$).} of $\Gamma=2$ and a Galactic neutral hydrogen column density of $9.1\times10^{20}$~cm$^{-2}$ \citep{HI4PI2016a}. The coadded count-rate limit translates to an unabsorbed flux of $<0.54\times10^{-13}~{\rm erg\,cm}^{-2}\,{\rm s}^{-1}$ in the range of $0.3-10$~keV and a luminosity of $<2\times10^{40}~{\rm erg\,s}^{-1}$.

\subsection{Spectroscopic Observations}
\label{ref:spectroscopy}
In the course of our follow-up campaign of SN~2023xgo we obtained 20 spectra ranging from $-$2.35 to 23.48~d post $r-$band maximum (in the observer frame of reference). We obtained eight epochs of spectroscopy with the Spectral Energy Distribution Machine \citep[SEDM;][]{Blagorodnova2018} mounted on the Palomar 60-inch telescope. The SEDM data were reduced using the pipeline described in \citet{Rigault2019,Kim2022}. We used the Double Beam Spectrograph (DBSP; \citealt{OkeandGunn,dbsp_drp:arxiv,dbsp_drp:joss,dbsp_drp:zenodo}) on the Palomar 200-inch telescope to obtain two spectra of SN~2023xgo. The data reduction was done using a custom DBSP data reduction pipeline \citep{Piascik2014,Roberson2022} relying on \texttt{Pypeit} \citep{Prochaska2019,pypeit:joss_pub,pypeit:zenodov_v1_6}. 
We obtained two epochs of spectroscopy with the Low-Resolution Imaging Spectrometer \citep[LRIS;][]{Oke1995} on the Keck I telescope, with data reduced using \texttt{lpipe} \citep{Perley2019}. We also got a single spectrum with the Spectrograph for the Rapid Acquisition of Transients (SPRAT; \citealt{Piascik2014}) on the 2.0~m Liverpool Telescope (LT). This spectrum was reduced and flux calibrated using a custom \texttt{Python} pipeline for the LT, utilising the packages \texttt{Astropy} (Astropy Collaboration 2022), \texttt{Matplotlib} \citep{4160265}, \texttt{NumPy} \citep{5725236} and \texttt{SciPy} \citep{2020zndo...4406806V}. Cosmic rays were corrected using the L.A.Cosmic algorithm \citep{lacosmic}. The standard star Feige 34 \citep{Oke1990}, observed on the same night, was used for flux calibration, with differences in the airmass of the standard star and science frames corrected by applying Table 1 from La Palma Technical Note No. 31\footnote{\url{https://www.ing.iac.es/Astronomy/observing/manuals/ps/tech_notes/tn031.pdf}}.
Observations using these telescopes/instruments were coordinated using the \texttt{FRITZ} data platform \citep{vanderWalt2019,Coughlin2023}. 

Additionally, spectroscopic observations were carried out using the Himalayan Faint Object Spectrograph and Camera (HFOSC) mounted on the Himalayan Chandra Telescope (HCT, \citealt{2010ASInC...1..193P}) and KOOLS-IFU \citep{2019PASJ...71..102M} on the Seimei Telescope. For HCT, we used a 2$\arcsec$ wide slit and Grisms Gr7/Gr8 for taking optical spectra. The spectra taken with HFOSC were reduced using the \textit{twodspec} package in \texttt{IRAF}, followed by wavelength and flux calibration. The spectra with KOOLS-IFU were taken through optical fibers and the VPH-blue grism. The data reduction was performed using the \texttt{Hydra} package in \texttt{IRAF} \citep{1994ASPC...55..130B} and a reduction software developed for KOOLS-IFU data\footnote{\url{http://www.o.kwasan.kyoto-u.ac.jp/inst/p-kools}}.  

\begin{figure}
	\begin{center}
	    \includegraphics[width=\columnwidth]{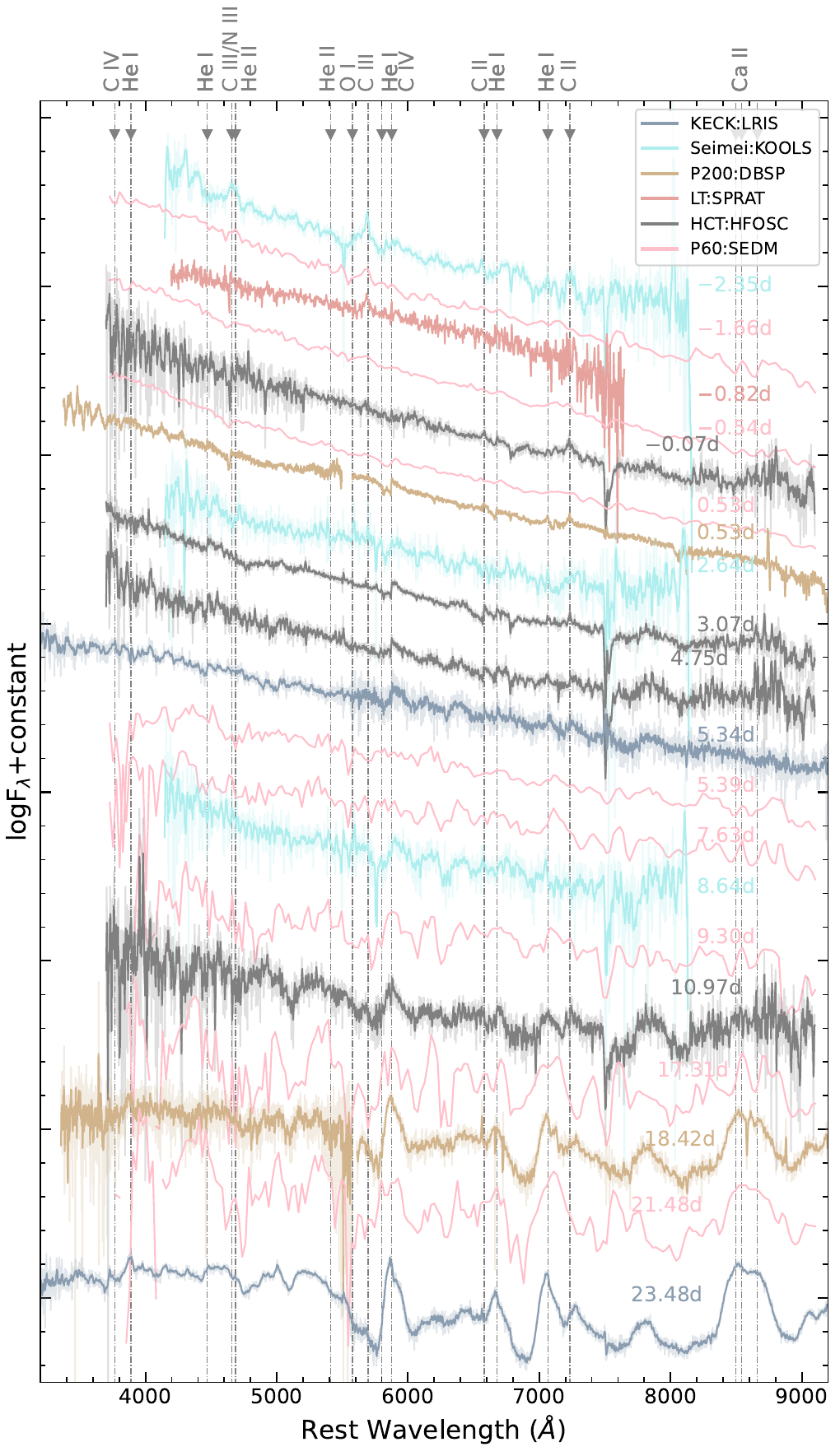}
	\end{center}
	\caption{The complete spectral evolution of SN~2023xgo from $-$2.35~d to 23.48~d post maximum. The phase is calculated with respect to the $r-$band maximum (MJD $=60262.86\pm0.46$). The colours represent different telescopes and instruments (Table~A\ref{tab:2023xgo_spec_obs}). The dark coloured spectrum is the smoothed version of the original spectrum shown in light-shaded colours.
    }
	\label{fig:spectralevolution}
\end{figure}

The slit loss corrections for all the spectra were done by scaling the spectra with respect to the SN photometry. The collage of all the spectra is shown in Figure~\ref{fig:spectralevolution}, and the spectroscopy log can be found in Table~A\ref{tab:2023xgo_spec_obs}.

\section{Spectroscopic Evolution}
\label{spectra}

The complete spectral evolution of SN~2023xgo from $-$2.35 d to 23.48~d post $r-$band maximum is shown in Figure~\ref{fig:spectralevolution}. The early spectral sequence of SN~2023xgo shows a blue continuum. The earliest spectrum of SN~2023xgo at $-$2.35~d shows a number of flash ionization lines from the recombination of the CSM. Three peaks are seen at 4655 \AA, 4685 \AA\ and 5680 \AA\ in the earliest spectrum of SN~2023xgo. The bluest peak arises due to a blend of \ion{C}{iii}/\ion{N}{iii}, and the peak at 4685 \AA\ is due to \ion{He}{ii} $\lambda$4686. The \ion{C}{iii}/\ion{N}{iii} and \ion{He}{ii} lines are common flash ionized lines for SNe~Icn \citep{GalYamNature2022Icn,Pellegrino2022Icn} and SNe~Ibn \citep{Gangopadhyay2020,Gangopadhyay2022}. However, although the third peak at 5696 \AA\ of \ion{C}{iii} is common for SNe~Icn, its strength is less for the SN~Icn 2023emq \citep{Pursiainen2023}, which transitioned to SN~Ibn. The feature is rarely seen in SNe~Ibn with prominence.  
SN~2023xgo exhibits one of the strongest \ion{C}{iii} $\lambda$5696 lines for the SN~Ibn subtype. The flash ionization lines are seen until 2.64~d in the spectral evolution.

\begin{figure}
	\begin{center}
	    \includegraphics[width=\columnwidth]{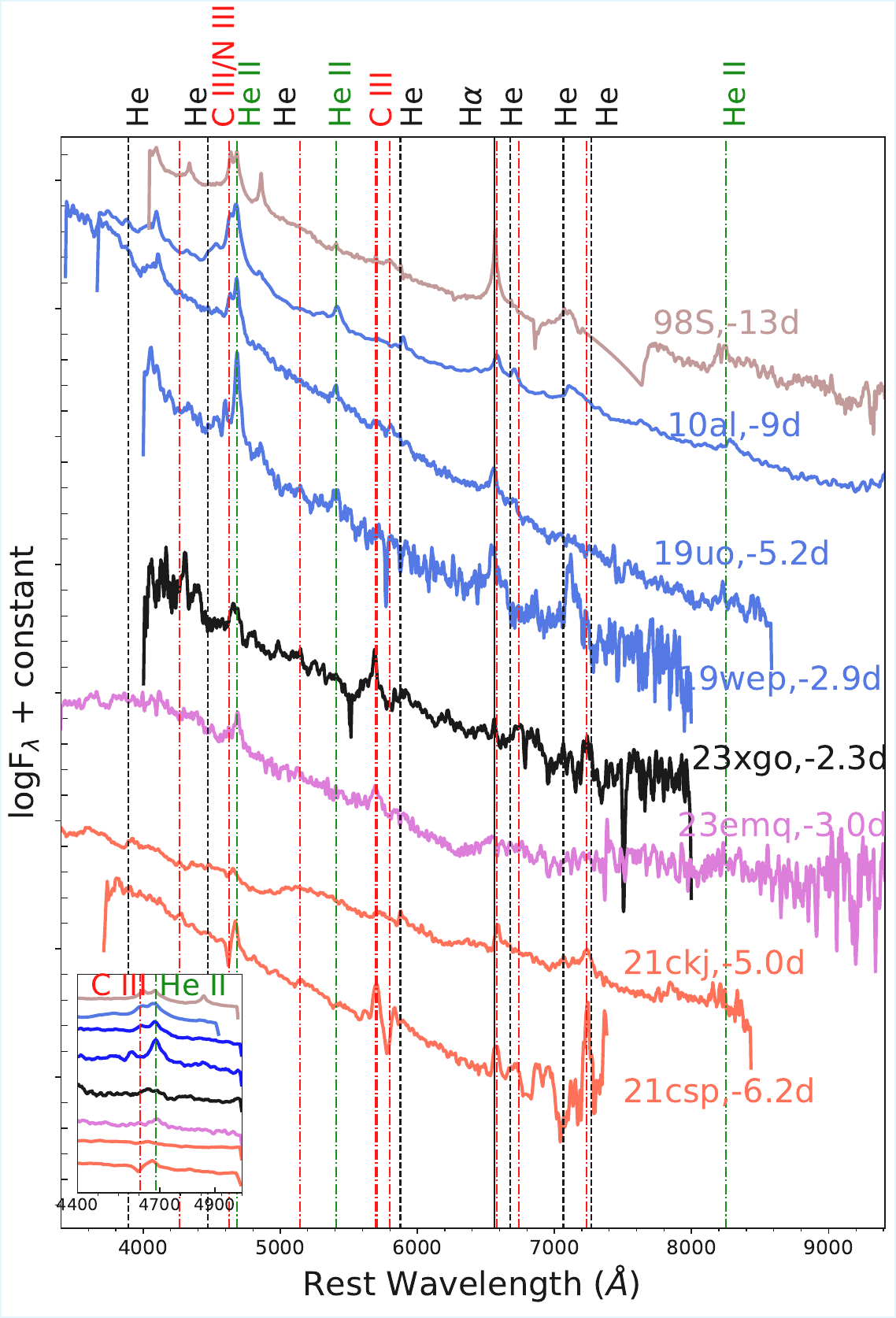}
	\end{center}
	\caption{The early spectral comparison of SN~2023xgo (in black) with a selected sample of SNe. The insets mark the unique characteristics of SN~2023xgo, where it shows \ion{C}{iii} $\lambda$5696 as the most prominent feature in the spectral evolution, unlike other SNe~Ibn with flash ionization signatures. The different colour coding are: Brown: Type IIn SNe; Blue: Ibn; Red: Icn, and Purple: transitional.}
	\label{fig:early}
\end{figure}

The overall spectral evolution of SN~2023xgo indicates that it could be a transitional SN~Ibn/Icn. The SN behaved like a SN~Icn early on with strong carbon features and then transformed to a SN~Ibn around maximum light with carbon disappearing and narrow helium lines appearing. It is not known whether most SNe~Ibn show early spectroscopic signatures like this, or to what time scales SNe~Ibn can show flash ionization signatures.

\subsection{Pre-maximum spectral evolution}
We proceeded to compare the earliest spectrum of SN~2023xgo with a sample of SNe~IIn, Ibn, and Icn, which showed signatures of flash ionization \citep{GalYam2014NatureFlash}
in their early spectra. Among the SNe~Ibn, we selected SNe~2010al \citep{PastorelloSN2010al}, 2018bcc \citep{Karamehmetoglu2021}, 2019uo \citep{Gangopadhyay2020} and 2019wep \citep{Gangopadhyay2022}. Among the SNe~Icn, we compared the earliest spectrum of SN~2023xgo with SNe~2021ckj \citep{Nagao2023} and 2021csp \citep{PerleyIcn}. We also compared SN~2023xgo with the flash ionized Type IIn SN~1998S \citep{Fassia2001} and with SN~2023emq \citep{Pursiainen2023}, where the latter is also a candidate to show transitional signatures between SN~Ibn and SN~Icn sub-classes. 

The early spectral comparison of SN~2023xgo is shown in Figure~\ref{fig:early}. All the comparison spectra were downloaded from Wiserep \citep{wiserep2012} and corrected for the redshift of the host galaxy. All the SNe~IIn, Ibn and SN~2023emq in this sample show an emission feature around 4650 \AA\ due to a blend of \ion{C}{iii}, \ion{N}{iii} and \ion{He}{ii}, similar to what we observe in SN~2023xgo (shown in the left inset of Figure~\ref{fig:early}). The two peaks are not well resolved for SN~2023xgo, but can be seen clearly for the other SNe~IIn, Ibn and SN~2023emq. These signatures in this phase are typical for young flash ionized SNe \citep{Khazov2016,Bruch2021,Bruch2023}. SNe~Icn at similar phases do not show a prominent flash ionized signature of these lines, however \cite{PerleyIcn} highlighted this as a ``narrow phase'' for Type Icn SN~2021csp where a P-Cygni feature can be seen at similar wavelengths. SN~2023xgo also shows a prominent emission around 5700 \AA\ due to \ion{C}{iii} $\lambda$5696. A very faint trace of that line is seen for SNe~2019uo, 2019wep, and the H-rich SN~1998S, but not prominent in SN 2010al, possibly due to comparison with an earlier phase spectrum. \cite{2016MNRAS.461.3057S} also noticed a line at 5700 \AA\ in the early spectrum of the Type Ibn SN~2015U, in addition to the blended line at 4650 \AA. They suggested that the line possibly arises from \ion{N}{ii} 
$\lambda$5680 because of the presence of more \ion{N}{ii} lines in their subsequent spectra. The \ion{C}{iii} $\lambda$5696 is one of the most prominent lines in our SN~Icn sample. Figure~\ref{fig:early} shows that SN~2023xgo displays the most prominent \ion{C}{iii} $\lambda$5696 line among SNe~Ibn, even more prominent than SN~2023emq, which also clearly showed this line but with less strength. 

Although \cite{Pursiainen2023} argue that the flash ionization signatures could possibly arise for both SNe~Ibn and SNe~Icn, the presence of long-lasting \ion{C}{iii} $\lambda$5696 signatures in the spectral evolution of SN~2023xgo as compared with other SNe~Ibn makes it different, which is probably a combined effect of ionization, abundance, density and luminosity.
A closer look at the feature at 4650 \AA\ shows that the \ion{He}{ii} peak for SNe~Ibn is more prominent than for SNe~Icn, which was also noticed by \cite{Pellegrino2022Icn}. 
SN~2023xgo shows a more symmetric, blended \ion{He}{ii} and \ion{C}{iii}/\ion{N}{iii} feature. Although the line at 5696 \AA~ can be reasonably well reproduced by \ion{N}{ii} $\lambda$5680, we do not see evidence for other \ion{N}{ii} lines in the spectra. In this regard, SN~2023xgo seems more similar to SNe~Icn, with stronger emission of \ion{C}{iii} $\lambda$5696 than SN~2023emq. 

\begin{figure}
	\begin{center}
	    \includegraphics[width=\columnwidth]{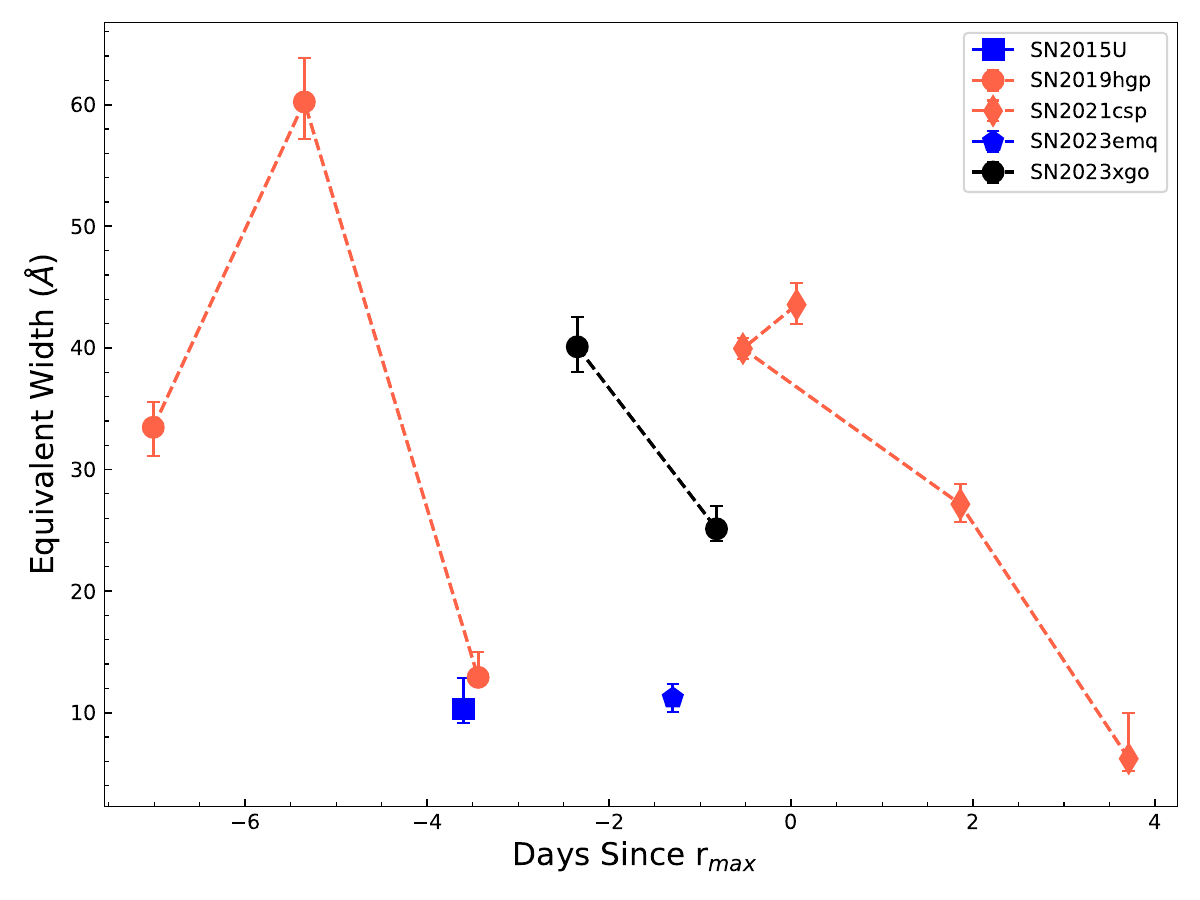}
	\end{center}
	\caption{
    Pseudo-Equivalent Width measurements of \ion{C}{iii} $\lambda$5696 for a group of SNe~Ibn and SNe~Icn compared to the values measured for SN~2023xgo. The equivalent width values of SN~2023xgo show similarity with those of SNe~Icn from the comparison sample. The red and the blue colours represent the SNe~Icn and SNe~Ibn, respectively.
    }
	\label{fig:eqw}
\end{figure}

\begin{figure*}
	\begin{center}
        \includegraphics[width=\columnwidth]{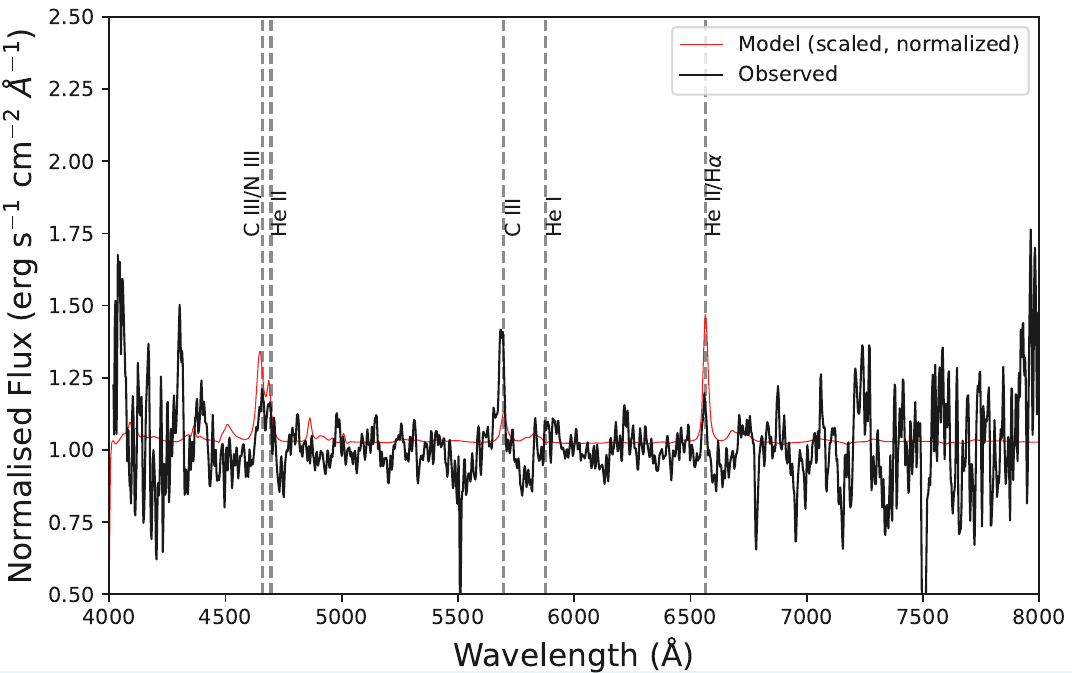}
	    \includegraphics[width=\columnwidth]{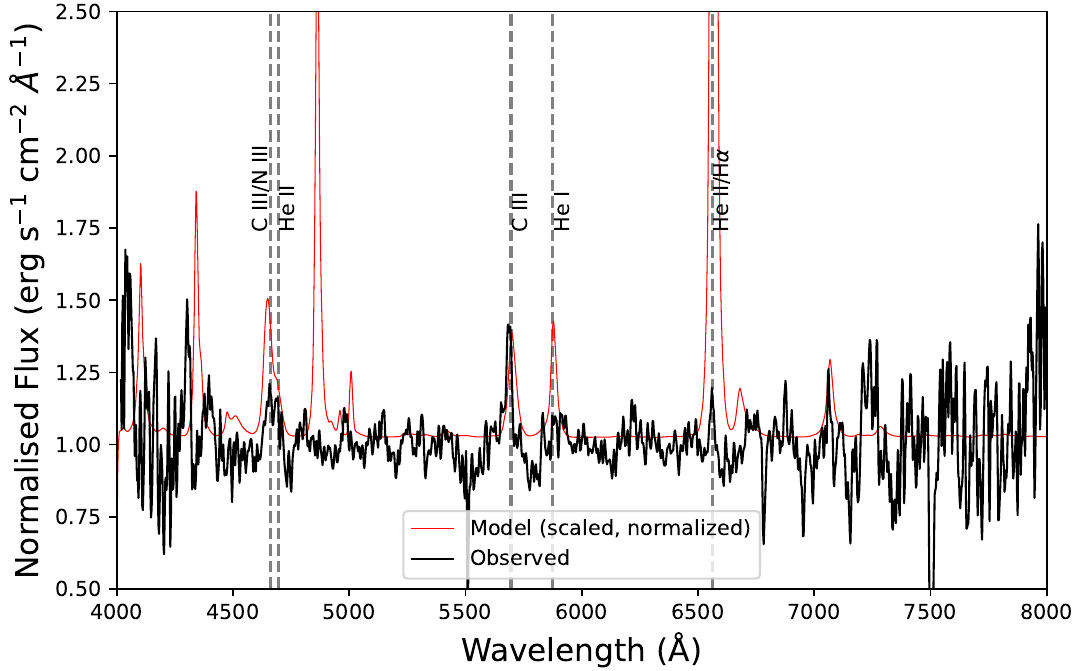}
	\end{center}
	\caption{Comparisons of the earliest spectrum having flash ionization signatures of SN~2023xgo with models by \protect\cite{BoianGroh2019}. These are a set of radiative transfer codes (\texttt{CMFGEN}) which generate a set of spectra over different luminosity and mass-loss rates 1 d after explosion. Our spectra are reproduced by spectral model of mass-loss rate 10$^{-3}$~M$_{\odot}\,{\rm yr}^{-1}$  and increase in flux reproduces the \ion{C}{iii} $\lambda$5696 feature.
    }
	\label{fig:boiangroh}
\end{figure*}

We measure the pseudo-equivalent width (EW) of the \ion{C}{iii} $\lambda$5696 feature for SN~2023xgo at $-$2.35 d and $-$0.82~d past maximum (see Figure~\ref{fig:eqw}). The estimated values are $40\pm5$ \AA\ and $25\pm5$ \AA. The pseudo-EW of the emission line for SN~2023xgo at this phase matches very well with the pseudo-EW measurements for SNe~Icn 2019hgp and 2021csp. In contrast, the pseudo-EWs of \ion{C}{iii} for both SNe~Ibn 2023emq and 2015U at similar phases are lower, at approximately 11 \AA. The prominence of this line could be the result of a combination of high ionization conditions, carbon abundance, and elevated temperatures and densities in the emitting region. At this phase, SN~2023xgo shows similarity with SNe~Icn.

\subsubsection{Comparing the flash ionized spectrum with Boain and Groh models}
\label{boaingroh}
Figure~\ref{fig:boiangroh} shows the comparison of the early time spectra of SN~2023xgo with a set of models by \cite{BoianGroh2019}. They use the radiative transfer code \texttt{CMFGEN} to build an extensive library of spectra simulating the interaction of SN ejecta with their progenitor’s wind or CSM. A range of progenitor mass-loss rates (${\dot{M}} = 5 \times 10^{-4} - 10^{-2}$~M$_{\odot}\,{\rm yr}^{-1}$ ), abundances (solar, CNO-processed, and He-rich), and SN luminosities ($L=1.9\times10^{8}-2.5\times 10^{10}~L_{\odot}$) were considered in these models. The models simulate events approximately one day after explosion, and assume a fixed location of the shock front at R$_{\rm in}$ = 8.6 $\times$ 10$^{13}$~cm. We have continuum-normalized our earliest spectrum ($-$2.35~d) and convolved it with the resolution of the available models (R $\sim$1000). The left panel of Figure~\ref{fig:boiangroh} shows that the \ion{C}{iii}/\ion{N}{iii}, the \ion{He}{ii} features and the \ion{He}{ii}/H are well reproduced around 4600~\AA\ and 6500~\AA\ assuming a luminosity of 1.5 $\times$ 10$^{9}$ $L_{\odot}$, a mass-loss rate of 10$^{-3}$~M$_{\odot}\,{\rm yr}^{-1}$ , radius of 16 $\times$ 10$^{13}$~cm and solar abundance. Since H is absent throughout the spectral evolution of SN~2023xgo, the feature at 6500~\AA\ is most likely due to \ion{He}{ii}.
This gives us good confidence in our line identifications. 
However, these set of values do not reproduce the strength of the \ion{C}{iii} $\lambda$5696 feature, which is stronger than in the model. When we tried to match our spectrum with the higher luminosity model of $L = 3.1\times 10^{9}~L_{\odot}$, a mass-loss rate of 3 $\times$ 10$^{-3}$~M$_{\odot}\,{\rm yr}^{-1}$  and radius of 32 $\times$ 10$^{13}$~cm, the strength of the \ion{C}{iii} $\lambda$5696 is well reproduced, but all other flash ionized lines are overestimated. 
The luminosity at which we underestimate the \ion{C}{iii}~$\lambda$5696 feature matches with our observed bolometric luminosity at this phase ($L=5.22 \times10^{42}~$erg~s$^{-1}$). Although a stronger \ion{C}{iii}$\lambda$5696 profile might suggest a luminosity effect, luminosity is most likely not the primary driver here, as this SN is the least luminous in the Ibn/Icn sample (see Section\ref{phot}). Instead, factors such as ionization, temperature, and density that themselves influence luminosity play a more significant role in shaping the observed carbon feature.

We want to remark that the models do not have radiation hydrodynamics that is post-processed by \texttt{CMFGEN}, so they assume a temperature of the radiation field and calculate the spectrum between R$_{\rm in}$ and R$_{\rm out}$ for a fixed shock velocity. So, the radius values can vary up to two orders in real time scale. However, these models typically give us a mass loss rate of 10$^{-3}$~M$_{\odot}\,{\rm yr}^{-1}$ for SN~2023xgo.

\subsection{Post-maximum spectral evolution}
 
Around maximum light, from 0.53~d to 4.75~d in observer frame, we see narrow P-Cygni lines of \ion{He}{i} $\lambda$5876 persisting in the spectral evolution of SN~2023xgo.
Narrow P-Cygni \ion{He}{i} $\lambda$5876 lines are a common feature seen in many SNe~Ibn \citep{Hosseinzadeh2017}. In addition, \ion{He}{i} 
$\lambda\lambda$6678,7065 start appearing at 4.75~d. From 5.34~d to about 10.97~d, the spectra show a transformation from narrow P-Cygni \ion{He}{i} $\lambda$5876 to an intermediate width feature. During this phase, the features of \ion{Ca}{ii}, \ion{Si}{ii}, and \ion{Na}{iD} also start appearing in the spectral evolution of SN~2023xgo (see Figure~\ref{fig:spectralevolution}).

\begin{figure}
	\begin{center}
	    \includegraphics[width=\columnwidth]{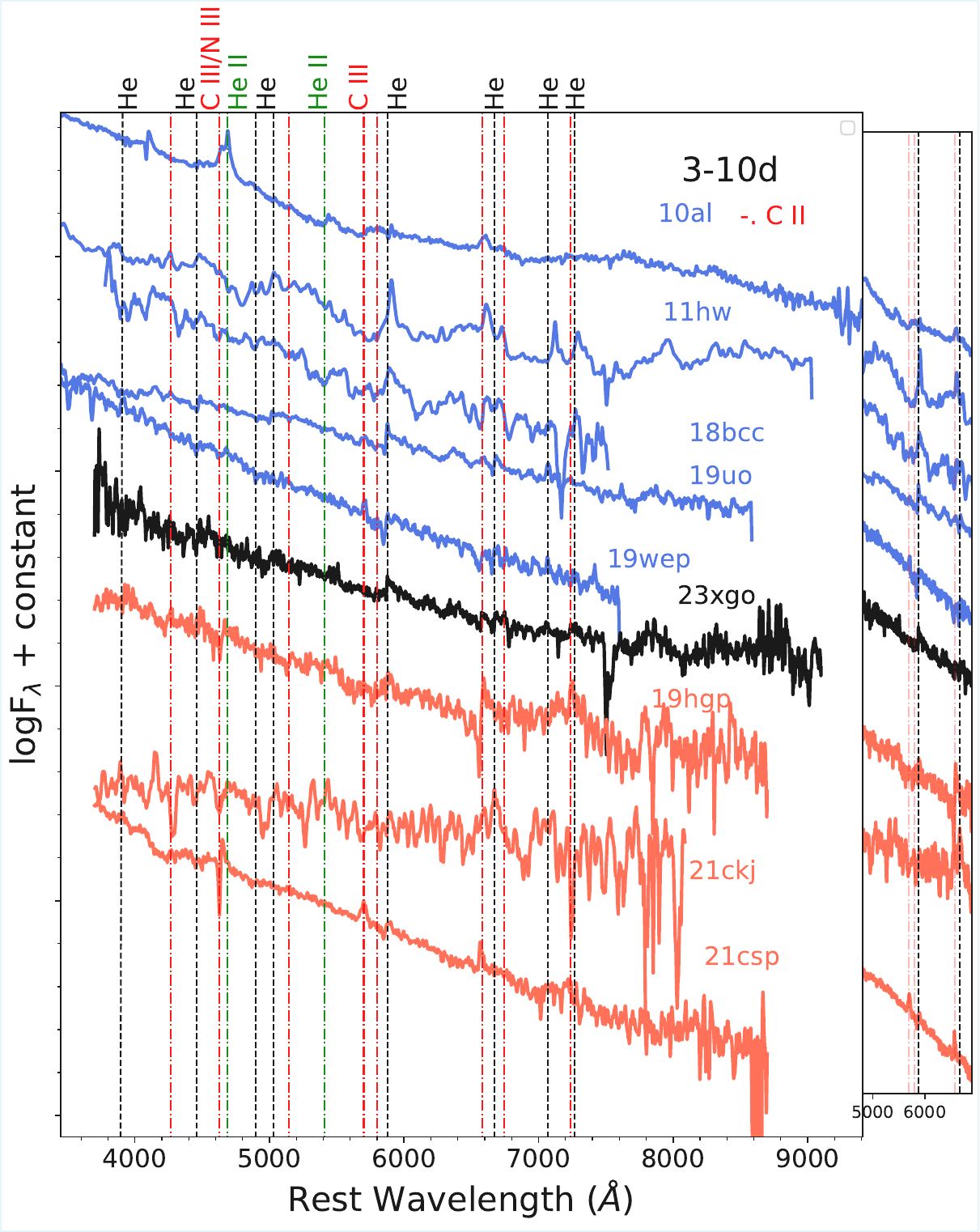}
	\end{center}
	\caption{The mid epoch ($3-10$~d) spectral evolution of SN~2023xgo, compared with a set of SNe~Ibn (blue) and SNe~Icn (red). The right panel of the figure shows the inset region of the 5000 to 6000 \AA\ region where we could detect both the ``P-Cygni" and ``emission" sub-class of \ion{He}{i} $\lambda$5876.}
	\label{fig:mid}
\end{figure}

Figure~\ref{fig:mid} shows the spectral comparison, after maximum, of SN~2023xgo with a number of SNe~Ibn (marked in blue) and SNe~Icn (marked in red). The comparison sample is similar to the one used in Figure~\ref{fig:early} complemented with other spectra of SNe~Ibn from  \cite{Hosseinzadeh2017}. The middle epoch comparison with SNe~Ibn and Icn shows that SN~2023xgo has developed narrow P-Cygni feature of \ion{He}{i} $\lambda$5876 at this phase, similar to SN~2019uo. The prominent lines of carbon have vanished during this phase. Although some very narrow helium is noticed for all the SNe~Icn, it is much more significant in all SNe~Ibn, as shown in the right zoomed-in panel of Figure~\ref{fig:mid}. 
Following the suggestion by \cite{Hosseinzadeh2017}, the line evolution of SN~2023xgo shows that it belongs to the ``P-Cygni'' subclass in the same way as SN~2019uo. The P-Cygni \ion{He}{i} features are narrow but gradually broaden over time. The physical explanation could possibly be that an optically thick shell is lit up by the explosion, and the narrow P-Cygni features transition to broader emission as the shell is swept up by the SN ejecta.  According to \cite{Hosseinzadeh2017}, this is mostly an effect of the optical depth giving rise to both the P-Cygni and emission line profiles for SNe~Ibn. The viewing angle dependence also determines whether a profile looks like ``P-Cygni'' or ``emission'', depending on whether it is viewed edge on or face on, which is not relevant if we already have an approximation of spherical symmetry. However, this scenario was questioned by \cite{Karamehmetoglu2021}, who suggest that the line formation is instead largely dependent on the density, temperature, and optical depths. In their spectral modelling of SN~Ibn 2018bcc (also shown in Figure~\ref{fig:mid}), they found that their observed profiles are a mix of both P-Cygni and emission profiles of helium as a result of high optical depths and densities in the CSM. They found that at late times, the emission lines of helium are still optically thick, but they are emission dominated because they lack other lines to branch into it. Helium ionization and recombination are mostly caused by UV and X-ray emission, occurring at the shock boundary deep in the interacting regions. Although most of the emission and the electron scattering are produced by the ionized region outside the shock, P-Cygni features usually originate from optical depths $\leq$ 1. X-rays penetrating further into the P-Cygni producing regions will fill in the absorption and lead to emission features. Thus, this provides an alternative scenario to the spectral profiles. This marks the phase when SN~2023xgo has transformed from a SN~Icn to a SN~Ibn with prominent lines of \ion{He}{i}.

\begin{figure}
	\begin{center}
	    \includegraphics[width=\columnwidth]{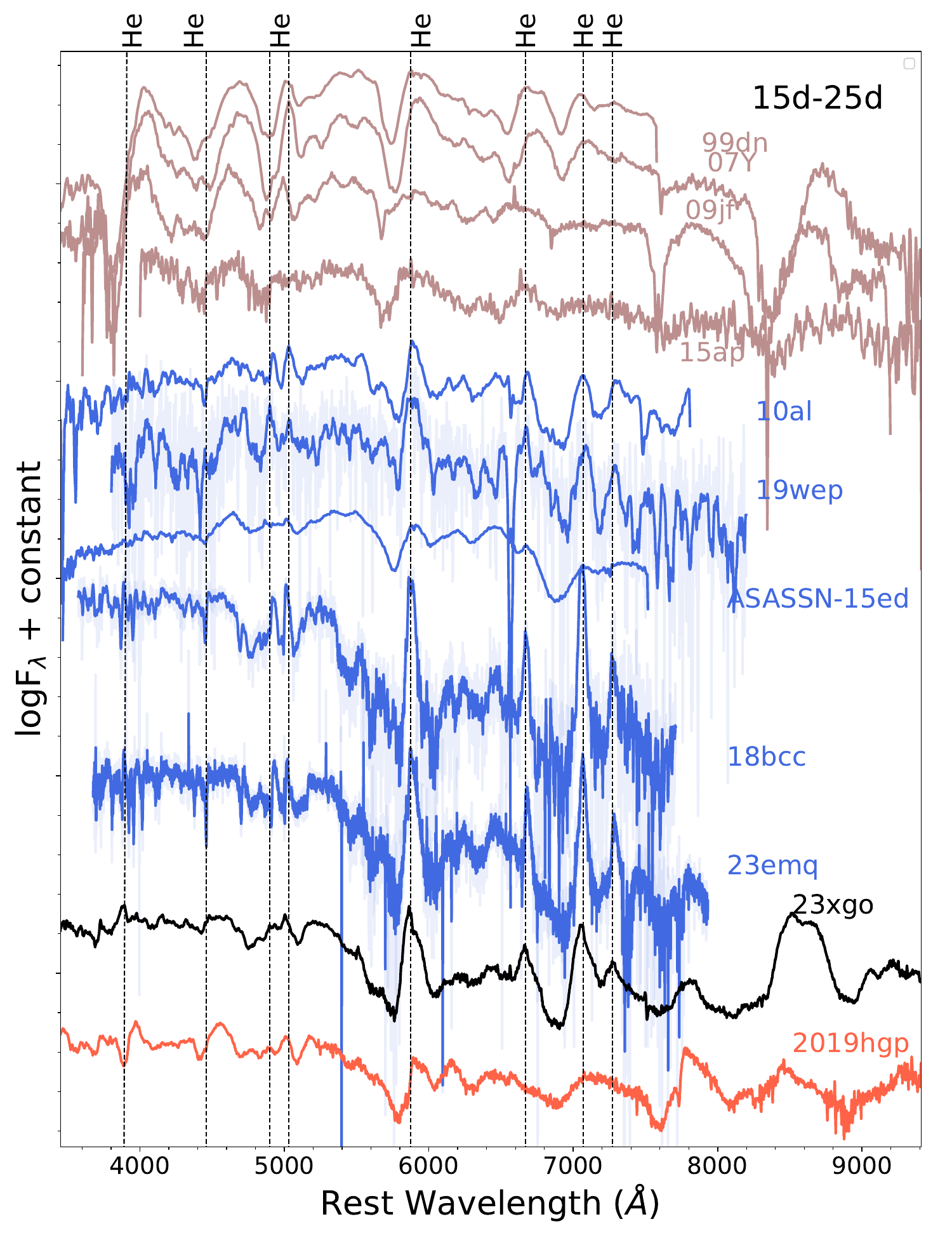}
	\end{center}
	\caption{The late epoch (23.48~d) spectral comparison of SN~2023xgo with a group of SNe~Ibn (blue), Icn (red), and Ib (brown). SN~2023xgo shows striking similarity with ASASSN-15ed at this phase, with broad P-Cygni features, and is also similar to some SNe~Ib.}
	\label{fig:late}
\end{figure}

\begin{figure}
	\begin{center}
	    \includegraphics[scale=0.5]{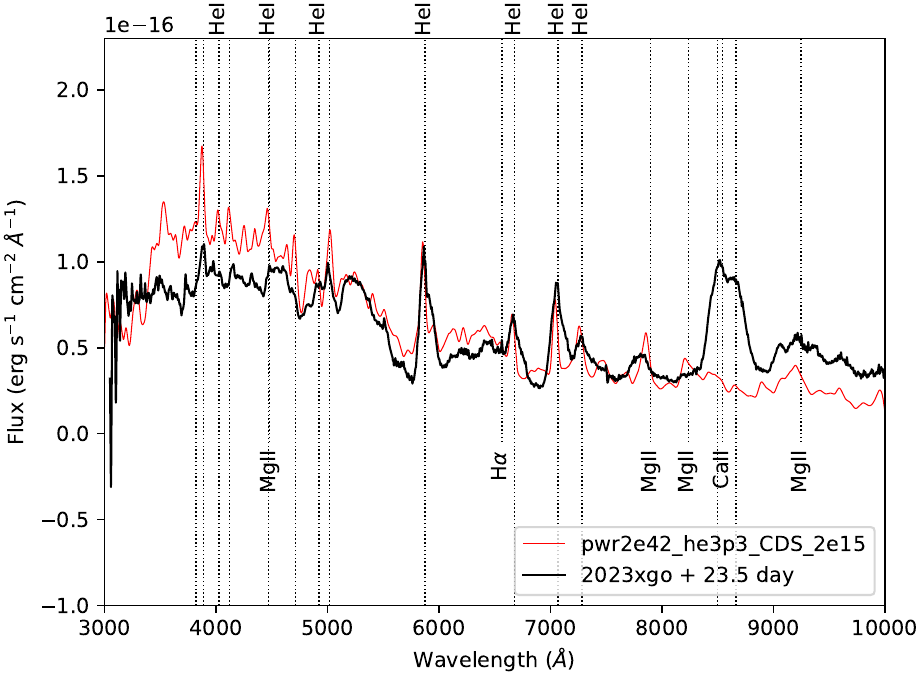}
	\end{center}
	\caption{The best match of our observed (23.48~d) spectrum with the models by \protect\cite{Dessart2022}. We find that a low mass He star model of 3~M$_{\odot}$ best reproduces our observed \ion{He}{i} spectral features.}
	\label{fig:dessartmatch}
\end{figure}

\subsection{Late-time evolution}
\label{late}

From 10.97~d to 23.48~d, the spectra of SN~2023xgo transform, and the \ion{He}{i} features have developed significantly (see Figure~\ref{fig:late}). Apart from the \ion{He}{i} $\lambda\lambda$5876, 6678 and 7065 features, some ejecta lines like the Ca NIR triplet are also seen in the spectral evolution of SN~2023xgo at this phase (see Figure~\ref{fig:spectralevolution}). Figure~\ref{fig:late} shows the comparison of SN~2023xgo with a group of SNe~Ibn (blue), SNe~Icn (red), and SNe~Ib (brown), at late epochs. The spectral evolution of SN~2023xgo shows remarkable similarity with ASASSN-15ed, which had Full Width at Half Maxima (FWHM) velocities of the \ion{He}{i} $\lambda$5876 line in the order of $5000-6000$~km~s$^{-1}$, whereas the FWHM velocity for SN~2023xgo is 6300 $\pm$ 1000~km~s$^{-1}$. This is less than the typical velocity for SNe~Ib which have absorption trough velocities between $7000-9000$~km~s$^{-1}$, as seen in the brown-coloured spectra of our comparison sample (Figure~\ref{fig:late}). For the other SNe~Ibn (2010al, 2018bcc, 2019wep and 2023emq), we do see absorption dips for \ion{He}{i} $\lambda$5876, but less prominent than for SN~2023xgo. It is interesting to note that SN~2019hgp, which is a SN~Icn, also shows a similar absorption profile at 5800 \AA\ as the helium feature in SNe~Ibn, but for the SN~Icn this is most likely due to an adjacent \ion{C}{ii} line. The major difference observed between SN~Ibn and SN~Ib spectra is that SNe~Ibn have more symmetric line profiles, while in SNe~Ib the absorption part dominates over the emission. We note that, the absorption components of the \ion{He}{i} lines in SN~2023xgo are similar to those in SN~Ib spectra, suggesting a higher kinetic energy per mass unit in the SN ejecta. This is seen especially in the late phases of SNe~Ibn that develop broad lines \citep{Pastorello2015-15ed}. 
The simultaneous presence of a very narrow feature at early phases and a transition to broader features later on suggests that these features arise from different emitting regions: the broader \ion{He}{i} P-Cygni features are likely a signature of the SN ejecta, while the narrow \ion{He}{i} P-Cygni lines are generated in the unperturbed, He-rich CSM.

\subsubsection{Late phase spectral matching}

To investigate the behavior of SN~2023xgo during the nebular phases, we compare our last epoch spectrum (23.48~d) with a set of nebular phase spectral models by \cite{Dessart2022}, who performed numerical simulations for radiation hydrodynamics of the ejecta and CSM as well as non-local thermodynamic equilibrium (non-LTE) radiative transfer in a H-free, He-rich dense shell powered by interaction. We compare the last spectrum of SN~2023xgo with the models by \cite{Dessart2022}, both for binary and single He star explosion models, with the mass of the star ranging between 2.9~M$_{\odot}$ and 12~M$_{\odot}$, and assuming two shells corresponding to ejecta and CSM, respectively, taken from the models by \cite{Ertl2020}. Comparing with a grid of He-star mass models having a CDS velocity of 2000~km~s$^{-1}$, a grid of luminosity between $2-10\times10^{42}$~erg~s$^{-1}$ and different mixing, we find the best fit match with a model of 3~M$_{\odot}$ He star. 
The best-matched model spectrum (Figure~\ref{fig:dessartmatch}) has a luminosity of $2 \times 10^{42}$~erg~s$^{-1}$, higher than our observed bolometric luminosity of $6.0 \times 10^{41}$~erg~s$^{-1}$ at similar phases. We applied Gaussian smoothing to match the model resolution to the spectral resolution of our data. The prominent \ion{He}{i} lines at $\lambda\lambda$4471, 5876, 6678, 7065, and 7281 are well reproduced. The model also captures several O, Si, and Mg emission features across the optical range. Despite strong emission in the blue, this region is heavily suppressed by Fe line blanketing \citep{Dessart2022}. Some of the Mg lines are well matched, while Ca is underestimated, possibly due to differences in the structure of the density of CDS influenced by turbulence, temperature and ionization, as seen also in SN~2006jc \citep{Dessart2022}. The mismatch also reflects uncertainties in the progenitor shell compositions. \cite{Dessart2022} have shown that Mg emission dominates in the ONeMg shells, whereas Ca dominates in the Si/S and Fe/He shells. Thus, \ion{O}{i}, \ion{Mg}{ii}, and \ion{Ca}{ii} line strengths are highly model dependent, distinguishing interacting SNe from normal Type Ibc models. Overall, the observed He features are consistent with those in the model of a 3~M$_{\odot}$ He star with a CDS at $2 \times 10^{15}$~cm. In contrast, SN~2020nxt showed a late-time increase in the \ion{Ca}{ii} NIR triplet strength as the shell ionization dropped \citep{Wang2024}, an effect not captured in \cite{Dessart2022}, suggesting an ionization rather than abundance origin. Notably, the models by \cite{Dessart2022} overpredict flux of the \ion{Mg}{ii} lines at 7896, 8234, and 9218 \AA\ at all epochs, which we do not observe prominently in our spectra.

\begin{figure*}
	\begin{center}
	    \includegraphics[scale=0.5]{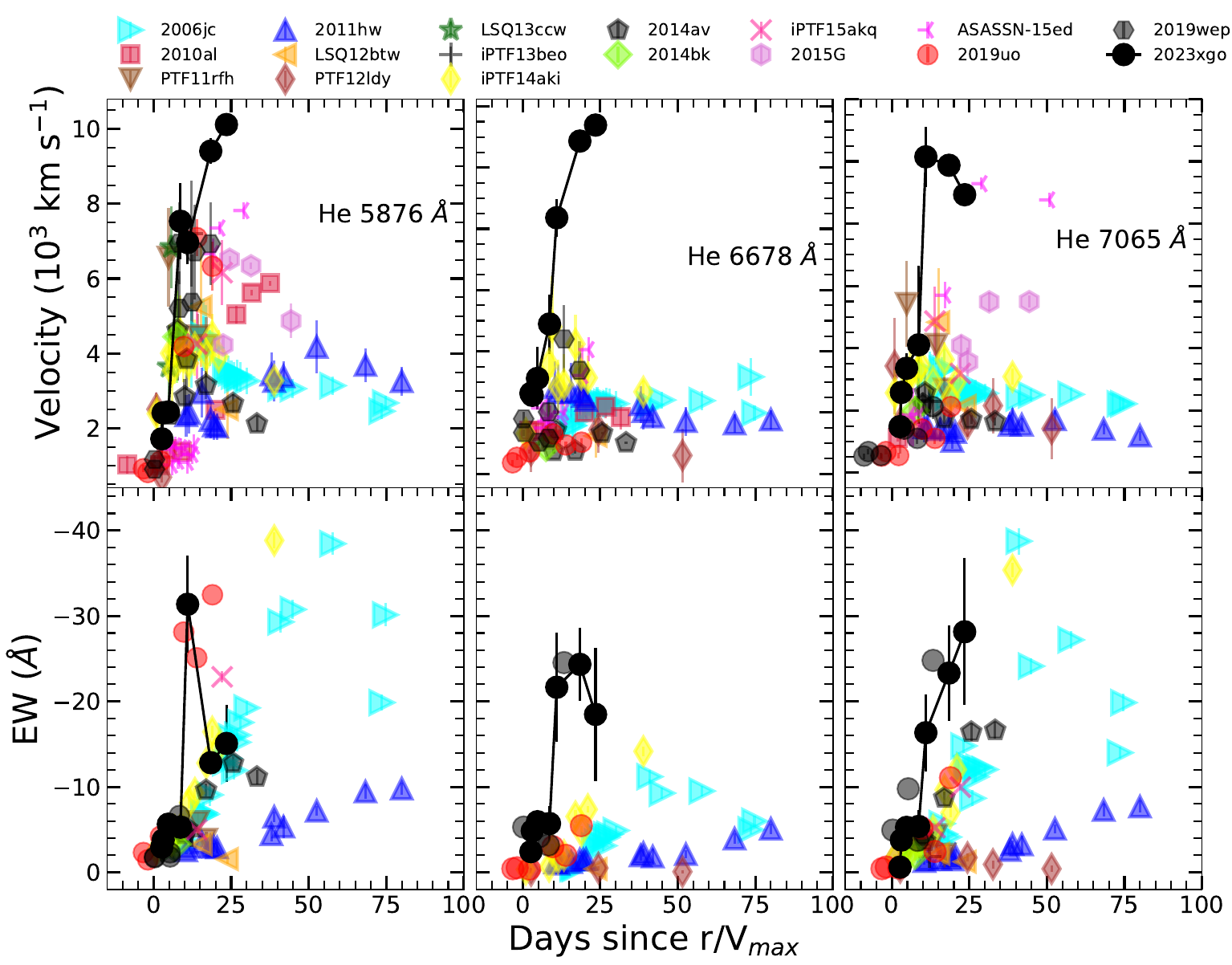}
	\end{center}
	\caption{The velocity and equivalent widths of the \ion{He}{i}~$\lambda\lambda$5876, 6678 and 7065 lines of SN~2023xgo compared with the values from a group of SNe~Ibn from \protect\cite{Hosseinzadeh2017}.}
	\label{fig:vel_eqw}
\end{figure*}

\subsection{Velocities and Pseudo-Equivalent Width Evolution}
\label{vel-eqw}
Figure~\ref{fig:vel_eqw} shows the pseudo-EW and the velocity evolution of three helium lines in the spectra of SN~2023xgo, compared with the measurements from other members of the SN~Ibn sample from \cite{Hosseinzadeh2017}. We see 
that the velocities and pseudo-EWs for SN~2023xgo increase drastically for all \ion{He}{i} lines.
The emission lines of \ion{He}{i} were fit using a Gaussian on a linear continuum. The estimates of the pseudo-EW involve the calculation of the integral of the flux normalized to the local continuum. In this estimate, we do not measure the pseudo-EW of the absorption component of the P-Cygni lines. The velocities are estimated from the FWHM of the emission component of the \ion{He}{i} profile. We see an increasing trend in both the line velocities and pseudo-EWs of the He lines. The velocity estimates of SN~2023xgo lie at the upper range of what is found for SNe~Ibn and show a faster evolution. In SN~2023xgo, the broad features seen are wider than typical emission velocities for SNe~Ibn around maxima.
The FWHM velocity of \ion{He}{i} $\lambda$5876 for SN~2023xgo is much broader than most of the SNe~Ibn of our comparison sample. It is important to note that the measured FWHM of \ion{He}{i} $\lambda$5876 may be affected by contamination from \ion{Na}{i}~D, particularly at certain phases, which could influence the line width measurements. In our comparison sample, ASASSN-15ed and SN~2019wep lie in the higher velocity end of SNe~Ibn; however, SN~2023xgo shows an even higher FWHM velocity at later phases. \cite{Hosseinzadeh2017} claimed that the ``P-Cygni'' subclass showed a faster evolution in line velocities reaching broader emission profiles while the emission subclass shows less evolution in line velocities. The pseudo-EWs of the \ion{He}{i} lines for SN~2023xgo are at an the very high end compared to the normal SNe~Ibn. This indicates that the late time behavior of SN~2023xgo is similar to what is seen in SNe~Ibn, which shows ejecta signatures at late phases reaching velocities of almost the same high velocities as SN~Ib do at this phase.

\section{Host Galaxy}
\label{host}
\subsection{Observations}
We retrieved science-ready coadded images from the Panoramic Survey Telescope and Rapid Response System (Pan-STARRS, PS1) DR1 \citep{Chambers2016a}, the Two Micron All Sky Survey \citep[2MASS;][]{Skrutskie2006a}, and WISE images \citep{Wright2010a} from the unWISE archive \citep{Lang2014a}\footnote{\href{http://unwise.me}{http://unwise.me}} that include WISE data from the NEOWISE-Reactivation mission $1-7$ \citep{Mainzer2014a, Meisner2017a}. We used the software package LAMBDAR (Lambda Adaptive Multi-Band Deblending Algorithm in R) \citep{Wright2016a} and tools presented in \citet{Schulze2021a}, to measure the brightness of the host galaxy. To expand the spectral energy distribution we use the \textit{Swift} data obtained between December 2023 and January 2024 and measured the brightness using a slightly larger aperture than in the optical to account for differences in the point spread function. All measurements are summarized in Table~\ref{tab:hostphot}. We also observed the host galaxy spectrum of the SN using KCWI. The observations made on August 10th, 2024, and January 1st, 2025. The original observations were reduced using the official pipeline. Cosmic ray rejection was performed using a modified version of the KcwiKit pipeline \footnote{https://github.com/yuguangchen1/KcwiKit}, sky background subtraction was done with ZAP and inverse sensitivity and telluric fitting were handled with KSkyWizard \footnote{https://github.com/zhuyunz/KSkyWizard}. The frames were then stacked using an inverse-variance weighted average. The elliptical aperture is defined using one-sigma isophotal fitting from photutils, and the host spectra were extracted using aperture sizes of 1$\times$, 1.5$\times$, and 2$\times$ the fitted size.

\begin{table}
\centering
\caption{Photometry of the host galaxy of SN~2023xgo}
\label{tab:hostphot}
\begin{tabular}{|l|l|c|c|}
\hline
\textbf{Survey/} & \textbf{Instrument} & \textbf{Filter} & \textbf{Brightness (mag)} \\
\textbf{Telescope} &                    &                 &                            \\
\hline
\textit{Swift} & UVOT & $w2$  & $20.49 \pm 0.12$ \\
\textit{Swift} & UVOT & $m2$  & $20.58 \pm 0.14$ \\
\textit{Swift} & UVOT & $w1$  & $19.86 \pm 0.17$ \\
\textit{Swift} & UVOT & $u$   & $18.85 \pm 0.17$ \\
\textit{Swift} & UVOT & $v$   & $16.96 \pm 0.13$ \\
PS1            &      & $g$   & $17.36 \pm 0.04$ \\
PS1            &      & $r$   & $16.79 \pm 0.05$ \\
PS1            &      & $i$   & $16.49 \pm 0.04$ \\
PS1            &      & $z$   & $16.30 \pm 0.04$ \\
PS1            &      & $y$   & $16.11 \pm 0.10$ \\
2MASS          &      & $J$   & $16.00 \pm 0.07$ \\
2MASS          &      & $H$   & $15.78 \pm 0.10$ \\
\textit{WISE}  &      & $W1$  & $16.91 \pm 0.09$ \\
\textit{WISE}  &      & $W2$  & $17.53 \pm 0.07$ \\
\hline
\end{tabular}
\vspace{0.2cm}
\begin{flushleft}
\textbf{Note.} All magnitudes are reported in the AB system and are not corrected for reddening.
\end{flushleft}
\end{table}

\subsection{Spectral Energy Distribution modelling}
We model the observed SED with the software package \texttt{Prospector} \citep{Johnson2021a} version 1.4.\footnote{\texttt{Prospector} uses the \texttt{Flexible Stellar Population Synthesis} (\texttt{FSPS}) code \citep{Conroy2009a} to generate the underlying physical model and \texttt{python-fsps} \citep{ForemanMackey2014a} to interface with \texttt{FSPS} in \texttt{python}. The \texttt{FSPS} code also accounts for the contribution from the diffuse gas based on the \texttt{Cloudy} models from \citet{Byler2017a}. We use the dynamic nested sampling package \texttt{dynesty} \citep{Speagle2020} to sample the posterior probability.} We assume a Chabrier initial-mass function (IMF; \citealt{Chabrier2003a}) and approximate the star-formation history (SFH) by a linearly increasing SFH at early times followed by an exponential decline at late times. 
To account for any reddening between the expected and the observed SED, we use the Calzetti attenuation model \citep{Calzetti2000a}. The priors of the model parameters are set identical to those used by \citet{Schulze2021a}.

\begin{figure}
   \begin{center}
		\includegraphics[width=\columnwidth]{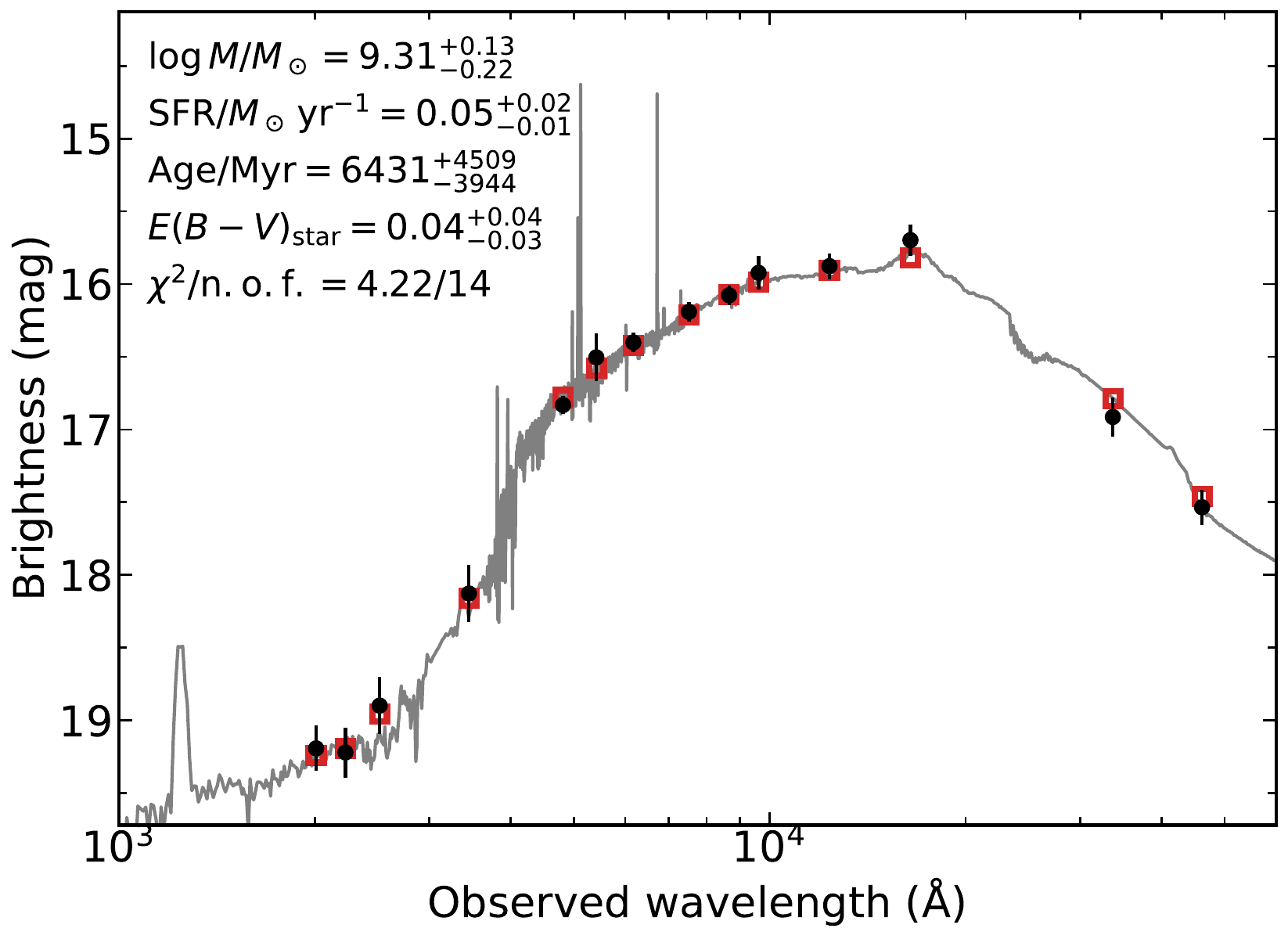}
	\end{center}
   \caption{Spectral energy distribution (SED) of the host galaxy of SN 2023xgo from 1000 to 60000 \AA\ (black data points). The solid line displays the best-fitting model of the SED. The red squares represent the model-predicted magnitudes. The fitting parameters are shown in the upper-left corner. The abbreviation ``n.o.f.'' stands for number of filters.}
    \label{fig:gal_sed}
\end{figure}

Figure~\ref{fig:gal_sed} shows the observed SED and its best fit. The SED is adequately described by a galaxy template with a mass of $10^{9.31^{+0.13}_{-0.22}}$~M$_{\odot}$, and a star-formation rate of $0.05^{+0.02}_{-0.01}~$M$_{\odot}\,{\rm yr}^{-1}$. The mass and star-formation rate are similar to those of the SN host galaxies from the Palomar Transient Factory (grey contours; 
\cite{Schulze2021a}), 
SNe~Ibn and SNe~Icn (colour-coded; the values of the SNe~Icn were taken from \citealt{GalYam2014NatureFlash, PerleyIcn, Pellegrino2022Icn}). 

\begin{table}
\centering
\caption{Emission line fluxes measured from the host galaxy spectrum of SN~2023xgo.}
\label{tab:emissionlineflux}
\begin{tabular}{lcc}
\hline
\textbf{Line} & \textbf{Flux} ($10^{-16}$~erg~cm$^{-2}$~s$^{-1}$) \\
\hline
{[O~III]}~$\lambda$4960 & $18.81 \pm 0.94$ \\
{[O~III]}~$\lambda$5007 & $59.10 \pm 1.59$ \\
H$\gamma$           & $8.82 \pm 1.27$ \\
H$\beta$            & $35.48 \pm 5.45$ \\
H$\alpha$           & $126.95 \pm 2.80$ \\
{[N~II]}~$\lambda$6549 & $6.24 \pm 1.01$ \\
{[N~II]}~$\lambda$6585 & $17.39 \pm 1.18$ \\
\hline
\end{tabular}
\end{table}

After tying the flux calibration of the host spectra to the host photometry, we could measure the line fluxes from a simple emission line analysis, measuring the line fluxes of [\ion{O}{iii}] $\lambda$4960, 5007, H$\alpha$, H$\beta$ and [\ion{N}{ii}] $\lambda$6549, 6585 lines reported in Table~\ref{tab:emissionlineflux}. The values reported are not corrected for extinction.

The MW-extinction corrected H$\beta$/H$\alpha$ flux ratios are $\approx3.6$, differing from the theoretically predicted value of 2.86, assuming Case B recombination, electron temperature of $10^4~\rm K$, and electron density of $10^2~\rm cm^{-3}$ \citep{Osterbrock2006a}. The excess in the flux ratio translates to $E_{\rm host}(B-V) = 0.06^{+0.13}_{-0.06}$~mag, assuming the Calzetti attenuation model with $R_V=4.05$. 
Using this O3N2 diagnostic together with the parameterisation from \citet{Curti2017a}, we infer a gas-phase metallicity of $0.7 \pm 0.1$ solar. The H$\alpha$ luminosity and the level of star formation are tightly correlated \cite{Kennicutt1998a}. The attenuation-corrected star-formation rate is $0.05 \pm 0.02~$M$_{\odot}\,{\rm yr}^{-1}$ using \citet{Kennicutt1998a} and \citet{Madau2014a} to convert from the Salpeter IMF (assumed in \citealt{Kennicutt1998a}) to the Chabrier IMF (assumed in our galaxy SED modelling).

\begin{figure}
   \begin{center}
		\includegraphics[width=\columnwidth]{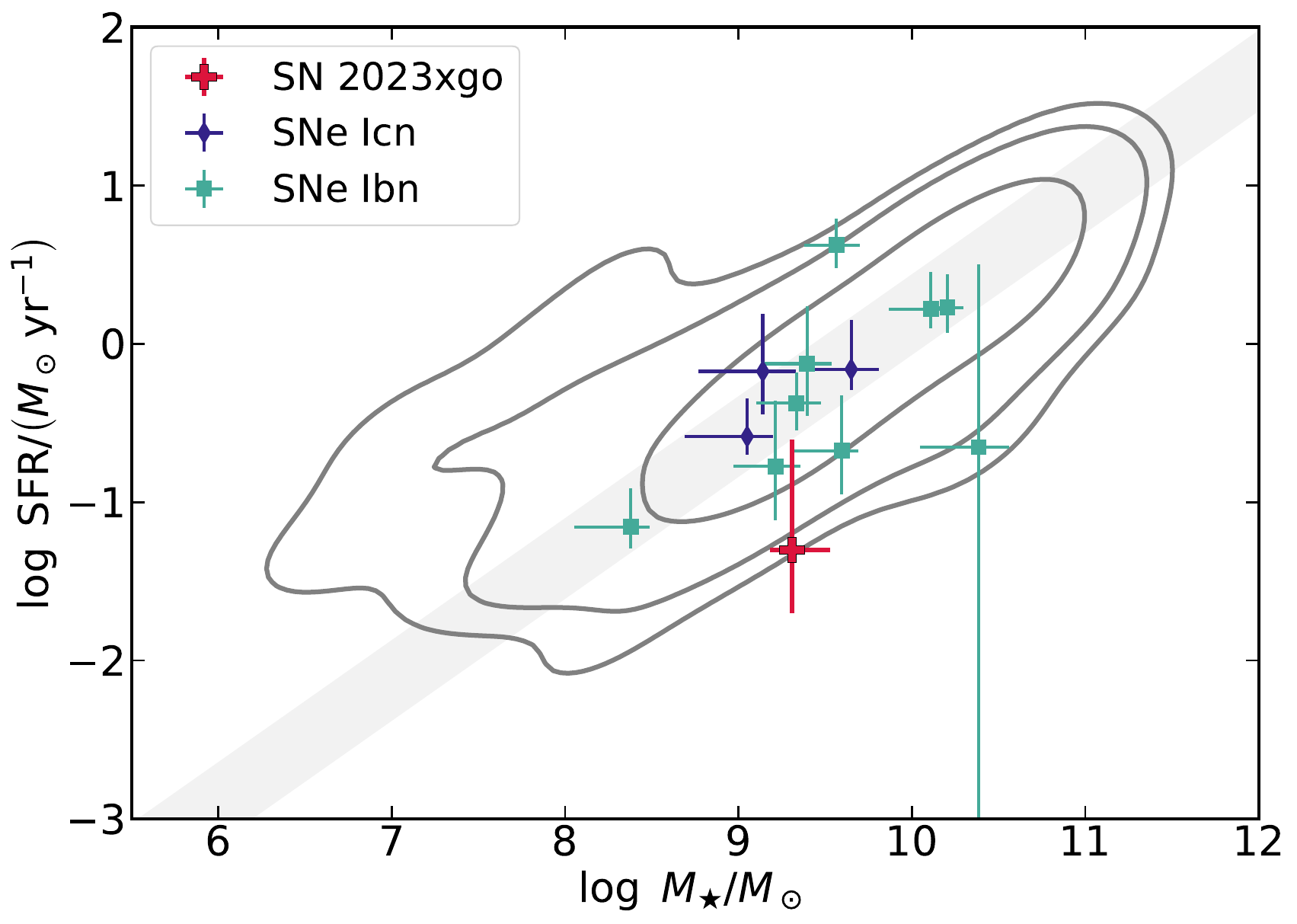}
	\end{center}
   \caption{The host galaxies of SN~2023xgo and the other Type Ibn/Icn SNe in the context of all core-collapse SN host galaxies from the Palomar Transient Factory (PTF). The grey contours are data from PTF \citep{Schulze2021a}, SNe~Ibn and SNe~Icn (colour-coded; the values of the SNe~Icn were taken from \citealt{GalYam2014NatureFlash, PerleyIcn, Pellegrino2022Icn}). The host of SN~2023xgo has fairly common properties for Type Ibn/Icn SNe.
}
    \label{fig:gal_sfr}
\end{figure}

\begin{figure*}
   \centering
   \includegraphics{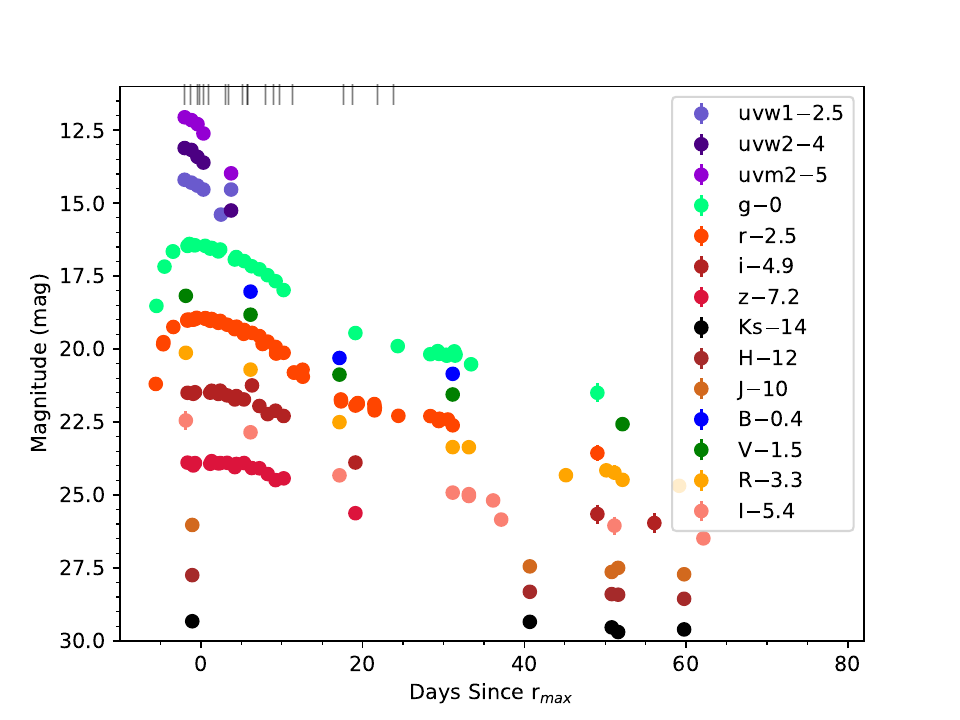}
   \caption{{\it uvBgVRIrizJHK} light-curve evolution of SN~2023xgo. The light-curves show a fast early rise to peak, a decline consistent with those of most SNe~Ibn, and then a flattening at late times. All magnitudes are plotted in AB. The vertical lines on top denote the phases at which spectra have been observed.}
    \label{fig:lc}
\end{figure*}

The host galaxy of SN~2023xgo is a low-mass ($\sim10^{9.3}\,\mathrm{M}_\odot$), moderately star-forming ($\sim0.05\,\mathrm{M}_\odot\,\mathrm{yr}^{-1}$), and subsolar-metallicity ($\sim0.7\,\mathrm{Z}_\odot$) dwarf system. These characteristics closely match observed host environments of Type Ibn SNe that have been linked to interacting binary progenitors via observational and modeling studies \citep[e.g.,][]{Sun2019, Wang2024, Moriya2025, QinZabludoff2024}. In particular, systems like SN~2006jc and SN~2015G show direct evidence for surviving companions in low-metallicity hosts, and population synthesis supports ultra-stripped helium stars in compact binaries as plausible progenitors in such galaxies.

%%%%%%%%%%%
%--------------------------------------------------------------------
\section{Photometric Evolution}
\label{phot}

\begin{figure}
	\begin{center}
		\includegraphics[width=\columnwidth]{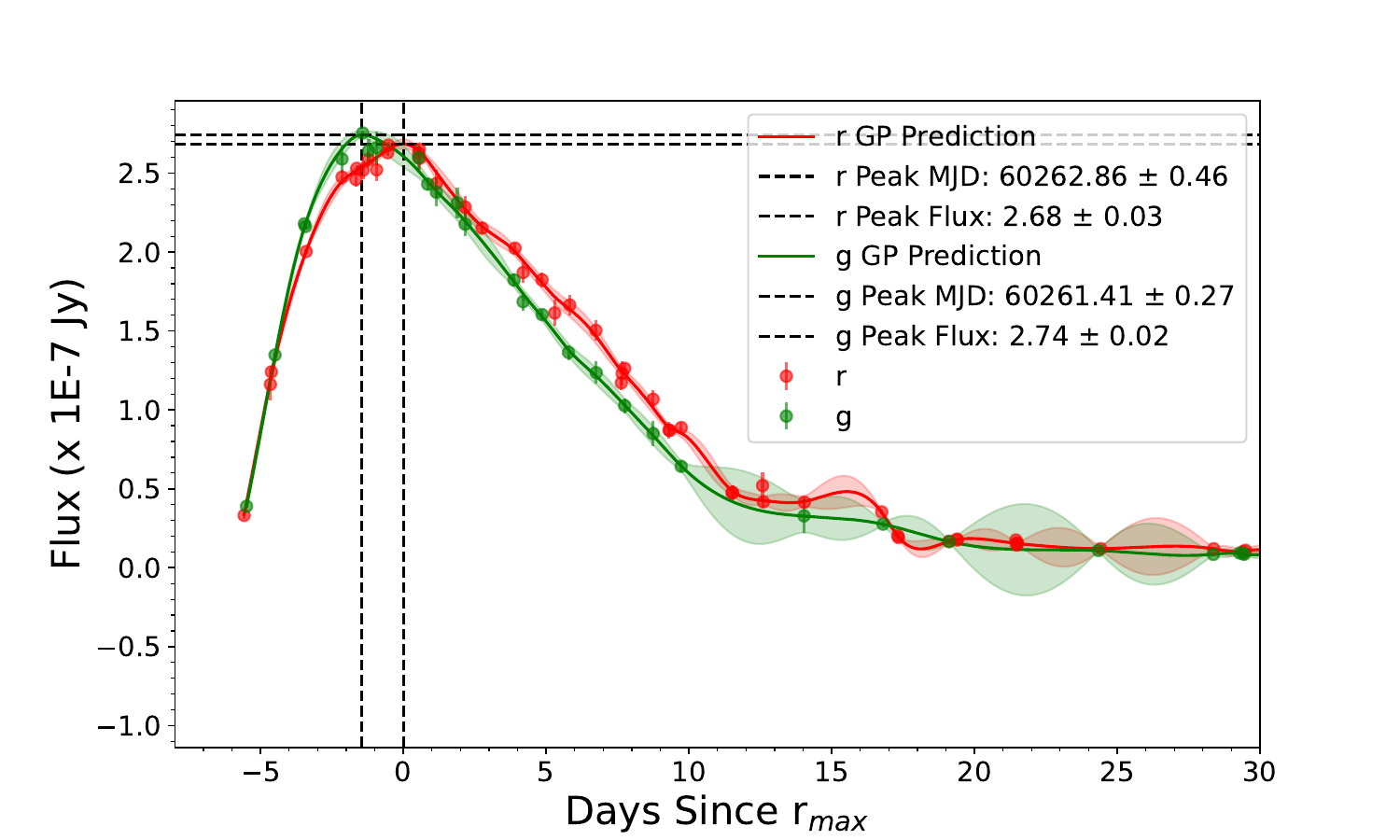}
	\end{center}
	\caption{Figure shows the Gaussian process interpolation to estimate the time of maximum and the peak magnitude in $g$ and $r$ bands.}
	\label{fig:lcfit_2023xgo}
\end{figure}

The complete light-curve evolution of SN~2023xgo is shown in Figure~\ref{fig:lc}. Although we probably missed the peak in the UV light curves and other broadband filters, we can trace the maxima in the $gri-$bands. 
The light curves of SN~2023xgo were observed in the near-infrared bands of $zJHK$ as well. To estimate the peak date and magnitude, we use the Gaussian Process (GP) interpolation, utilizing the \texttt{PYTHON} package \texttt{GEORGE} \citep{Ambikasaran2015} with a Matern 3/2 kernel (see Figure~\ref{fig:lcfit_2023xgo}). We used the interpolated $r-$band light curve to estimate the peak magnitude and to define the rise time as the time between the explosion epoch and the epoch of maximum. The $r-$band maximum is estimated to be MJD 60262.86 $\pm$ 0.46 at 16.43 $\pm$ 0.12~mag. Assuming the explosion epoch as inferred in Section~\ref{expl}, we find that the rise time of SN~2023xgo from explosion to maximum is 5.14 $\pm$ 2.30 days. This is well within the estimated values of rise times seen for a sample of SNe~Ibn \citep{Hosseinzadeh2017} and SNe~Icn \citep{Pellegrino2022Icn}. 

The $g$ and $r-$band light curves between $0-30$ days showed decline rates of $0.141 \pm 0.003$~mag~d$^{-1}$ and $0.139 \pm 0.004$~mag~d$^{-1}$, respectively. The sample of SNe~Ibn in \cite{Hosseinzadeh2017} are fast-evolving with a typical decline rate of 0.1~mag~d$^{-1}$ during the first month post-maximum. SN~2023xgo follows the same decay rate, but on the faster end of the distribution. We also notice a late-time flattening in all the light curves of SN~2023xgo, which is most likely due to a combination of CSM interaction and radioactivity (or some central powering source) contributing at late times.

\begin{figure}
	\begin{center}
	    \includegraphics[width=\columnwidth]{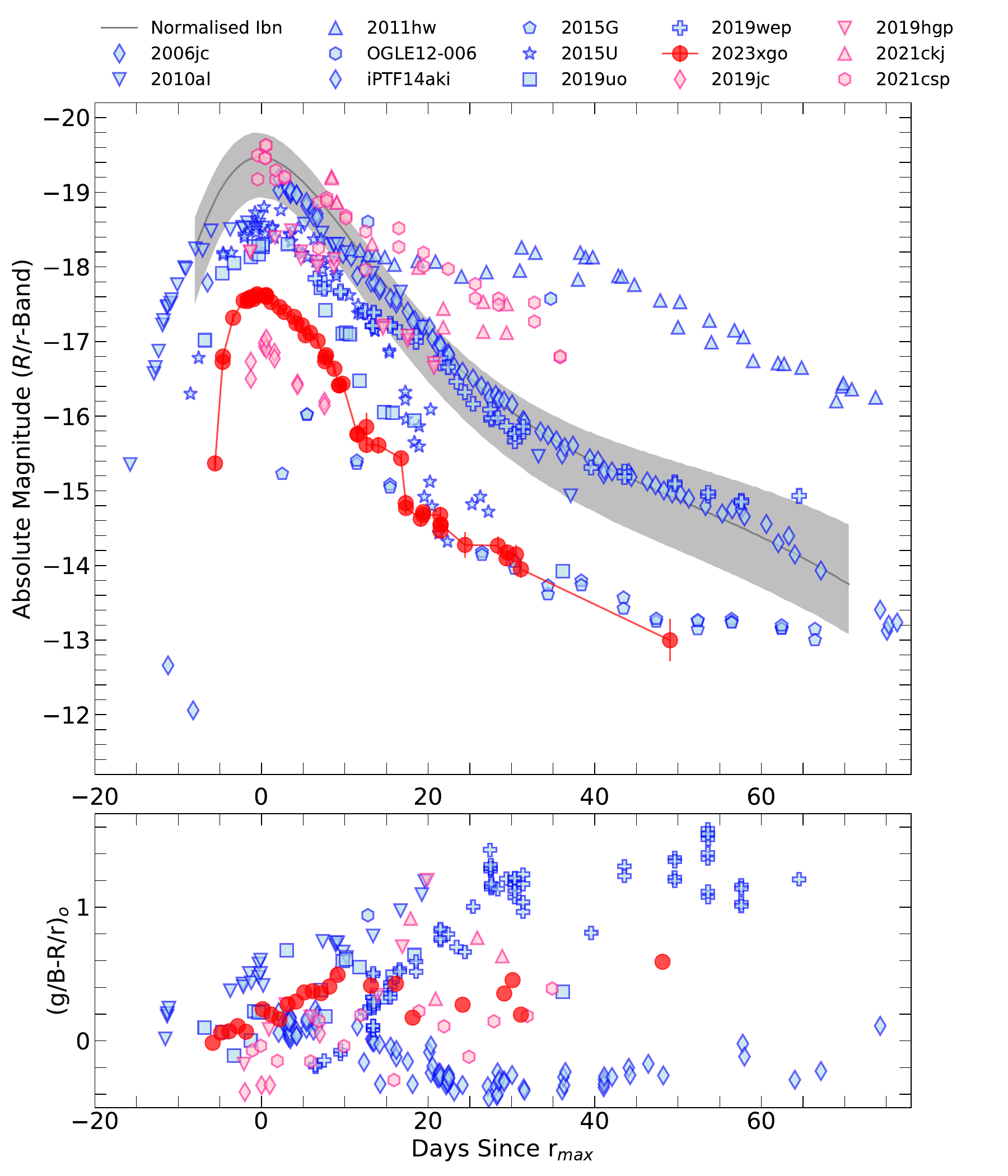}
	\end{center}
	\caption{Upper panel of the Figure shows the $r-$band light curve comparison of SN~2023xgo with a sample of SNe~Ibn (blue)\citep{Hosseinzadeh2017}. The SN~Ibn comparison sample includes SNe~2006jc \protect\citep{2007Natur.447..829P,2007ApJ...657L.105F}, 2010al \protect\citep{PastorelloSN2010al}, OGLE-SN-006 \protect\citep{Pastorello2015-OGLE}, 2011hw \protect\citep{PastorelloSN2010al}, iPTF14aki \protect\citep{Hosseinzadeh2017}, 2015U \protect\citep{2016MNRAS.461.3057S,Hosseinzadeh2017}, 2015G \protect\citep{Hosseinzadeh2017}, 2019uo \protect\citep{Gangopadhyay2020} and 2019wep \protect\citep{Gangopadhyay2022}. Phase is plotted in observer-frame days with respect to maxima for all the panels. The SN~Icn sample (red) has been taken from \protect\cite{Pellegrino2022Icn,PerleyIcn,Nagao2023}. Lower panel shows the colour of SN~2023xgo plotted with the comparison sample.}
	\label{fig:abscolor}
\end{figure}

Figure~\ref{fig:abscolor} shows the absolute magnitude light curve of SN~2023xgo along with those of other SNe~Ibn and Icn.
The absolute magnitude of SN~2023xgo is $-$17.65 $\pm$ 0.04~mag (corrected for host and galactic extinction), making it one of the faintest members among the SN~Ibn and SN~Icn comparison sample used in the paper. It is also $\sim$1.9~mag fainter than the normalized SN~Ibn light curve. The blue band in Figure~\ref{fig:abscolor} shows the average light curve (comprising 95$\%$ of the SN~Ibn data) taken from \cite{Hosseinzadeh2017}. The average light curve was generated using a Gaussian process to fit a smooth curve to the combined light curves of the sample of \cite{Hosseinzadeh2017}. The fit was performed in log–log space to ensure consistency and smoothness between the early and late-time light-curves.

\begin{figure*}
\begin{center}
	    \includegraphics[scale=0.50]{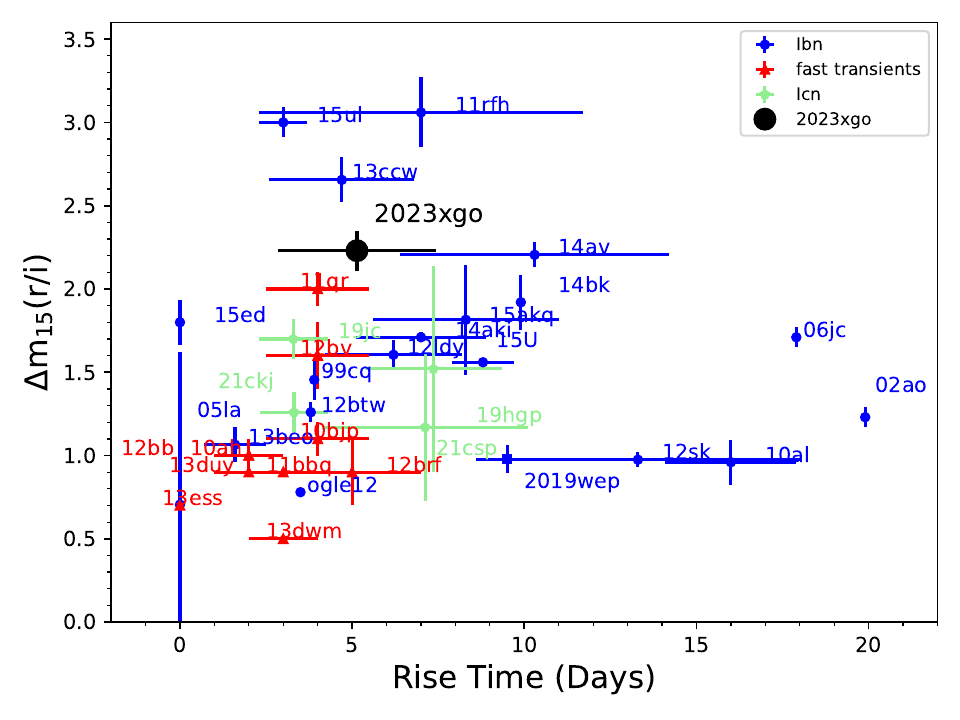}  
		\includegraphics[scale=0.50]{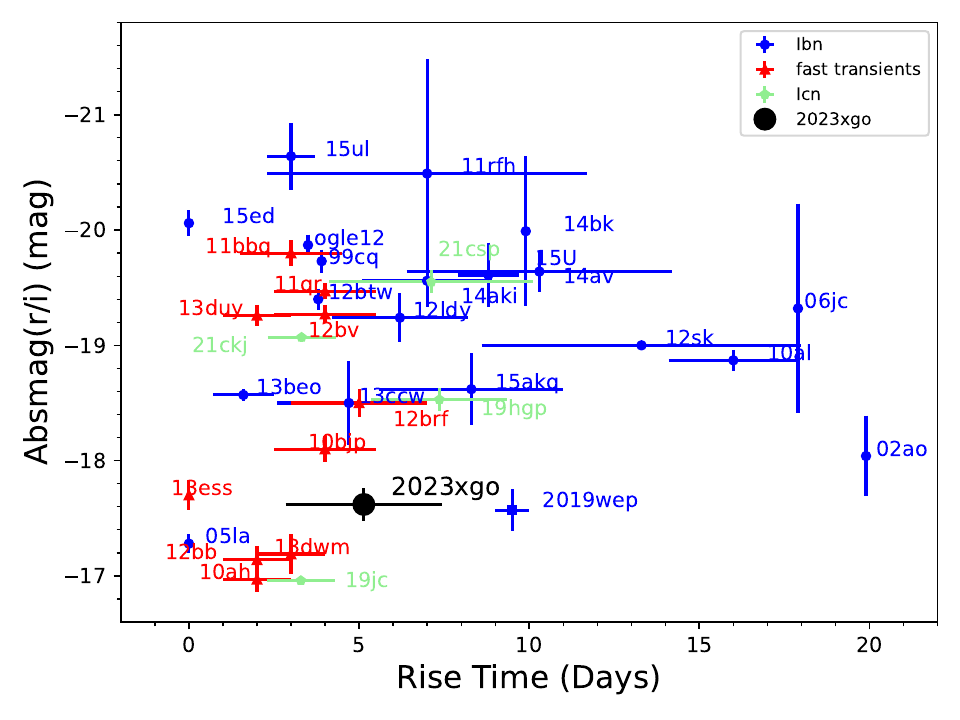}
\end{center}
	\caption{The $\Delta$m$_{15}$ vs rise time (left panel) and absolute magnitude vs rise time (right panel) correlation plots of SN~2023xgo with a sample of SNe~Ibn, Icn and fast transients. The sample of SNe~Ibn is from \protect\cite{Hosseinzadeh2017}, the SNe~Icn are taken from \protect\cite{Pellegrino2022Icn}, and the sample of fast transients is taken from \protect\cite{Drout2014}. The objects with ``0" values, have possibly no estimation of rise times.}
	\label{fig:correlationplots}
\end{figure*}

The lower panel of Figure~\ref{fig:abscolor} shows the colour evolution of SN~2023xgo compared with all other SNe~Ibn/Icn in the comparison sample. SN~2023xgo shows an initial blue-to-red transition in the colours. 
The ($g/B - R/r$) colour evolution of SN~2023xgo increases from $-0.01$~mag up to 0.59~mag $\sim$15 days post $r-$band max, subsequently remains flatter till $\sim$50 days. Even though the early colour evolution of SN~2023xgo is similar to most SNe~Ibn, the trend in the colour evolution of SN~2023xgo at late phases is similar to that of SN~2019wep, which is a SN showing Ibn-like features early on but looks like SN~Ib at later stages. The colour behaviour of SN~2023xgo is quite different from that of SN~2006jc, which shows a late-time flattening in the colour curve. All SNe~Icn in our sample also show almost similar evolution in the colour as does SN~2023xgo.

\begin{figure}
\begin{center}
	    \includegraphics[scale=0.55]{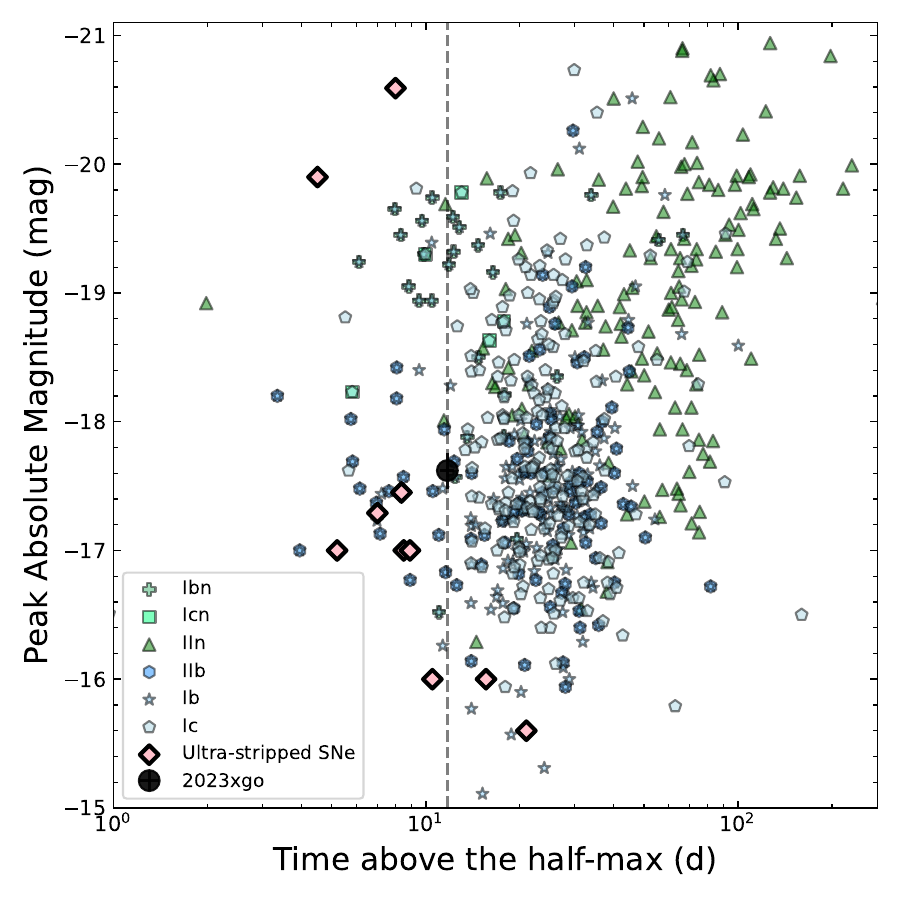} 
\end{center}
	\caption{We compare the time above half maximum and peak absolute magnitude of SN~2023xgo with all other interacting and stripped-envelope SNe from the BTS experiment \protect\citep{Perley2020}. For sources with two peaks in the $r-$band light curve, the second peak is used. We also compare SN~2023xgo with a group of fast-evolving transients that are thought to be ultra-stripped SNe, and the comparison sample used is same as from \protect\cite{Das2024_zaw}. The dashed vertical line indicates SN~2023xgo in that phase-space.}
	\label{fig:correlationplot2}
\end{figure}

Fast transients are a group of objects that rise and fade in brightness on time scales much shorter than those of typical SNe \citep{Drout2014,Pellegrino2022Icn,MaedaMoriya2022}. The progenitor systems include massive WR stars undergoing core collapse, compact object mergers such as neutron star–black hole interactions, and ultra-stripped SNe resulting from binary evolution. These scenarios reflect the diversity of mechanisms leading to rapidly evolving luminous events, and they display spectra with a featureless blue continuum similar to those of most SNe~Ibn/Icn. \cite{Ho2023}, using the sample of fast transients from ZTF, found an interesting connection between these fast transients and SNe~Ibn/Icn. We therefore made the comparison of SNe~Ibn with a group of fast transients. Figure~\ref{fig:correlationplots} shows the time scales and luminosities of SNe~Ibn/Icn and fast transients, where $\Delta$m$_{15}$ (which is the decay rate from peak to 15 days post maximum) and absolute magnitude are plotted against the rise time. The rise time, here, is calculated as the time between explosion and maximum light. The SNe for which explosion epochs are not constrained, we estimate the time as the difference between discovery and the maximum light, which gives a lower limit on the rise times typically. The plot indicates that the rise time of SN~2023xgo is longer than the fast transients but is well consistent with the rise time estimates of other SNe~Ibn and SNe~Icn. The $\Delta$m$_{15}$ value matches well with SN~2014av and is on the higher end of SNe~Ibn/Icn. Although rise times indicate that SN~2023xgo lies at the faster end of the SNe~Ibn/Icn sample, still, the most remarkable feature is still that it is among the faintest of our SNe~Ibn/Icn comparison sample. In fact, SN~2023xgo is fainter than for most of the fast transients. SN~2023xgo, thus, stands out to be least luminous but fast decaying member of SNe~Ibn/Icn among our comparison sample.

We also compare the time above the half-maximum with the peak absolute magnitudes for all the groups of SNe from the Bright Transient Survey sample taken from the BTS explorer (SNe~IIn, IIb, Ib, Ic, Ibn, Icn; \citealp{Perley2020}) to see where it stands out in that space (see Figure~\ref{fig:correlationplot2}). Motivated by the fact that the SN~2023xgo shows a fast rise and decay observationally, and also shows low ejecta masses and Ni masses (from light-curve modelling; see Section~\ref{section:LC_fit}), we also compare it with the sample of fast transients that could be ultra-stripped from \cite{Das2024_zaw}. SN~2023xgo also shows a very unique behaviour in this space. Even though it is less luminous than all the interacting SNe of our comparison sample, the time above maximum are similar to those of SNe~Ibn/Icn except SN~2005la. SN~2023xgo is slower than most of the usual ultra-stripped fast transients, but is faster than almost the entire stripped envelope population and most SNe~IIn. This hints towards the possibility that even if the SN is powered by CSM interaction, the contribution of CSM is not as much as seen in other SNe~Ibn, and it could also be a little bit of radioactivity contributing to the light curve of SN~2023xgo.

\section{Bolometric \& Multi-band Light-curve modelling}\label{section:LC_fit}

To construct the (pseudo-)bolometric light curve of SN~2023xgo from UV to IR, we used the \texttt{SuperBol} code \citep{nicholl2018_superbol}. The missing UV and NIR data were supplemented by extrapolating the Spectral Energy Distributions (SEDs) using the blackbody approximation, and then we used the direct integration method as described in \cite{2017PASP..129d4202L}. The UV extrapolation on the blue end was restricted up to a wavelength of 3000 \AA.

\begin{figure}
	\begin{center}
		\includegraphics[width=\columnwidth]{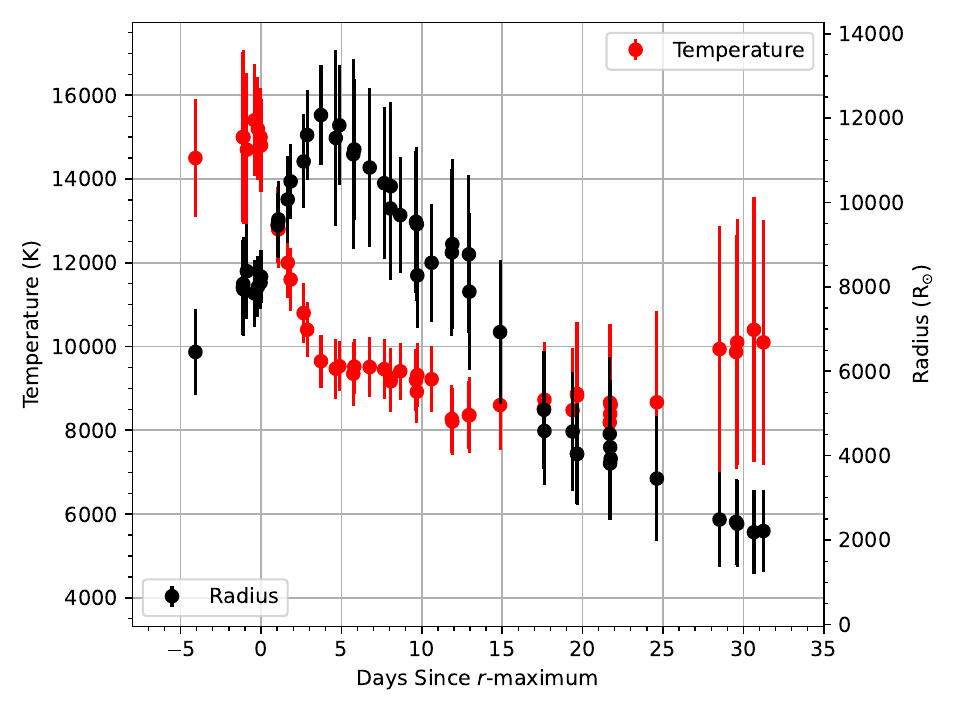}
	\end{center}
	\caption{The radius and temperature evolution of SN~2023xgo. SN~2023xgo shows a temperature variation between 14500~K to 15500~K, till maximum light, and then the temperature decreases to 9500~K and becomes flatter thereafter. The radius follows the same pattern as light-curve evolution with some phase offset, i.e., increasing to maximum light and decreasing thereafter.}
	\label{fig:rad-temp}
\end{figure}

Typically, as the SN ejecta expand, the shock breakout from the surface of the progenitor is followed by a rapid cooling phase due to the rapid expansion driven by the shock \citep{falk1977}. This would lead to a increasing photospheric radius and a decrease in the temperature of the SN ejecta within a couple of hours of the explosion. 
However, for interacting SNe this is not always the case as the ejecta are masked by the CSM \citep{Irani2024}. Figure~\ref{fig:rad-temp} shows the radius and temperature evolution of SN~2023xgo. The temperature and radius are obtained from the blackbody fits, including the UV to NIR data. The Stefan-Boltzmann law was used to estimate the luminosity, temperature and radius. SN~2023xgo first shows a constant temperature varying between 14500~K and 15500~K till maximum light. Following maximum light, the temperature declines to approximately 9500~K and subsequently remains roughly constant within the uncertainties. However, the increasing scatter and deviation from a smooth evolution suggest that the spectrum deviates from a simple blackbody approximation at these phases. The early high temperature indicates an injection of energy in the cooling ejecta, perhaps due to interaction with a dense CSM.

The blackbody radius of SN~2023xgo increases from approximately 6200~$R_{\odot}$ to 12000~$R_{\odot}$ at maximum light. From maximum light until 24 days past maximum, the radius decreases from about 12000 $R_{\odot}$ to 5000 $R_{\odot}$.

\subsection{Semi-analytical modelling}

While the decay of radioactive nickel synthesized in the explosion is the dominant powering mechanism for the light curves of most SESNe, that is not always the case for SNe~Ibn. This is obvious for objects with extremely long-lived light-curves \citep[e.g.,][]{Karamehmetoglu2017,Kool2021}, but is also generally true for the more luminous and rapidly evolving population \citep{Ho2023,Pellegrino2022Icn}. SN~2023xgo is, however, among the least luminous SNe Ibn/Icn, and therefore an investigation on how much radioactivity or interaction contributes to the overall light curve is warranted.
We fitted the multi-band light-curves of SN~2023xgo with a suite of different models, to probe the powering mechanisms at play. The modelling was carried out with \texttt{REDBACK}, the Bayesian inference package for modelling transients \citep{sarin_redback}. In the subsequent modelling, we used a Gaussian likelihood. We fitted a CSM interaction-powered semi-analytic model \citep{Chatzopoulos12,Chatzopoulos2013}, as well as a combination of that model with a radioactivity-powered analytic model \citep{Arnett1982}. 
Each model was fitted to the entire photometric data-set (UV/optical/IR), corrected for extinction using the estimated A$_{V}=0.49\pm0.01$ mag 
(Section~\ref{sec:ext}). We explored the parameter space with the nested sampling package \texttt{dynesty} \citep{Ashton2019,Speagle2020}. 

\subsubsection{Radioactivity}
\label{arnett}
The main source driving the luminosity of a SN is often the radioactive decay of $^{56}$Ni to $^{56}$Co to $^{56}$Fe. We started by assuming the radioactive decay of $^{56}$Ni as the dominant source of energy. We fitted an \cite{Arnett1982} model to the multiband lightcurves. The Arnett model assumes spherical symmetry and a homologous expansion of the ejecta.
%, spherical symmetry and density proportional to the radius of the ejecta. 
Table~\ref{table:Arnett_fitting} contains the priors used for the inference. We fixed the optical (electron scattering), and gamma-ray opacities to the fiducial values of $\kappa_{o}=0.07$~cm$^{2}$~g$^{-1}$ \citep{Valenti2008} and $\kappa_{\gamma}=0.03$~cm$^{2}$~g$^{-1}$ \citep{Colgate1969}, respectively. 

\begin{figure*}
   \centering
   \includegraphics[scale=0.5]{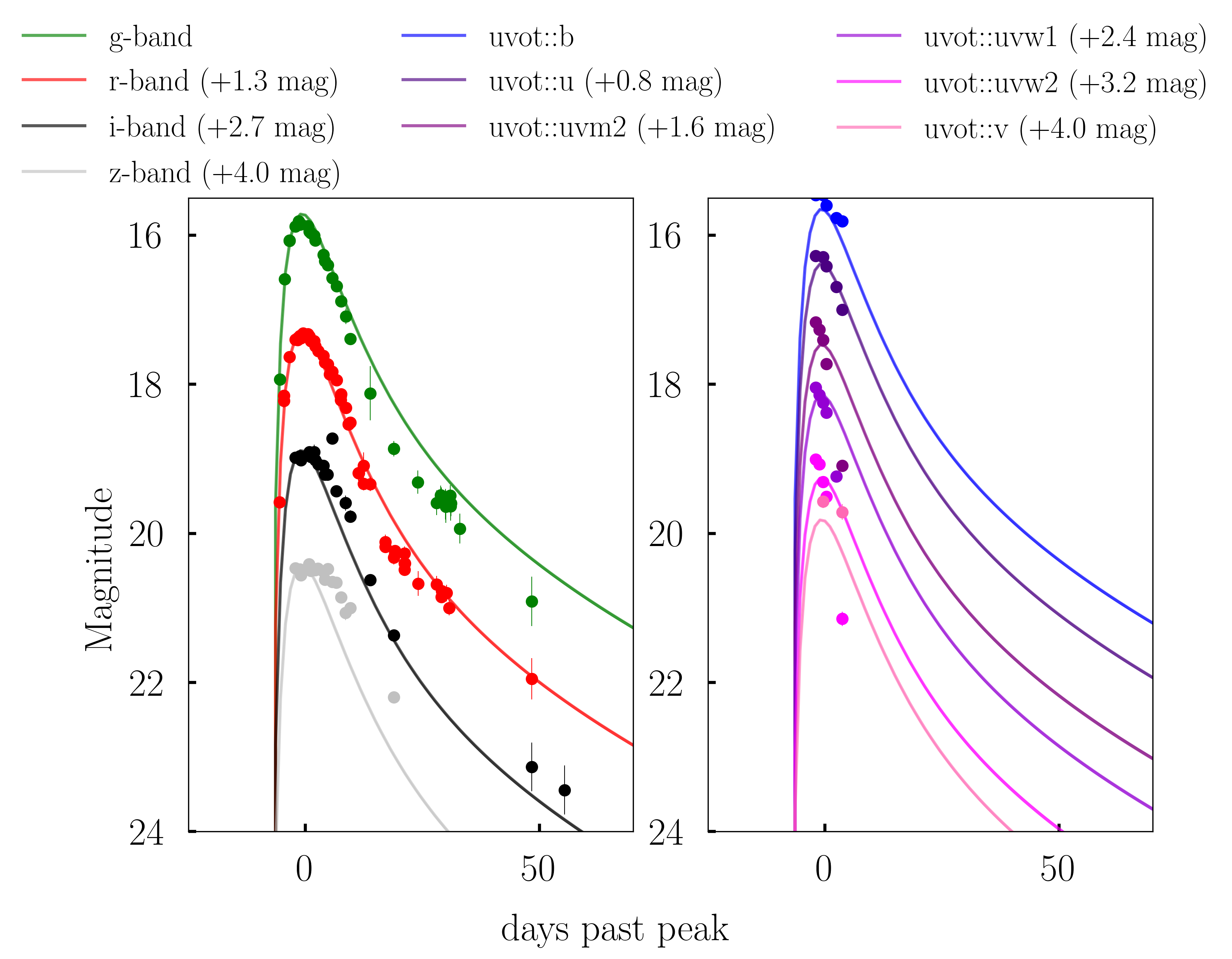}
   \caption{Multi-band lightcurve fitting of SN~2023xgo with radioactivity powered model by \protect\cite{Arnett1982} using \texttt{REDBACK}. We see that radioactivity fairly reproduces our observed lightcurve evolution.}
   \label{fig:Arnett_model_lc}
\end{figure*}

The %employed 
free parameters of the $^{56}$Ni model are the ejecta mass M$_{\rm ej}$ and the $^{56}$Ni mass M$_{\rm Ni}$. Following \cite{Arnett1982}, $\tau_m$ determines the width of the bolometric light curve, which is also a free parameter and can be expressed in terms of the optical opacity $\kappa_{o}$, ejecta mass M$_{ej}$, and the photospheric velocity at luminosity peak v$_{ph}$: \\
\begin{equation}
\tau_m = \sqrt{2} \left( \frac{\kappa_{o}}{\beta c} \right)^{\frac{1}{2}} \left( \frac{M_{\text{ej}}}{v_{\text{ph}}} \right)^{\frac{1}{2}}
\end{equation}

where $\beta=13.8$ is a constant of integration \citep{Arnett1982} and $c$ is the speed of light. The kinetic energy for spherically symmetric ejecta with a uniform density is 

\begin{equation}
E_k = \frac{3}{10} M_{\text{ej}} v_{\text{ph}}^2.
\end{equation}

Figure~\ref{fig:Arnett_model_lc} shows the best fitted Arnett model to the multiband lightcurves of SN~2023xgo and the cornet plot is shown in Figure~\ref{fig:arnett_model_cornerplot}. It can account for the peak but slightly overestimates the decline rate. Nevertheless, the model of plain radioactive decay is good enough to approximate the entire lightcurve. Under the assumptions of the model, the inferred explosion parameters are listed in Table~\ref{table:Arnett_fitting}. The Arnett model yields an ejecta mass of M$_{\mathrm{ej}}\simeq0.35~\rm M_{\odot}$ with M$_{\mathrm{Ni}}\simeq0.29~\rm M_{\odot}$. A significant fraction of the ejected mass is thus required to be nickel (exceeding 50\%) for the Arnett model alone to reproduce the lightcurve. Although the multiband light curves are fairly well reproduced by this $^{56}$Ni model, this very high fraction of nickel and the fact that we do see interaction signatures in the spectral evolution of SN~2023xgo make us consider other models. In particular, we also fitted a model with CSM-interaction in addition to radioactivity to the multiband light curves.

\begin{table}
\caption{Table shows the best-fit values obtained by fitting the equational form of \protect\cite{Arnett1982} assuming blackbody approximation. We see that radioactivity fairly reproduces our observed light curves in all the bands.}
\label{table:Arnett_fitting}
\centering
\begin{tabular}{lll}
\hline
\multicolumn{1}{c}{Parameters} & \multicolumn{1}{c}{Priors} & \multicolumn{1}{c}{Best fitted values} \\[2pt]
\hline
z & 0.01325 & 0.01325 \\[2pt] 
$\kappa_{o}$~[cm$^{2}$~s$^{-1}$] & 0.07 & 0.07 \\[2pt]
$\kappa_{\gamma}$~[cm$^{2}$~s$^{-1}$] & 0.03 & 0.03 \\[2pt]
f$_{\mathrm{Ni}}$ & $\log \mathcal{U}~[10^{-3}, 0.999]$ & $0.83^{+0.01}_{-0.01}$ \\[5pt]
M$_{\mathrm{ej}}~ [M_{\odot}]$ & $\log \mathcal{U}~[0.25, 15]$ & $0.35^{+0.01}_{-0.01}$\\[5pt]
v$_{\mathrm{ej}}$$~${[}km$~$s$^{-1}${]} & $\log \mathcal{U}~[10^{3}, 5 \times 10^{4}]$ & $17241.5^{+210.3}_{-206.5}$ \\[5pt]
T$_{\mathrm{floor}}$~[K] & $\log \mathcal{U}~[10^{3},  10^{5}]$ & $15215.8^{+56.7}_{-53.2}$  \\[5pt]
t$_{0}$~[MJD] & $\mathcal{U}~[t_{\mathrm{detection}}-70,  t_{\mathrm{detection}}-1]$ & $60255.81^{+0.02}_{-0.02}$   \\[5pt] 
\hline
\end{tabular}
\end{table}

\subsubsection{CSM}
\label{csm}
Since we wanted to infer whether our light curves are driven by CSM interaction only, we try to model our light curves with the semi-analytic model for CSM interaction from \cite{Chatzopoulos12}. We restricted the parameter space of our Bayesian inference in accordance with the following constraints: (i) the CSM photosphere would always be inside the CSM envelope and (ii) the diffusion time scale would always be smaller than the shock crossing time scale \citep[two central assumptions in the work by][]{Chatzopoulos12}. 

\begin{figure*}
   \centering
   \includegraphics[scale=0.5]{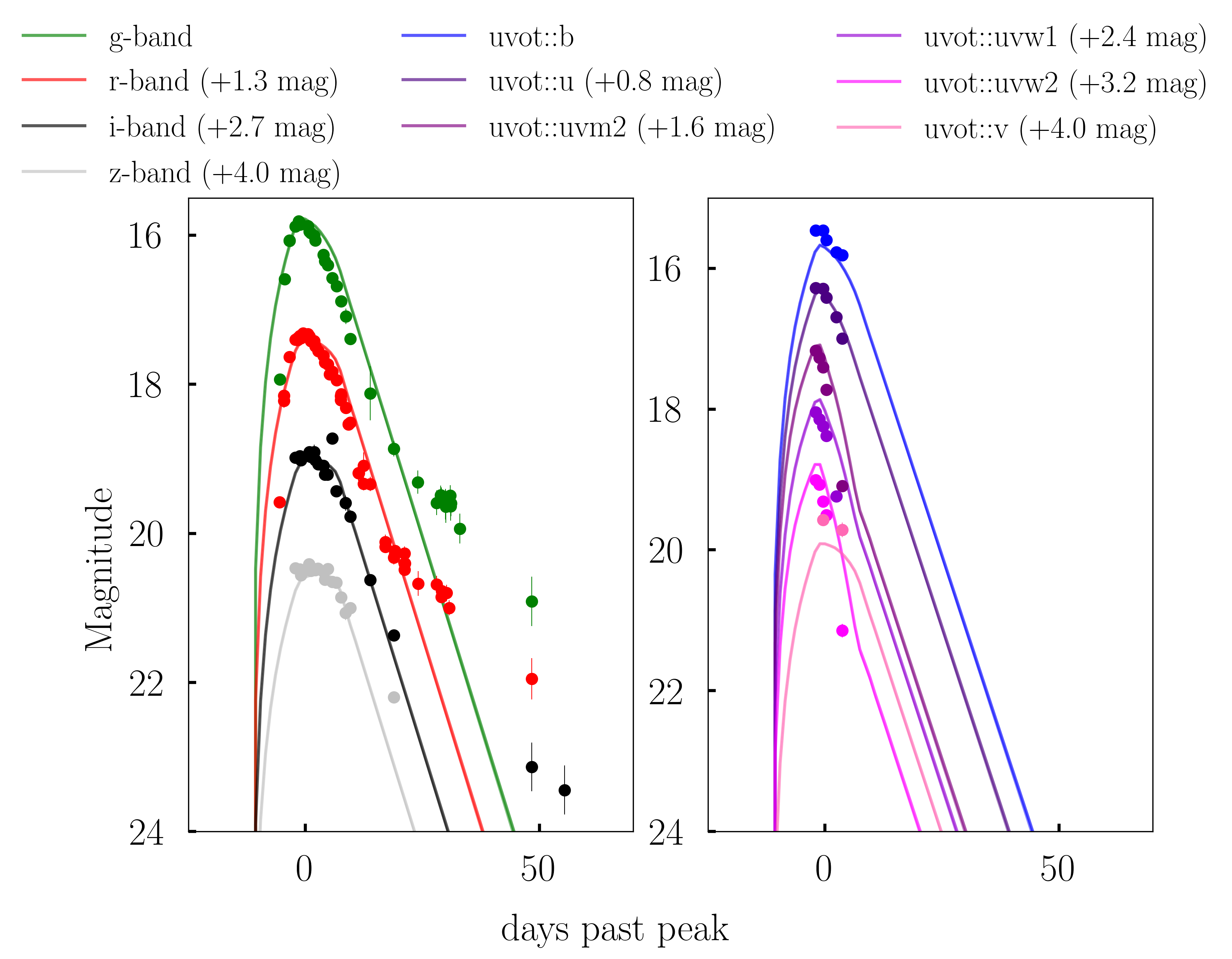}
   \caption{Multi-band light-curve modelling using the CSM model by \protect\cite{Chatzopoulos12} as implemented in \texttt{REDBACK}. We see that SN~2023xgo is well reproduced using a CSM mass of 0.43~M$_{\odot}$ and a ejecta mass of 0.10~M$_{\odot}$. However, the model was not able to reproduce the late-time light-curve after 30 d past maximum.}
   \label{fig:CSM_model_lc}
\end{figure*}

In the \texttt{csm\_interaction} model (the \texttt{REDBACK} implementation of the \citealt{Chatzopoulos12} treatment), it is assumed that the SN luminosity results from the conversion of kinetic energy from both the forward and reverse shock into heating. We fixed the opacity $\kappa_{o}=0.07$~cm$^2$~g$^{-1}$ and the gamma-ray opacity to $\kappa_{\gamma}=0.03$~cm$^2$~g$^{-1}$. We take the rest of the parameters as fitting parameters. 

For the ejecta mass (CSM mass), we consider uniform priors from $0.1-15$~M$_{\odot}$ ($0.001-15$~M$_{\odot}$) based on the values obtained by \cite{Pellegrino2022Ibn} for a sample of SNe~Ibn. For the inner radius of the CSM (R$_0$), we consider a uniform prior from $0.1-100$~AU, and for the CSM density at the inner radius ($\rho$), we consider a log-uniform prior for densities from $10^{-15} - 10^{-9}$ g cm$^{-3}$. The prior on the CSM density profile index, $\eta$, was chosen as a uniform distribution between $\eta=0$ (shell-like) and $\eta=2$ (wind-like). 

The best fitted parameters are listed in \autoref{table:CSM_fitting}. The $1\sigma$ errors listed represent only the statistical uncertainty associated with the fitting and therefore are not indicative of the model uncertainties. It is important to note that some parameters (like M$_{\mathrm{ej}}$ and r$_{0}$) are ``pushing'' the limits of the parameter space (see Table~\ref{table:CSM_fitting} and Figure~\ref{fig:csm_model_cornerplot}). The inference prefers a model with parameters outside the realistic ranges provided in the priors. Nevertheless, interpretation of the inferred parameters should be taken with caution in such cases.

\begin{table}%[]
\caption{Table shows the best fit light-curve parameters from the CSM model.}
\label{table:CSM_fitting}
\centering
\begin{tabular}{lll}
\hline
\multicolumn{1}{c}{Parameters} & \multicolumn{1}{c}{Priors} & \multicolumn{1}{c}{Best fitted values} \\ [2pt]
\hline
M$_{\mathrm{ej}}$~[M$_{\odot}]$ & $\log \mathcal{U}~[0.1, 15]$ & $1.0005^{+0.0001}_{-0.0001}\times 10^{-1}$ \\[5pt]
v$_{\mathrm{ej}}$~[km~s$^{-1}$] & $\log \mathcal{U}~[10^{3}, 5\times 10^{4}]$ & $4966.2^{+7.1}_{-17.6}$ \\[5pt]
M$_{\mathrm{CSM}}$~[M$_{\odot}]$ & $\log \mathcal{U}~[0.001, 15]$ & $0.435^{+0.005}_{-0.001}$ \\[5pt]
T$_{\mathrm{floor}}$~[K] & $\log \mathcal{U}~[10^{2},  10^{4}]$ & $9971.3^{+22.2}_{-56.8}$ \\[5pt]
t$_{0}$~[MJD] & $\mathcal{U}~[t_{\mathrm{detection}}-6,  t_{\mathrm{detection}}]$ & $60251.220^{+0.001}_{-0.001}$ \\[5pt]
$\eta$ & $\mathcal{U}~[0, 2]$ & $1.7^{+2.3}_{-1.2}\times 10^{-3}$ \\[5pt]
$\rho$~[g~cm$^{-3}$] & $\log \mathcal{U}~[10^{-15}, 10^{-9}]$ & $6.58^{+0.07}_{-0.04} \times 10^{-12}$ \\[5pt]
r$_{0}$~[AU] & $\log \mathcal{U}~[10^{-1}, 10^{2}]$ & $1.01^{+0.01}_{-0.01}\times 10^{-1}$ \\[5pt]
\hline
\end{tabular}
\end{table}

We see that about 0.435 M$_{\odot}$ of CSM mass reproduces our observed light curves with an ejecta mass of 0.10 M$_{\odot}$ and a radius of CSM at 0.10~AU. Previous theoretical and observational studies of ultra-stripped SNe have shown such low ejecta and CSM masses, as well as small CSM radii mostly for low-mass He stars in binary companion \citep{Das2024_zaw,Wu2022}

The mass-loss rate estimated using the best fit parameters for our model is given by \begin{equation}
    \dot{M} = 4 \pi R_{\text{CSM}}^2 \rho_{\text{CSM}} v_w
\end{equation}

where \( \dot{M} \) is the mass-loss rate, \( R_{\text{CSM}} \) is the radius of the CSM, \( \rho_{\text{CSM}} \) is the density of the CSM, and \( v_w \) is the wind velocity. Using the values from Section~\ref{vel-eqw}, the v$_{w}=1800$~km~s$^{-1}$ as measured from the spectrum around maximum light (see subsection~\ref{vel-eqw}), the mass-loss rate is estimated to be 5.3 $\times$ 10$^{-4}$~M$_{\odot}\,{\rm yr}^{-1}$ . This is consistent with what we obtain by matching our values from the models of \cite{BoianGroh2019} in subsection~\ref{boaingroh}. The light-curve modelling favours a shell like scenario (see $\eta$ from Table~\ref{table:CSM_fitting}) here with shell expelled few hours before explosion. This timescale is estimated from the radius and velocity listed in Table~~\ref{table:CSM_fitting}. Since this fit was not able to reproduce the late-time light-curve, and we see ejecta signatures start appearing in the spectra at 23.48~d, we further investigated whether radioactivity also plays a role in governing the light-curve evolution of SN~2023xgo. 

\subsubsection{CSM+Ni}
\label{csmni}
Given the emergence of broad P-Cygni features in the spectral evolution of SN~2023xgo and the inability of a pure CSM interaction model to reproduce the late-time light-curve, we adopted a hybrid approach combining the Arnett model \citep{Arnett1982} with the semi-analytic CSM interaction framework from \cite{Chatzopoulos12}. The Arnett model assumes that the main source driving the luminosity of a SN is the radioactive decay of $^{56}$Ni to $^{56}$Co to $^{56}$Fe, assumes spherical symmetry, and a homologous expansion of the ejecta. Thus, along with the assumptions referred in subsection~\ref{csm}, we included radioactivity as a powering source to reproduce the overall light curve of SN~2023xgo.

While fitting the combined radioactivity and the CSM interaction model, we find that when the data are fitted with prior parameters not restricted further than the fiducial ranges that they can take, the radioactive model dominates the luminosity output. We therefore decided to constrain $M_{\mathrm{Ni}}$ to allow for some appreciable contribution from the \cite{Chatzopoulos12} model. We assumed that M$_{\mathrm{Ni}}$ cannot take values larger than 0.04~M$_{\odot}$, motivated by the typical values of M$_{\rm Ni}$ reported for the sample of \cite{Pellegrino2022Ibn}. 
The chosen priors and the best-fit values of the model's free parameters are listed in the second and third columns of Table~\ref{table:CSMplusNi_fitting}, respectively.

We fitted the multi-band light-curves of SN~2023xgo~with the \texttt{csm\_nickel} model. All the priors for the CSM interaction models are same as in subsection~\ref{csm}. In addition, we consider a uniform prior from $0.002-0.6$ for the nickel fraction in the ejecta (f$_{\mathrm{Ni}}$ = M$_{\mathrm{Ni}}$ / M$_{\mathrm{ej}}$).

\begin{figure*}
   \centering
   \includegraphics[scale=0.5]{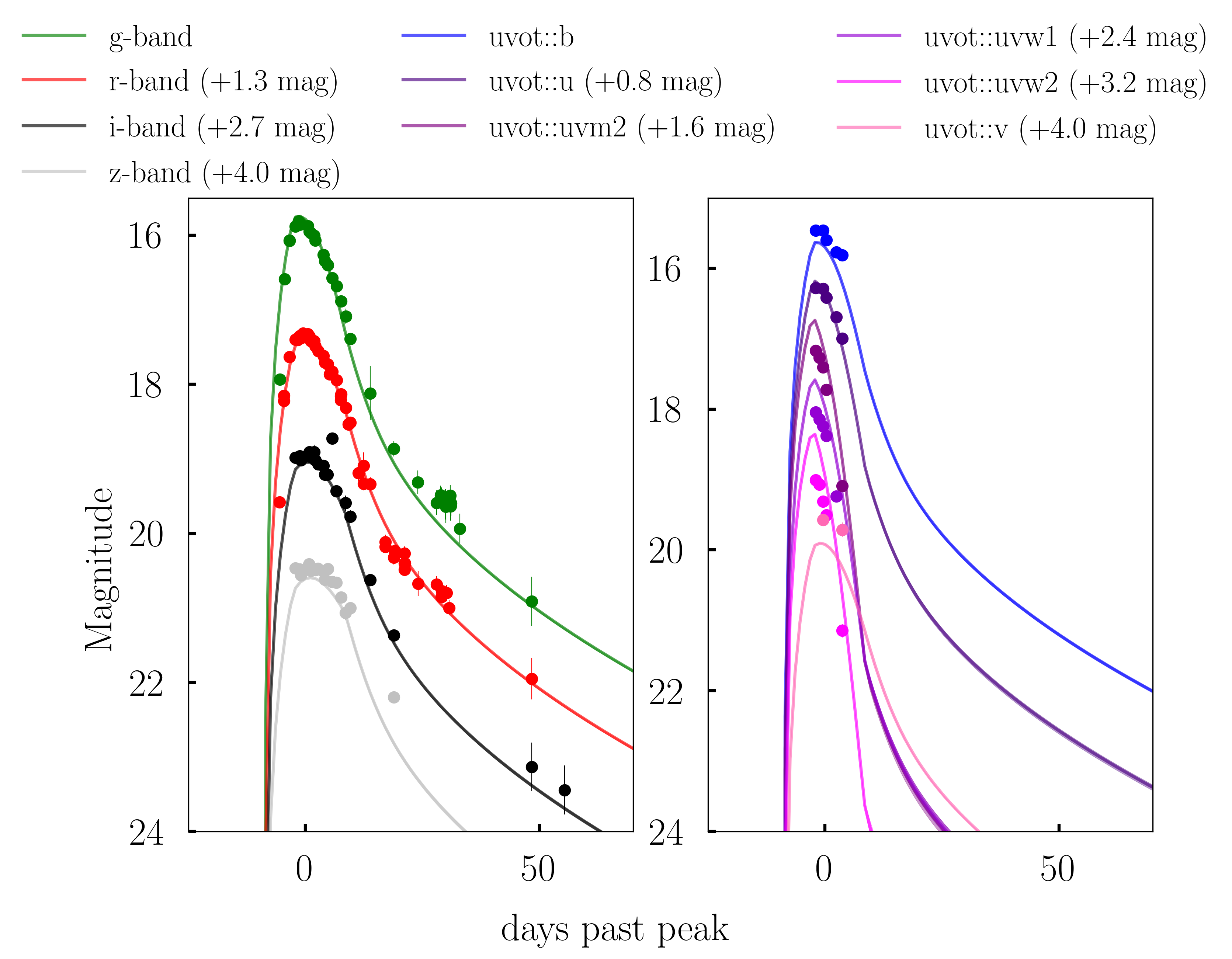}
   \caption{Multi-band light-curve modelling using CSM+Ni by \protect\cite{Chatzopoulos12} using \texttt{REDBACK}. We see that SN~2023xgo is well reproduced using a $^{56}$Ni mass of 0.04~M$_{\odot}$ (fixed), a CSM mass of 0.22~M$_{\odot}$ and a ejecta mass of 0.12~M$_{\odot}$.}
   \label{fig:CSM-Ni_model_lc}
\end{figure*}

The best-fitting parameters are listed in \autoref{table:CSMplusNi_fitting} and the best-fit light-curves are shown in Figure~\ref{fig:CSM-Ni_model_lc}. The $1\sigma$ errors listed represent only the statistical uncertainty associated with the fitting and therefore are not indicative of the model uncertainties.

\begin{table}%[]
\caption{Table shows the best-fit light-curve parameters from the CSM+Ni model. We note that the light curves are fairly reproduced by a high Ni mass and low ejecta and CSM masses than other SNe~Ibn/Icn.}
\label{table:CSMplusNi_fitting}
\centering
\begin{tabular}{lll}
\hline
\multicolumn{1}{c}{Parameters} & \multicolumn{1}{c}{Priors} & \multicolumn{1}{c}{Best fitted values} \\ [2pt]
\hline
f$_{\mathrm{Ni}}$ & $\log \mathcal{U}~[0.002, 0.6]$ & $0.324^{+0.001}_{-0.002}$ \\[5pt]
M$_{\mathrm{ej}}~$[M$_{\odot}]$ & $\log \mathcal{U}~[0.1, 15]$ & $0.1230^{+0.0002}_{-0.0002}$ \\[5pt]
f$_{\mathrm{Ni}} \cdot$ M$_{\mathrm{ej}}~$[M$_{\odot}]$ & $\log \mathcal{U}~[0.001, 0.04]$ & $3.99^{+0.01}_{-0.03}\times 10^{-2}$ \\[5pt]
E$_{\mathrm{kin}}$~[erg] & $\log \mathcal{U}~[5 \times10^{49}, 10^{52}]$ & $5.005^{+0.068}_{-0.037} \times 10^{49}$ \\[5pt]
M$_{\mathrm{CSM}}$~[M$_{\odot}]$ & $\log \mathcal{U}~[0.001, 15]$ & $22.147^{+0.003}_{-0.004} \times 10^{-2}$ \\[5pt]
T$_{\mathrm{floor}}$~[K] & $\log \mathcal{U}~[10^{2},  10^{4}]$ & $7127.0^{+90.6}_{-98.1}$ \\[5pt]
t$_{0}$~[MJD] & $\mathcal{U}~[t_{\mathrm{detection}}-6,  t_{\mathrm{detection}}]$ & $60254.03^{+0.01}_{-0.01}$ \\[5pt]
$\eta$ & $\mathcal{U}~[0, 2]$ & $3.3^{+4.4}_{-2.4}\times 10^{-3}$ \\[5pt]
$\rho$~[g~cm$^{-3}$] & $\log \mathcal{U}~[10^{-15}, 10^{-9}]$ & $4.8^{+0.1}_{-0.1} \times 10^{-12}$ \\[5pt]
r$_{0}$~[AU] & $\log \mathcal{U}~[10^{-1}, 10^{2}]$ & $0.104^{+0.006}_{-0.003}$ \\[5pt]
\hline
\end{tabular}
\end{table}

From our estimated parameters, with the maximum allowed Ni mass of $\sim$0.04~M$_{\odot}$, we obtain a CSM mass of $\sim$0.22~M$_{\odot}$. We also obtain a small ejecta mass of 0.12~M$_{\odot}$. The constraint on the nickel ($^{56}$Ni) mass was overly prescriptive, resulting in an inferred value that was not constrained by the observational data. Consequently, the rest of the inferred parameters should be taken with a grain of salt. Nonetheless, this modelling can serve as a comparison with studies using the same prescription for a combination of CSM and radioactivity powered SNe. The mass-loss rates estimated using the R$_\mathrm{CSM}$ of 0.10~AU, v$_{w}$ of 1800~km~s$^{-1}$ and density of $\rho$$_\mathrm{CSM}$ of 4.8 $\times$ 10$^{-12}$~g~cm$^{-3}$, is 3.9 $\times$ 10$^{-4}$~M$_{\odot}\,{\rm yr}^{-1}$ . 
In their sample study of five SNe~Ibn, \cite{Pellegrino2022Ibn} found that their light curves were well reproduced by ejecta masses between $0.7-3$~M$_{\odot}$, CSM masses of $0.2-1$~M$_{\odot}$, and CSM radii between $20-65$~AU. Our obtained values of CSM mass of 0.22~M$_{\odot}$ is consistent with the lower end of the CSM masses obtained by \cite{Pellegrino2022Ibn}, but our ejecta mass and radius are much lower than those found for the sample of SNe~Ibn in \cite{Pellegrino2022Ibn}. 
The fits associated with both models infer that for the case of SN~2023xgo, $^{56}$Ni plays a role in powering the late-time light-curve. Since we do observe interaction signatures throughout the spectral evolution of SN~2023xgo, we introduce the CSM interaction in the light-curve modelling and find that both the CSM mass and the radius of the CSM are the lowest among the group of interacting SNe~Ibn \citep{Pellegrino2022Ibn}. As mentioned earlier, M$_{\mathrm{Ni}}$ was constrained to values $\leq$ 0.04~M$_{\odot}$ in order for the \texttt{csm\_interaction} model to become non-neglible. The corner plots for both the CSM and the Ni+CSM models are shown in Figure~\ref{fig:csm_model_cornerplot} and Figure~\ref{fig:CSM-Ni_model_cornerplot}, respectively. 

Very low ejecta masses have also been seen for ultra-stripped SNe \citep{Das2024_zaw,Wu2022}, but the rise time for SN~2023xgo is slower than for ultra-stripped SNe (see Figure~\ref{fig:correlationplot2}). Recently, \cite{Moriya2025}, through their theoretical modeling have found that in some cases ultra-stripped SNe can result in Type Ibn SNe. The progenitors of the ultra-stripped SNe can induce violent silicon burning phase just before the core-collapse that results in a dense CSM. They suggest that because the dense CSM is more massive than the SN ejecta, the ejecta is immediately decelerated and the light curve is powered mainly by the interaction due to which the ultra-stripped SN is observed as a Type Ibn SN.

There are some caveats associated with the semi-analytical light-curve modelling and the parameters estimated. 
The simplifications associated with the model and their possible implications are discussed further in Section~\ref{caveats}. 
Also, the models by \cite{Chatzopoulos12} assume a constant efficiency. 
Subsequently, we employ a non-equipartition light-curve model that incorporates the time-dependent evolution of radiative efficiency, accounting for the varying fraction of kinetic energy converted into observable radiation throughout the post-shock interaction and expansion phases.

\subsection{Non-equipartition light-curve model}

\begin{figure}
	\begin{center}
		\includegraphics[width=\columnwidth]{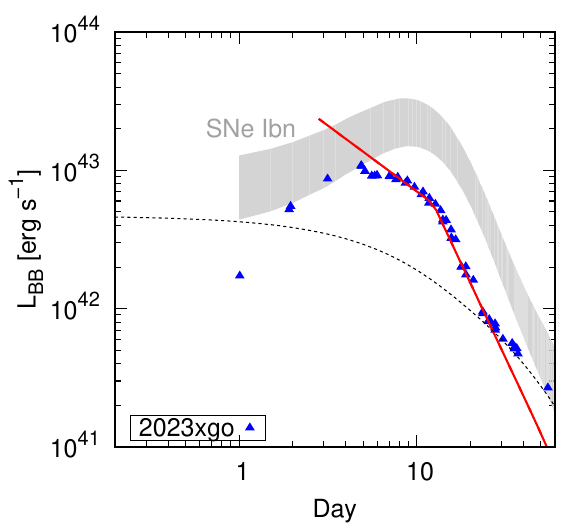}
	\end{center}
	\caption{Fitting the light curve of SN~2023xgo with the models by \protect\cite{MaedaMoriya2022}. The grey shaded region shows the template light curve of SNe~Ibn \protect\citep{Hosseinzadeh2017} while the red line shows the best fit model to the data of SN~2023xgo. The maximally allowed $^{56}$Ni contribution is described by the black dashed line.}
	\label{fig:lc-fitting-hydro}
\end{figure}

Even though the semi-analytical models did give us an estimate of CSM mass, radius and mass-loss rates, it did not take into account a time-varying efficiency conversion. We therefore also adopt a non-equipartition analytical light-curve model based on the formulation by \cite{MaedaMoriya2022} to reproduce our observed light curve.

The hydrodynamic evolution of the SN-CSM interaction, characterized by a forward shock (FS), contact discontinuity (CD), and a reverse shock (RS), is a classical problem with a well-developed solution. The evolution can be described by a self-similar solution as long as each of the ejecta and CSM structures is described by a single power law as a function of radius (which is the case when the reverse shock has not yet reached the inner, flat part of the ejecta). This hydrodynamic behavior, or a simplified version thereof, is frequently used in light-curve calculations, where it is also often assumed that the post-shock regions quickly reach thermal equilibrium. \cite{MaedaMoriya2022} pointed out that the latter is not always the case; they modelled the light curves of a sample of SNe~Ibn, including the physical processes in which high-energy photons created initially at the post-shock regions are eventually degraded to optical photons. They found that the flatter light-curve evolution is followed by a steeper evolution in the post-peak LCs, characterized by the transition of the shock properties from radiative to adiabatic. They consider the processes by which high-energy photons produced at the shocks are degraded to optical wavelengths, incorporating the effects of radiative cooling and the evolving optical depth of the CSM. In general, for a given CSM density the optical luminosity in the model by \cite{MaedaMoriya2022} is smaller than in models like the one by \cite{Chatzopoulos12,Chatzopoulos2013} (i.e., the derived mass-loss rate is higher for any given LC data) since the conversion efficiency in most situations is found to be much lower than often assumed. 

Given the limitations of the analytical CSM+Ni model, we are motivated to apply this more physically self-consistent LC model to SN~2023xgo. 
In the model, the SN ejecta are described by a broken power law in the density structure, where the inner density is constant and the outer part is given as $\rho_{\rm SN}$ $\propto$ $\nu$$^{-n}$ (e.g., \citealt{Chevalier1982,Moriya2013}). Here, $n=7$ is fixed in the model, which is within the range expected for an explosion of a compact He star progenitor \citep{Matzner1999}. 
The ejecta properties were described by the ejecta mass (M$_{\rm ej}$) and the kinetic energy (E$_{\rm k}$), which are varied as input parameters. For the CSM structure, a power-law distribution is assumed and the normalisation constant of the distribution is set to 1 for a density of $\rho_{\text{CSM}}$ = 10$^{-14}$~g~cm$^{-3}$ and a radius of 5 $\times$ 10$^{14}$~cm. The detailed set of equations relating the densities to the mass-loss rates of the progenitor is explained in \citet[][their section 3.1]{MaedaMoriya2022}.

The models assume a He-rich composition for both SN ejecta and the CSM.
Our best matched model parameters using the equations of the modelling section of \cite{MaedaMoriya2022} are 
\begin{align*}
    M_{\text{ej}} &= 2.0 \, \mathrm{M}_{\odot} \quad (\text{fixed}), \\
    E_{\text k} &= 1.7 \times 10^{51} \text{erg}, \\
    \text{Ejecta slope} \ (n) &= 7 \quad (\text{fixed}), \\
    \text{CSM slope} \ (s) &= 2.9, \\
    D' &= 1.5.
\end{align*}
where $D^\prime$ is the normalisation constant in the best matched density profile given by
\begin{equation}
    \rho_{\text{CSM}} = 10^{-14} D' \left(\frac{r}{5 \times 10^{14} \, \text{cm}}\right)^{-s} \quad \text{g cm}^{-3}.
\end{equation}

The model LC is shown in Figure~\ref{fig:lc-fitting-hydro}. 
We do not fit the early part of the light curve of SN~2023xgo, as diffusion is not taken into account while modelling the early part of the light curve of SN~2023xgo. The characteristic break at $\sim 10-20$ days since the explosion is explained by a transition in the shock properties (radiative to adiabatic) in this model, without introducing a change in the CSM density slope. The derived CSM density slope $s \sim3$ is typical for SNe~Ibn (as seen from the comparison of observational parameters by \citealt{MaedaMoriya2022}), indicating increasing mass-loss rate close to the explosion. 

We note that the sample used in \cite{MaedaMoriya2022} is somewhat under-luminous as compared to the larger SNe~Ibn sample of \cite{Hosseinzadeh2017}. SN~2023xgo falls in the low luminosity regime of the models by \cite{MaedaMoriya2022}. Considering the best fit value of $D^\prime$ and 
using v$_{w}$ = 1800~km~s$^{-1}$, the mass-loss rate is estimated to be very high at $\sim$2.7~M$_{\odot}\,{\rm yr}^{-1}$  (at 0.082 yrs or 32.2 days before the explosion). 

We note that the flatter light-curve tail could be fit by the $^{56}$Co decay represented by the dashed line in Figure~\ref{fig:lc-fitting-hydro}, assuming M$_{\rm Ni}$ = 0.05~M$_{\odot}$ and taking into account the gamma-ray escape. Although this light-curve flattening could also be attributed to the CSM distribution being flattened in the outer regions, we mention the Co decay interpretation because of the spectral transition of SN 2023xgo from a SN Icn/Ibn with clear interaction signatures in the early phases (powered by the interaction) to a `SN Ibn' without narrow lines in the late phase and even broad features (could be powered by radioactive decay) at even later times as discussed in the late phase spectral comparison in subsection~\ref{late}. 
The mass of $^{56}$Ni in this scenario, 0.05~M$_{\odot}$, is similar to the values found for SNe~IIP \citep{Hamuy_2003}, but smaller than those found for SNe~Ibc \citep{MaedaMoriya2022}, and larger than the sample of SNe~Ibn \citep{Pellegrino2022Ibn}.  
We see that both our semi-analytical and non-equipartition light-curve model use $0.04-0.05$~M$_{\odot}$ of Ni to reproduce the late-time light-curve. SN~2023xgo may thus be a link to investigate further a relation between interacting and non-interacting SESNe. However, 
%we want to remark that 
using the semi-analytical model, the nickel mass (0.04~M$_{\odot}$) is 
%basically
set by the typical value from \citet{Pellegrino2022Ibn}. In the non-equipartition model by \citet{MaedaMoriya2022}, the nickel mass is only estimated from the tail. This introduces a large uncertainty, so the interpretation of the nickel mass should be treated with caution.
 
%--------------------------------------------------------------------
\section{Discussion}
\label{discussion}

We have presented the photometric and spectroscopic analysis of SN~2023xgo, and hereafter we discuss the main properties of the SN and summarize our results. We divide the SN evolution into three different phases: a) pre-maximum time ($-$2.35 d to 0 d), b) maximum to late (before 23.48 d), and c) late time (after 23.48 d), and briefly describe in each subsection the behavior of the SN.

\subsection{Pre-maximum times} SN~2023xgo is spectrally similar to a SN~Icn, with prominent narrow \ion{C}{iii} lines at pre-maximum times. The early time spectral comparison of SN~2023xgo with other SNe~Ibn and SNe~Icn shows that the \ion{C}{iii} emission is stronger than for all SNe~Ibn and also stronger than in the transitional SN~2023emq. While flash ionization may account for the observed features in SN~2023xgo, such an interpretation would require it to exhibit the strongest carbon signatures observed among all known SNe~Ibn. The pseudo-EW measurements of the \ion{C}{iii} feature also highlights that this SN most likely belongs to the transitional SN~Icn to SN~Ibn category and is similar to a SN~Icn at this early phase. 

\cite{Blinnikov2010} showed that the light curves of some Type I SNe can be modelled as a shock wave propagating in C-O-rich CSM material. SN~2023xgo probably has a C-O-He rich CSM. Assuming that part of the \ion{C}{iii} emission line enhancement arises from recombination or cascading transitions from \ion{C}{iv}, we note that such processes typically involve ionization potentials exceeding 47~eV. This corresponds to a characteristic temperature above 50000~K, indicating the presence of a high-energy radiation field or a hot ionizing source. This is the maximum temperature with the blackbody approximation. In reality, photoionization from non-thermal radiation fields (like UV/X-rays from shocks or interaction zones) can ionize carbon even at lower gas temperatures \citep{1989ApJ...343..323F}. Thus, in realistic SN environments, ionization can be driven by non-thermal radiation fields, allowing for significant \ion{C}{iii} populations even when the local gas temperature is substantially lower than 50000~K. \cite{Martins2012} investigated the formation processes of \ion{C}{iii} $\lambda$5696 and \ion{C}{iii} $\lambda$4650 in the atmosphere of O stars using the non-LTE atmosphere code \texttt{CMFGEN} at two effective temperatures (30500~K and 42000~K). The formation of the \ion{C}{iii} $\lambda$5696 line is governed by radiative interactions with nearby UV transitions (\ion{C}{iii} $\lambda$386, 574, and 884), which regulate the population levels of the lower and upper states involved. Line overlaps with metallic species such as \ion{Fe}{iv} $\lambda$386, 574 and \ion{S}{v} $\lambda$884 strongly influence emission strength through photon draining or enhancement. Variations in effective temperature, surface gravity, mass loss rate, and metallicity further modulate the line profile, making \ion{C}{iii} $\lambda$5696 a very critical line profile derived from abundance diagnostic.
 
From Figure~\ref{fig:rad-temp} we see that the temperature for SN~2023xgo at early times is quite high, varying from 14500~K to 15500~K, and then becomes constant at 9500~K a few days after maximum light, albeit with larger error bars associated with it. The temperature around maximum for SN~2023xgo is consistent with most SNe~Ibn, but, still on the higher end \citep{Gangopadhyay2020,Brennan2024,Ben-Ami2023,Dong2024}. The prolonged presence of carbon features in the spectra of SN~2023xgo, compared to typical SNe~Ibn, may be attributed to a combination of elevated carbon abundance and sustained ionization driven by high temperatures in the early ejecta or CSM. The models by \cite{BoianGroh2019} show that the \ion{C}{iii} feature is better matched when the luminosity increases from 1.9 to 3.1 $\times$ 10$^{9}$ $L_{\odot}$. But, being a low luminosity member among the class (see Section~\ref{phot}), the temperature, elemental abundance and density must play the most important role for the case of SN~2023xgo in generating the \ion{C}{iii} $\lambda$5696 feature. The mass-loss rate of 10$^{-3}$~M$_{\odot}\,{\rm yr}^{-1}$  reproduces our observed spectral feature of \ion{C}{iii} at this phase.

Photometrically, SN~2023xgo behaves at this phase like a fast rising SN~Ibn/Icn. The rise time of 5 days is consistent with all the SNe~Ibn/Icn and shorter than that of Type II and Type Ib/c SNe. 
However, even if it photometrically resembles both SNe~Ibn and SNe~Icn at this phase, spectroscopically it is more similar to a SN~Icn. If instead interpreted as a flash ionized SN~Ibn, SN~2023xgo would be unique among its class with such a strong carbon emission.

\subsection{From maximum to about 20 days past max} From maximum light, we see a transformation in the spectrum of SN~2023xgo. The \ion{C}{iii} $\lambda$5696 signature completely vanish and narrow \ion{He}{i} lines develop throughout the evolution. The spectral comparison at this epoch shows a strong similarity with SNe~Ibn \citep{Hosseinzadeh2017} and the spectra are also very similar to the transitional Type~Ibn/Icn SN~2023emq \citep{Pursiainen2023}. This phase 
shows increasing velocities of the emission lines of \ion{He}{i} $\lambda$5876, from 1800~km~s$^{-1}$ to about 10000~km~s$^{-1}$. This is also the phase when the P-Cygni \ion{He}{i} signatures broaden and are taken over by emission with time. 

At peak, SN~2023xgo is among the faintest events (M$_{r}$ = $-$17.65 $\pm$ 0.04~mag) within the SN~Ibn/Icn population, exhibiting a decline rate of 0.14~mag~day$^{-1}$, which is typical of this class. In Section~\ref{phot} we show that SN~2023xgo is one of the lowest luminosity members of the SN~Ibn/Icn family. As the luminosity declines, the temperature decreases accordingly, leading to recombination of carbon and the subsequent emergence of emission features of \ion{He}{i} characteristic of typical SNe~Ibn. At this phase, temperature also drops to 9500~K, leading to simultaneous emergence of emission signatures from \ion{He}{i}, which require lower ionization energies. This spectroscopic transition, along with the coexistence of ionized and neutral features over time, may suggest a CSM with varying density and temperature structures.

\subsection{Late-time emission}

After 24 d, SN~2023xgo enters a phase in which both the ejecta and the CSM is optically thin. Spectral comparison with the models by \cite{Dessart2022} %helps us 
reproduce our late-phase spectrum with a low-mass He-star model of 3~M$_{\odot}$ assuming the CDS is at 2 $\times$ 10$^{15}$~cm and we see spectral signatures of both ejecta and CSM at this phase. We also see a possible evidence of eruptive mass loss activity from late-time light-curve and spectral modelling. Concurrently, the light curve exhibits a late-phase flattening, which might be explained by radioactive decay from $\sim$0.04 – 0.05~M$_{\odot}$ of synthesized $^{56}$Ni. This estimated Ni mass is comparable to values typically inferred for core-collapse SNe~II and SNe~Ib/c, indicating a substantial contribution from radioactive heating. However, additional contributions from ongoing ejecta–CSM interaction cannot be ruled out, particularly given the asymmetry suggested by the spectral modelling. The spectral and photometric features at these phases collectively suggest the presence of a dense, extended CSM component indicative of an eruptive pre-SN mass-loss episode, further supported by the high mass-loss rates required by both light-curve and spectral models.

\subsection{Caveats associated with the light-curve and spectral models}
\label{caveats}

We have used a number of spectral and photometric models to try to draw conclusions about the nature of the transitioning SN~Ibn/Icn 2023xgo. 
From the \cite{BoianGroh2019} spectral models (see subsection~\ref{boaingroh}) were able to reproduce the observed flash ionized lines of carbon, nitrogen, and helium. Increase in luminosity strengthened the \ion{C}{iii} $\lambda$5696 emission line.
Since SN~2023xgo is one of the lowest luminosity members in the subclass (see Section~\ref{phot}), there must be other factors like density, temperature, and abundance which play an important role in the formation of the strong carbon signature.
The model with a mass-loss rate of 10$^{-3}$~M$_{\odot}\,{\rm yr}^{-1}$ best matches our observed spectrum when the inner radius of the CSM is fixed at 8.6 $\times$ 10$^{13}$~cm. However, the models by \cite{BoianGroh2019} are not radiatively post-processed and assume a fixed CSM inner radius at a fixed time. Thus, the estimates are valid only for these simplistic assumptions.

Late-time spectral comparison with models of \cite{Dessart2022} indicates that the model from a low mass He star of 3~M$_{\odot}$ and a spectral model luminosity of 2 $\times$ 10$^{42}$~erg~s$^{-1}$ best matches our observed spectral feature of \ion{C}{iii}, while the observed luminosity is lower ($6.0 \times 10^{41}$~erg~s$^{-1}$) at similar phases. However, this assumes that the CDS is located at 2 $\times$ 10$^{15}$ cm. Also, 
the best matched model by \citet{Dessart2022} assume a CDS mass of $\sim$0.97~M$_{\odot}$ and an ejecta mass of 1.5~M$_{\odot}$. This corresponds to a mass-loss rate of approximately 0.16~M$\odot$ yr$^{-1}$, assuming v$_{w}$=1800~km~s$^{-1}$, consistent with an eruptive mass-loss scenario \citep{Smith2017}. 

We perform semi-analytical light-curve modelling using \texttt{redback} in subsections~\ref{arnett}, \ref{csm} and \ref{csmni}, assuming radioactivity, CSM interaction and both radioactivity + CSM driving the light-curve evolution of SN~2023xgo. The mass-loss rates (10$^{-4} - 10^{-3}$~M$_{\odot}\,{\rm yr}^{-1}$ ) and the radius of the CSM ($\sim$10$^{12} - 10^{13}$~cm) that we get from semi-analytical models are similar to the values in the \cite{BoianGroh2019} models. If we consider an eruptive mass-loss case, then assuming a wind velocity of 1800 km s$^{-1}$ and radius from Table~\ref{table:CSMplusNi_fitting}, we would find that the shell would be ejected approximately few hours before the explosion. The semi-analytical models, however, assume spherical symmetry and homologous expansion, which may not fully represent the complexity of the CSM geometry. Viewing angle effects and 2D/3D structure could significantly alter the obtained parameters \citep{Suzuki2019,VanBaal2023}.

In the semi-analytical light-curve modelling, the time-varying conversion of efficiency is not taken into account. So, we also perform late-time light-curve modelling following \cite{MaedaMoriya2022}, which calculates a time varying efficiency throughout the light-curve evolution. For a given CSM density, the optical luminosity in the model by \cite{MaedaMoriya2022} is smaller than in models by \cite{Chatzopoulos12,Chatzopoulos2013}. This gives us a higher mass-loss rate since the conversion efficiency at each time interval is much lower than often assumed in the semi-analytical approach.
%We could reproduce part of our light curve post-peak with a density exponent of 2.9 and CSM extending up to 10$^{15}$~cm (assuming a fixed M$_{\rm ej}$ of 2~M$_{\odot}$). 
Using this model, we obtain a mass-loss rate of 2.7~M$_{\odot}$~yr$^{-1}$ which suggests intense eruptive mass-loss occurring shortly before core collapse. While this value aligns with the upper limits of the eruptive scenarios modelled by \citet{Dessart2022}, it still represents one of the most extreme cases in terms of pre-SN mass-loss rates. This model is not able to reproduce the early part of the light curve as diffusion was not taken into account (pre-maximum times).
%Both the semi-analytical and the non-equipartition late-time model indicate that not only CSM interaction, but also a $^{56}$Ni mass of $0.05$~M$_{\odot}$  is needed to reproduce the light curve after 30 d.

\cite{Sorokina2016} performed \texttt{STELLA} models for the light curves of superluminous SNe~I and found significant differences in the numerical model parameters as compared to \cite{Chatzopoulos12} (see Figure 14 of \citealt{Sorokina2016}). They quote that this is probably due to very different diffusion time-scales and envelope masses in their light-curve modelling.
We note that neither the CSM nor the SN ejecta can be approximated by a single zone velocity \citep{Moriya2017}. Thus, there are caveats associated with these models, but we do infer some constraints on the geometry and the physical parameters of SN~2023xgo by combining them which we describe in detail in subsection~\ref{geometry}.

\subsection{Implications on geometry and possible progenitor}
\label{geometry}
The early-time characteristics, when interpreted using the flash ionization models of \citet{BoianGroh2019} for the spectra and the semi-analytical framework of \citet{Chatzopoulos12} for the light curve, indicate the presence of a compact CSM confined within a radius of $\sim$10$^{12}$–10$^{13}$~cm. The inferred mass-loss rates in this phase are modest, on the order of 10$^{-3}$–10$^{-4}$~M$_{\odot}$ yr$^{-1}$, consistent with a relatively quiescent, wind-driven phase prior to explosion. In contrast, modelling of the late-time light-curve following the approach of \citet{MaedaMoriya2022}, along with spectral synthesis model comparisons by \citet{Dessart2022}, reveals the presence of an extended, dense CSM reaching out to $\sim$10$^{15}$ cm, with significantly higher mass-loss rates ranging from 0.1 to 2.7 M$_{\odot}$ yr$^{-1}$. These findings may suggest a structured and possibly asymmetric CSM, composed of both compact, low-density material and a more extended, high-density component. However, such interpretations are model-dependent and could also arise from variations in the mass-loss history or radial density structure of the CSM, without necessarily invoking strong asymmetry. Multiple mass-loss episodes with differing energetics and geometries—potentially shaped by binary interaction, rotationally driven outflows, or eruptive events—could contribute to this complex CSM morphology.

The high mass-loss rates inferred from late-time modelling are consistent with eruptive phases in luminous blue variable (LBV) stars transitioning to the WR phase \citep{Smith2017}, which could indicate a single massive star as the progenitor of SN~2023xgo. However, if the CSM is asymmetric as indicated above with potential disk- or torus-like morphology, then it may also point toward a binary system. The modest helium-rich shell mass, likely originating from He-C and He-N layers, aligns with the 0.97~M$_{\odot}$ He-shell mass typically assumed in the models of \citet{Dessart2022}. These constraints limit the pre-SN mass to $3-4$~M$_{\odot}$ and suggest a ZAMS mass not exceeding 18~M$_{\odot}$. Given that the 3~M$_{\odot}$ model best matches the nebular spectrum, a binary origin involving low-mass stars such as He stars, becomes more plausible, as stellar winds from red supergiants at these masses are insufficient to fully strip the hydrogen envelope \citep{Beasor2020}.

Furthermore, theoretical predictions suggest that standard WR explosions with decelerated inner shells (2000~km~s$^{-1}$) should yield superluminous SNe \citep{Dessart2016,Dessart2022}. However, as shown in Section~\ref{phot}, SN~2023xgo is among the least luminous events within the SN~Ibn/Icn population.
This again supports progenitor scenarios involving binary systems, such as nuclear flashes in low-mass helium stars \citep{Woosley2019,Dessart2022,Wang2024}, or repeated mass-transfer episodes in compact binaries \citep{Langer2012,Tauris2013}. Furthermore, the similarity in ejecta mass between SN~2023xgo and ultra-stripped SNe reinforces the hypothesis of a low-mass helium star (ZAMS mass $\leq$10 $M_{\odot}$) in a close binary configuration, as proposed for SN~2023zaw \citep{Das2024_zaw}. The progenitor properties are further supported by the host galaxy’s stellar mass and star formation rate, which are similar to that of the general properties observed for the broader SN~Ibn/Icn population.

\section{Summary}
\label{summarize}
\begin{enumerate}
\item SN~2023xgo is a unique member of the interacting SN~Ibn/Icn class. It shows a prominent \ion{C}{iii} $\lambda$5696 spectral feature until maximum light, similar to what is seen for SNe~Icn, but spectrally behaves like a typical SN~Ibn with narrow helium lines post maximum light. 
\item Comparison of the early spectrum ($-$2.35~d) with models by \cite{BoianGroh2019} indicates that even if the other flash features matches when $L = 1.5 \times$ 10$^{9}$~$L_{\odot}$, the \ion{C}{iii} $\lambda$5696 feature only matches if L is increased to 3.1 $\times$ 10$^{9}$~$L_{\odot}$, with a mass-loss rate of 10$^{-3}$~M$_{\odot}\,{\rm yr}^{-1}$ . 
\item Early spectral comparison ($-$2.35~d) and pseudo-equivalent width measurement of this \ion{C}{iii} $\lambda$5696 line indicates that the strength and evolution of this line behaves similarly to what is seen for other SNe~Icn. If SN~2023xgo is classified as a SN~Ibn with flash ionized signatures, the sustained detection of \ion{C}{iii} emission until maximum light suggests either a comparatively enhanced carbon abundance or temperatures capable of sustaining significant carbon ionization throughout the early evolution
\item The mid-spectral epoch (just after maximum light) marks the phase when SN~2023xgo behaves like a normal SN~Ibn with narrow P-Cygni \ion{He}{i} $\lambda$5876 and other helium lines with typical FWHM velocities starting around 1800~km~s$^{-1}$. Spectral comparison at this phase with other SNe~Ibn and SNe~Icn also shows that it belongs to the ``P-Cygni'' subclass according to \cite{Hosseinzadeh2017}. 
\item After 10~d, ejecta signatures start appearing in the spectral evolution of SN~2023xgo. Late-time spectral comparison shows similarities with SN~2019wep, ASASSN-15ed, and some other SNe~Ib.
\item Comparison of the 23.48~d spectrum with the models by \cite{Dessart2022} 
favors a 3~M$_{\odot}$ He-star model with the CDS located at 2.1 $\times$ 10$^{15}$~cm.
\item The pseudo-equivalent widths of the \ion{He}{i} lines in SN~2023xgo shows an increase up to 30 \AA\ in 10 d,  
and velocities varying between 1800 km s$^{-1}$ $-$ 10000~km~s$^{-1}$ in 25 d. This value is higher than SN~2019wep and ASASSN-15ed and is similar to the velocity measurements for some SNe~Ib at this phase.
\item The host galaxy mass of $10^{9.31^{+0.13}_{-0.22}}$~M$_{\odot}$, metallicity of $0.7\pm0.1$ solar and the star-formation rate of $0.05 \pm 0.02$~M$_{\odot}\,{\rm yr}^{-1}$ is similar to that of the sample of SNe~Ibn/Icn.
\item Photometrically, SN~2023xgo behaves like a traditional SN~Ibn/Icn with a rise time of 5.14 $\pm$ 2.30 days and decline rate ($0-30$~d) of 0.14~mag~d$^{-1}$, similar to that of the values estimated from the larger sample of SNe Ibn/Icn \citep{Hosseinzadeh2017,Pellegrino2022Ibn,Pellegrino2022Icn}.
\item 
Comparisons between rise times, $\Delta$m$_{15}$ and absolute magnitudes with a group of fast transients, SNe~IIn, Ibn, Icn, and ultra-stripped SNe indicate that the decline rates of SN 2023xgo are similar to that of those of SNe~Ibn/Icn and are slower than the decline rates of fast transients and ultra-stripped SNe.
\item The absolute magnitude of M$_{r}$ = $-$17.65 $\pm$ 0.04 places SN~2023xgo among
the faintest members of the Type Ibn/Icn sub-class, similar to SNe~Ib.
\item Multi-band modelling (semi-analytic) of SN~2023xgo indicates that CSM interaction fairly reproduces the early light curve, but radioactivity plays a role in reproducing the overall light curve of SN~2023xgo, specially at late times. The light curve seems to be driven by a CSM mass of $\sim$0.22~M$_{\odot}$, and a small ejecta mass of 0.12~M$_{\odot}$, assuming its radioactivity driven with Ni mass of 0.04~M$_{\odot}$. A shell-like CSM with a radius of 10$^{12}$~cm is favoured by our observed data.
\item Late-time bolometric light-curve modelling fairly reproduces the light curve with a density exponent of 2.9, CSM extent at 10$^{15}$~cm and mass-loss rates of 2.7~M$_{\odot}\,{\rm yr}^{-1}$ . A Ni mass of 0.05~M$_{\odot}$ reproduces the late-time light-curve if we assume radioactivity to be powering the light curve.
\item Overall,the inferred mass-loss rates suggest two distinct CSM components: a compact, low-density region ($\sim$10$^{-3}$~M$_{\odot}$ yr$^{-1}$ at 10$^{12}$–10$^{13}$~cm) and an extended, dense region ($0.1-2.7$~M$_{\odot}$~yr$^{-1}$ at $\sim$10$^{15}$~cm). This structure may result from asymmetric geometry or reflect temporal and radial variations in mass-loss history, though such interpretations remain model-dependent.
\item The combination of a likely asymmetric CSM structure and spectral characteristics is similar to that of a binary progenitor system for SN~2023xgo, potentially involving a $\sim$3~M$_{\odot}$ helium star. Nonetheless, the observed indications of eruptive mass-loss do not exclude a single-star origin, especially in the context of advanced evolutionary phases that may produce non-steady outflows.
\end{enumerate}

\section*{Acknowledgements}
This work has been enabled by support from the research project grant “Understanding the Dynamic Universe” funded by the Knut and Alice Wallenberg under Dnr KAW 2018.0067.
Based on observations obtained with the Samuel Oschin Telescope 48-inch and the 60-inch Telescope at the Palomar Observatory as part of the Zwicky Transient Facility project. ZTF is supported by the National Science Foundation under Grant No. AST-2034437 and a collaboration including Caltech, IPAC, the Oskar Klein Center at Stockholm University, the University of Maryland, University of California, Berkeley, the University of Wisconsin at Milwaukee, University of Warwick, Ruhr University Bochum, Cornell University, Northwestern University and Drexel University. Operations are conducted by COO, IPAC, and UW. SED Machine is based upon work supported by the National Science Foundation under Grant No. 1106171. The ZTF forced-photometry service was funded under the Heising-Simons Foundation grant $\#$12540303 (PI: Graham). The Gordon and Betty Moore Foundation, through both the Data-Driven Investigator Program and a dedicated grant, provided critical funding for SkyPortal. 
We thank Takashi J. Moriya and Masaomi Tanaka for insightful discussion about the paper. 
K.M. acknowledges support from the Japan Society for the Promotion of Science (JSPS) KAKENHI grant JP24KK0070 and JP24H01810. R.D. acknowledges funds by ANID grant FONDECYT Postdoctorado Nº 3220449. M.S. acknowledges financial support provided under the National Post Doctoral Fellowship (N-PDF; File Number: PDF/2023/002244) by the Science $\&$ Engineering Research Board (SERB), Anusandhan National Research Foundation (ANRF), Government of India. N. Sarin and A. Singh acknowledges support from the Knut and Alice Wallenberg Foundation through the ``Gravity Meets Light" project and by the research environment grant ``Gravitational Radiation and Electromagnetic Astrophysical Transients'' (GREAT) funded by the Swedish Research Council (VR) under Dnr 2016-06012.

The spectrum taken by the Seimei telescope was obtained under the KASTOR (Kanata And Seimei Transient Observation Regime) project, specifically under the following program for the Seimei Telescope at the Okayama observatory (23B-K-0013 and 23B-N-CT11). The Seimei telescope at the Okayama Observatory is jointly operated by Kyoto University and the National Astronomical Observatory of Japan (NAOJ), with assistance provided by the Optical and Infrared Synergetic Telescopes for Education and Research (OISTER) program.

%%%%%%%%%%%%%%%%%%%%%%%%%%%%%%%%%%%%%%%%%%%%%%%%%%
\section*{Data Availability}
The data presented in this paper will be provided upon request. All the spectra will be made publicly available in WiseRep and Zenodo.



\bibliographystyle{mnras}
\bibliography{example} % if your bibtex file is called example.bib

%\begin{thebibliography}{99}
%\end{thebibliography}

%%%%%%%%%%%%%%%%%%%%%%%%%%%%%%%%%%%%%%%%%%%%%%%%%%

%%%%%%%%%%%%%%%%% APPENDICES %%%%%%%%%%%%%%%%%%%%%

\appendix

\input{2023xgo_phot}

\input{Table_spex_final}

\section{Light-curve modelling results}

\begin{figure*}
   \centering
   \includegraphics[scale=0.3]{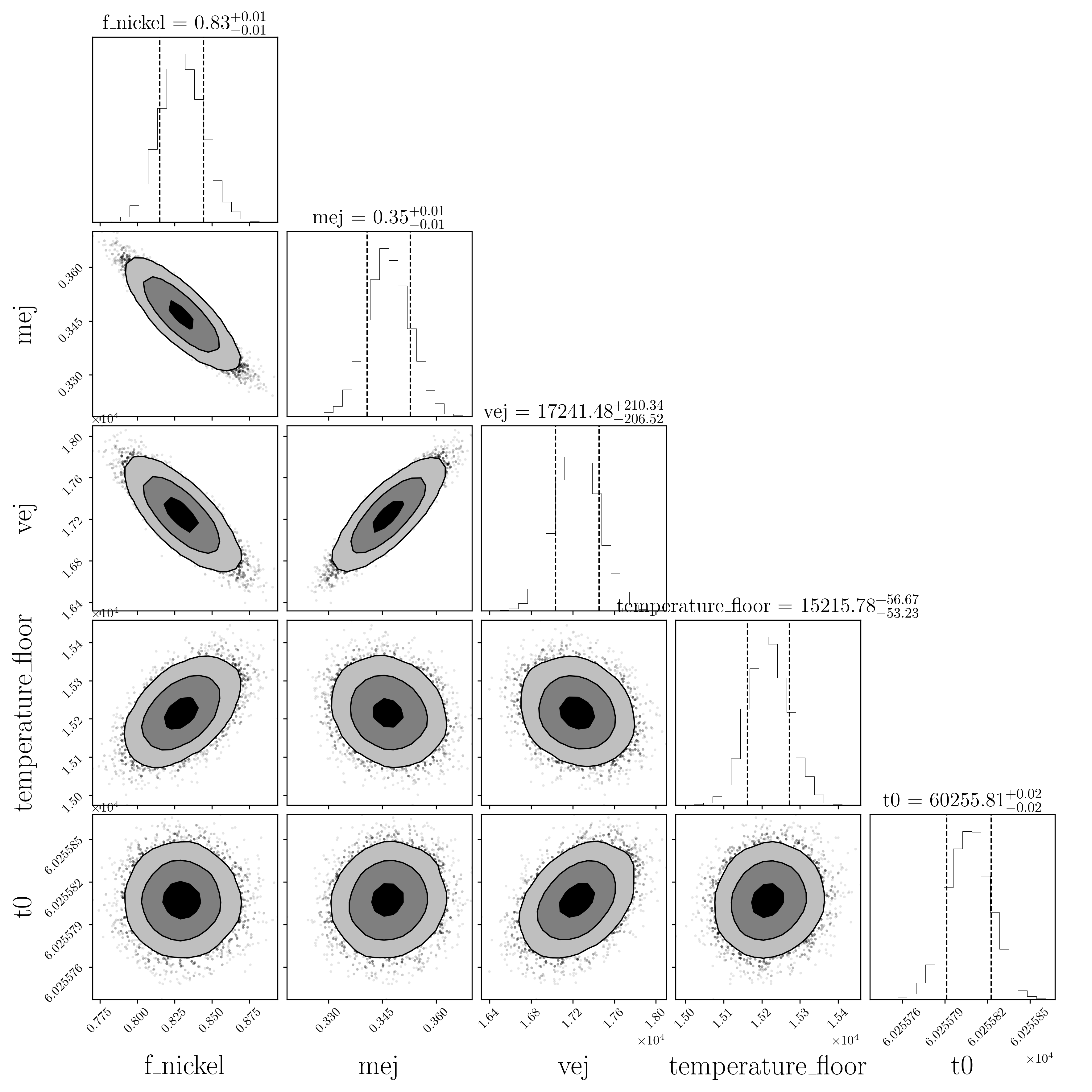}
   \caption{The corner plot showing the posterior of the parameters of SN~2023xgo in the radioactively powered model \citep{Arnett1982}}
   \label{fig:arnett_model_cornerplot}
\end{figure*}

\begin{figure*}
   \centering
   \includegraphics[scale=0.3]{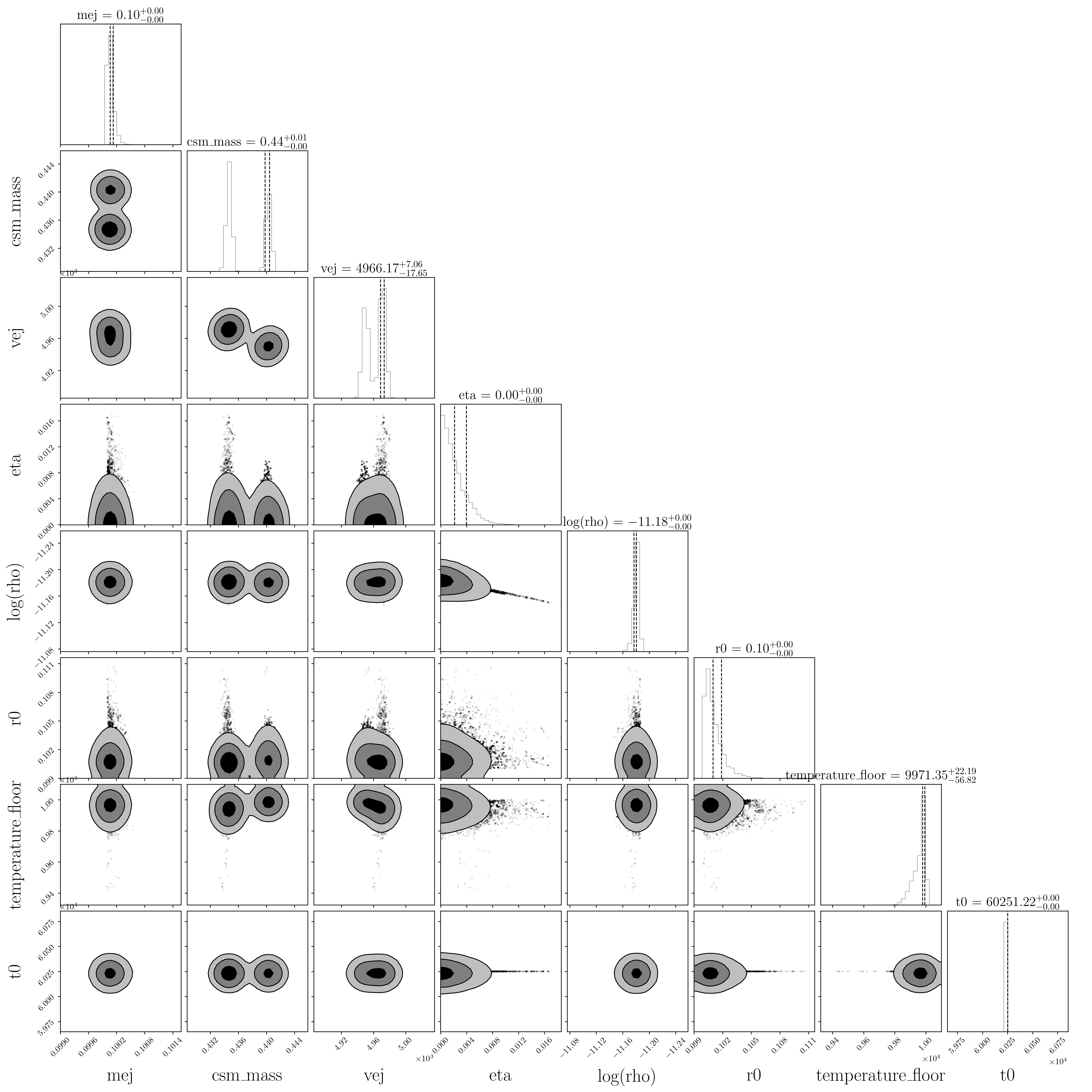}
   \caption{The corner plot showing the posterior of the parameters of SN~2023xgo in the CSM interaction model \citep{Chatzopoulos12,Chatzopoulos2013}}
   \label{fig:csm_model_cornerplot}
\end{figure*}

\subsection{CSM+Ni model}

\begin{figure*}
   \centering
   \includegraphics[scale=0.23]{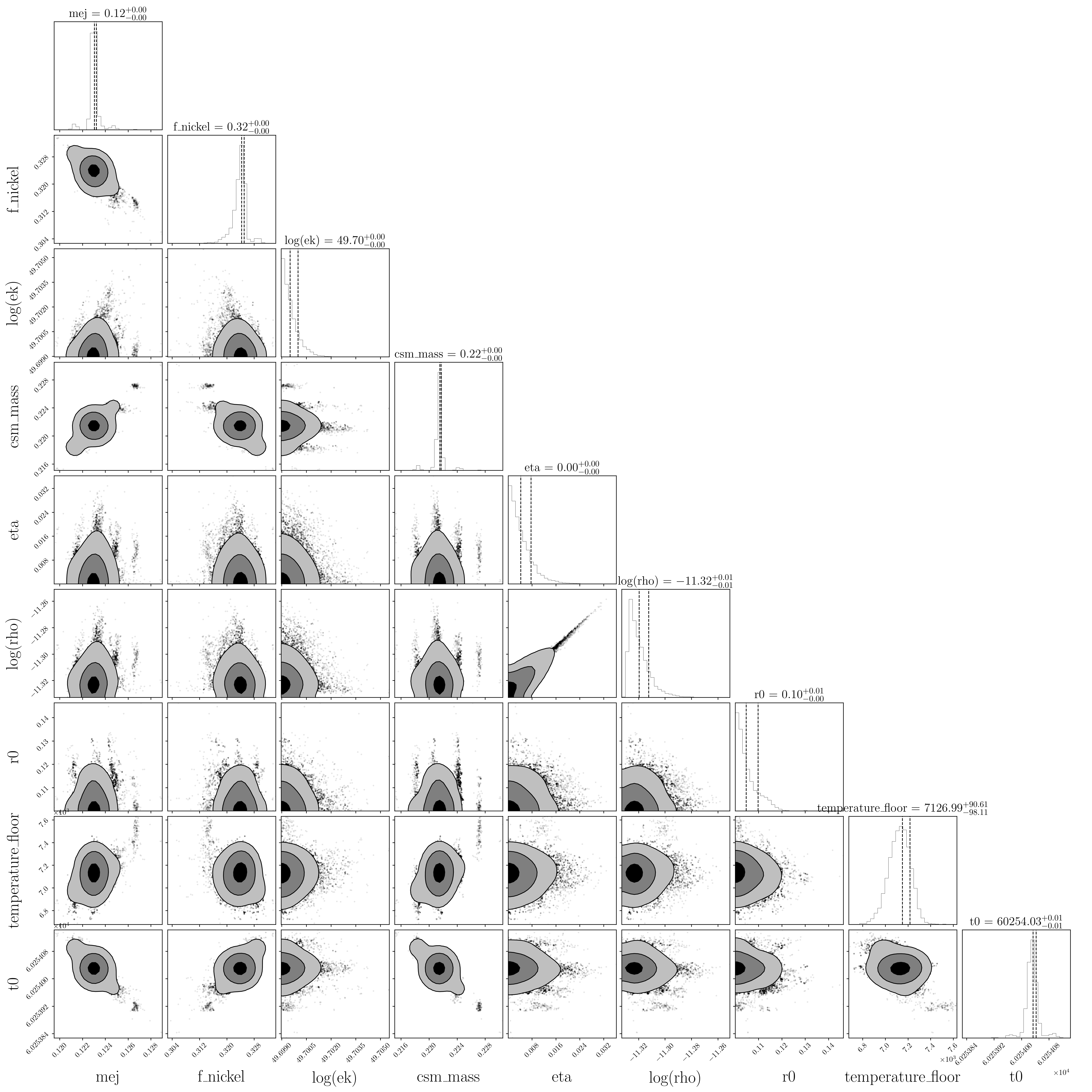}
   \caption{The corner plot showing the posterior of the parameters of SN~2023xgo in the Ni + CSM model \citep{Arnett1982,Chatzopoulos12}.}
   \label{fig:CSM-Ni_model_cornerplot}
\end{figure*}

%%%%%%%%%%%%%%%%%%%%%%%%%%%%%%%%%%%%%%%%%%%%%%%%%%

% Don't change these lines
\bsp	% typesetting comment
\label{lastpage}
\end{document}

%% file: 2023xgo_phot.tex
\begin{table*}
\caption {A small section of the photometric observations of SN~2023xgo. The full table is added in machine readable format in supplementary material.}
\label{tab:2023xgo_phot_obs}
\begin{center}
%\smallskip
%\small\addtolength{\tabcolsep}{-2pt}
\begin{tabular}{cccc}
\hline \hline
MJD & Mag & Filter & Instrument/Telescope \\
(day) & (mag) &  &   \\
\hline
60257.28 & 18.69 $\pm$ 0.07 & r & ZTF/P48 \\ \\
60257.37 & 18.52 $\pm$ 0.03 & g & ZTF/P48 \\ \\
60258.20 & 17.34 $\pm$ 0.09 & r & SEDM/P60 \\ \\
60258.24 & 17.26 $\pm$ 0.03 & r & SEDM/P60 \\ \\
60258.36 & 17.18 $\pm$ 0.01 & g & ZTF/P48 \\ \\
60259.39 & 16.65 $\pm$ 0.01 & g & ZTF/P48 \\ \\
60259.43 & 16.66 $\pm$ 0.01 & g & ZTF/P48 \\ \\
60259.45 & 16.74 $\pm$ 0.01 & r & ZTF/P48 \\ \\
60260.71 & 16.47 $\pm$ 0.04 & g & GIT \\ \\
60260.71 & 16.52 $\pm$ 0.01 & r & GIT \\ \\
60260.71 & 16.60 $\pm$ 0.03 & i & GIT \\ \\
60260.72 & 16.69 $\pm$ 0.06 & z & GIT \\ \\
60260.85 & 16.69 $\pm$ 0.03 & uvot:uvw1 & UVOT/Swift \\ \\
60260.85 & 16.26 $\pm$ 0.04 & uvot:u & UVOT/Swift \\ \\
60260.85 & 16.11 $\pm$ 0.05 & uvot:b & UVOT/Swift \\ \\
60260.85 & 17.11 $\pm$ 0.03 & uvot:uvw2 & UVOT/Swift \\ \\
60260.86 & 16.68 $\pm$ 0.09 & uvot:v & UVOT/Swift \\ \\
60260.86 & 17.05 $\pm$ 0.03 & uvot:uvm2 & UVOT/Swift \\ \\
60261.19 & 16.52 $\pm$ 0.02 & r & SEDM/P60 \\ \\
60261.22 & 16.49 $\pm$ 0.01 & r & SEDM/P60 \\ \\
60261.43 & 16.39 $\pm$ 0.01 & g & ZTF/P48 \\ \\
60261.45 & 16.49 $\pm$ 0.02 & r & ZTF/P48 \\ \\
60261.63 & 16.71 $\pm$ 0.07 & z & GIT \\ \\
60261.64 & 16.58 $\pm$ 0.03 & i & GIT \\ \\
60261.64 & 16.47 $\pm$ 0.02 & r & GIT \\ \\
60261.64 & 16.44 $\pm$ 0.03 & g & GIT \\ \\
\hline
\end{tabular}
\end{center}
\end{table*}

%% file: Table_spex_final.tex
\begin{table}
\caption {Log of spectroscopic observations of SN~2023xgo. The phase (rest frame) is measured with respect to $r$-band maximum (MJD$_{rmax}=60262.86$).}
\label{tab:2023xgo_spec_obs}
\begin{center}
%\smallskip
%\small\addtolength{\tabcolsep}{-2pt}
\begin{tabular}{c c c c c c}
\hline \hline
Date    & Phase     &    Instrument &    Telescope  &		Range  \\
        & (days)    &               &                 &     (\AA)   \\
\hline    
	20231112 &  -2.35  & KOOLS-IFU  &  3.8m Seime  &  4000-8000       \\  
20231113 &  -1.66  & SEDM       &  Palomar 60          &  3650-10000      \\
20231113 &  -0.82  & SPRAT      &  Liverpool Telescope &  4020-8100 \\
20231114 &  -0.54  & HFOSC      &  HCT                 &  3800-8350 \\
20231114 &  -0.07  & SEDM       &  Palomar 60          &  3650-10000      \\
20231115 &   0.53  & SEDM       &  Palomar 60          &  3650-10000      \\
20231115 &   0.53  & DBSP       &  Palomar 200         &  3000-11000      \\
20231117 &   2.64  & KOOLS-IFU  &  3.8m Seimei         &  4000-8000 \\
20231117 &   3.07  & HFOSC      &  HCT                 &  3800-8350 \\
20231119 &   4.75  & HFOSC      &  HCT                 &  3800-8350  \\
20231120 &   5.34  & LRIS       &  KECK                &  3500-11000      \\
20231120 &   5.39  & SEDM       &  Palomar 60          &  3650-10000      \\
20231122 &   7.63  & SEDM       &  Palomar 60          &  3650-10000      \\
20231123 &   8.64  & KOOLS-IFU  &  3.8m Seimei         &  4000-8000       \\         
20231124 &   9.30  & SEDM       &  Palomar 60          &  3650-10000      \\
20231125 &   10.97 & HFOSC      &  HCT                 &  3800-8350     \\
20231202 &   17.31 & SEDM       &  Palomar 60          &  3650-10000      \\
20231203 &   18.42 & DBSP       &  Palomar 200         &  3000-11000      \\
20231206 &   21.48 & SEDM       &  Palomar 60          &  3650-10000      \\
20231207 &   23.48 & LRIS       &  KECK                &  3500-11000      \\
\hline
\end{tabular}
\end{center}
\end{table}

%% file: main_revised.bbl
\begin{thebibliography}{}
\makeatletter
\relax
\def\mn@urlcharsother{\let\do\@makeother \do\$\do\&\do\#\do\^\do\_\do\%\do\~}
\def\mn@doi{\begingroup\mn@urlcharsother \@ifnextchar [ {\mn@doi@}
  {\mn@doi@[]}}
\def\mn@doi@[#1]#2{\def\@tempa{#1}\ifx\@tempa\@empty \href
  {http://dx.doi.org/#2} {doi:#2}\else \href {http://dx.doi.org/#2} {#1}\fi
  \endgroup}
\def\mn@eprint#1#2{\mn@eprint@#1:#2::\@nil}
\def\mn@eprint@arXiv#1{\href {http://arxiv.org/abs/#1} {{\tt arXiv:#1}}}
\def\mn@eprint@dblp#1{\href {http://dblp.uni-trier.de/rec/bibtex/#1.xml}
  {dblp:#1}}
\def\mn@eprint@#1:#2:#3:#4\@nil{\def\@tempa {#1}\def\@tempb {#2}\def\@tempc
  {#3}\ifx \@tempc \@empty \let \@tempc \@tempb \let \@tempb \@tempa \fi \ifx
  \@tempb \@empty \def\@tempb {arXiv}\fi \@ifundefined
  {mn@eprint@\@tempb}{\@tempb:\@tempc}{\expandafter \expandafter \csname
  mn@eprint@\@tempb\endcsname \expandafter{\@tempc}}}

\bibitem[\protect\citeauthoryear{{Ambikasaran}, {Foreman-Mackey}, {Greengard},
  {Hogg}  \& {O'Neil}}{{Ambikasaran} et~al.}{2015}]{Ambikasaran2015}
{Ambikasaran} S.,  {Foreman-Mackey} D.,  {Greengard} L.,  {Hogg} D.~W.,
  {O'Neil} M.,  2015, \mn@doi [IEEE Transactions on Pattern Analysis and
  Machine Intelligence] {10.1109/TPAMI.2015.2448083}, \href
  {https://ui.adsabs.harvard.edu/abs/2015ITPAM..38..252A} {38, 252}

\bibitem[\protect\citeauthoryear{{Arnett}}{{Arnett}}{1982}]{Arnett1982}
{Arnett} W.~D.,  1982, \mn@doi [\apj] {10.1086/159681}, \href
  {https://ui.adsabs.harvard.edu/abs/1982ApJ...253..785A} {253, 785}

\bibitem[\protect\citeauthoryear{{Ashton}, {H{\"u}bner}, {Lasky}, {Talbot}  \&
  et al.}{{Ashton} et~al.}{2019}]{Ashton2019}
{Ashton} G.,  {H{\"u}bner} M.,  {Lasky} P.~D.,  {Talbot} C.,   et al. 2019,
  \mn@doi [\apjs] {10.3847/1538-4365/ab06fc}, \href
  {https://ui.adsabs.harvard.edu/abs/2019ApJS..241...27A} {241, 27}

\bibitem[\protect\citeauthoryear{{Balcon}}{{Balcon}}{2023}]{2023TNSCR2931....1B}
{Balcon} C.,  2023, Transient Name Server Classification Report, \href
  {https://ui.adsabs.harvard.edu/abs/2023TNSCR2931....1B} {2023-2931, 1}

\bibitem[\protect\citeauthoryear{{Barden}}{{Barden}}{1994}]{1994ASPC...55..130B}
{Barden} S.~C.,  1994, in {Pyper} D.~M.,  {Angione} R.~J.,  eds,  Astronomical
  Society of the Pacific Conference Series Vol. 55, Optical Astronomy from the
  Earth and Moon. pp 130--138

\bibitem[\protect\citeauthoryear{{Beasor}, {Davies}, {Smith}, {van Loon},
  {Gehrz}  \& {Figer}}{{Beasor} et~al.}{2020}]{Beasor2020}
{Beasor} E.~R.,  {Davies} B.,  {Smith} N.,  {van Loon} J.~T.,  {Gehrz} R.~D.,
  {Figer} D.~F.,  2020, \mn@doi [\mnras] {10.1093/mnras/staa255}, \href
  {https://ui.adsabs.harvard.edu/abs/2020MNRAS.492.5994B} {492, 5994}

\bibitem[\protect\citeauthoryear{{Becker}}{{Becker}}{2015}]{hotpants}
{Becker} A.,  2015, {HOTPANTS: High Order Transform of PSF ANd Template
  Subtraction}, Astrophysics Source Code Library, record ascl:1504.004
  (\mn@eprint {ascl} {1504.004})

\bibitem[\protect\citeauthoryear{{Bellm} et~al.,}{{Bellm}
  et~al.}{2019}]{Bellm2019}
{Bellm} E.~C.,  et~al., 2019, \mn@doi [\pasp] {10.1088/1538-3873/ab0c2a}, \href
  {https://ui.adsabs.harvard.edu/abs/2019PASP..131f8003B} {131, 068003}

\bibitem[\protect\citeauthoryear{{Ben-Ami} et~al.,}{{Ben-Ami}
  et~al.}{2023}]{Ben-Ami2023}
{Ben-Ami} T.,  et~al., 2023, \mn@doi [\apj] {10.3847/1538-4357/acb432}, \href
  {https://ui.adsabs.harvard.edu/abs/2023ApJ...946...30B} {946, 30}

\bibitem[\protect\citeauthoryear{{Bertin}}{{Bertin}}{2013}]{psfex_ascl}
{Bertin} E.,  2013, {PSFEx: Point Spread Function Extractor}, Astrophysics
  Source Code Library, record ascl:1301.001

\bibitem[\protect\citeauthoryear{{Blagorodnova} et~al.,}{{Blagorodnova}
  et~al.}{2018}]{Blagorodnova2018}
{Blagorodnova} N.,  et~al., 2018, \mn@doi [\pasp] {10.1088/1538-3873/aaa53f},
  \href {https://ui.adsabs.harvard.edu/abs/2018PASP..130c5003B} {130, 035003}

\bibitem[\protect\citeauthoryear{{Blinnikov} \& {Sorokina}}{{Blinnikov} \&
  {Sorokina}}{2010}]{Blinnikov2010}
{Blinnikov} S.~I.,  {Sorokina} E.~I.,  2010, \mn@doi [arXiv e-prints]
  {10.48550/arXiv.1009.4353}, \href
  {https://ui.adsabs.harvard.edu/abs/2010arXiv1009.4353B} {p. arXiv:1009.4353}

\bibitem[\protect\citeauthoryear{{Boian} \& {Groh}}{{Boian} \&
  {Groh}}{2019}]{BoianGroh2019}
{Boian} I.,  {Groh} J.~H.,  2019, \mn@doi [\aap] {10.1051/0004-6361/201833779},
  \href {https://ui.adsabs.harvard.edu/abs/2019A&A...621A.109B} {621, A109}

\bibitem[\protect\citeauthoryear{Bradley et~al.,}{Bradley
  et~al.}{2024}]{photutils}
Bradley L.,  et~al., 2024, astropy/photutils: 1.13.0,
  \mn@doi{10.5281/zenodo.12585239}, \url
  {https://doi.org/10.5281/zenodo.12585239}

\bibitem[\protect\citeauthoryear{{Brennan} et~al.,}{{Brennan}
  et~al.}{2024}]{Brennan2024}
{Brennan} S.~J.,  et~al., 2024, \mn@doi [\aap] {10.1051/0004-6361/202349036},
  \href {https://ui.adsabs.harvard.edu/abs/2024A&A...690A.259B} {690, A259}

\bibitem[\protect\citeauthoryear{{Bruch} et~al.,}{{Bruch}
  et~al.}{2021}]{Bruch2021}
{Bruch} R.~J.,  et~al., 2021, \mn@doi [\apj] {10.3847/1538-4357/abef05}, \href
  {https://ui.adsabs.harvard.edu/abs/2021ApJ...912...46B} {912, 46}

\bibitem[\protect\citeauthoryear{{Bruch} et~al.,}{{Bruch}
  et~al.}{2023}]{Bruch2023}
{Bruch} R.~J.,  et~al., 2023, \mn@doi [\apj] {10.3847/1538-4357/acd8be}, \href
  {https://ui.adsabs.harvard.edu/abs/2023ApJ...952..119B} {952, 119}

\bibitem[\protect\citeauthoryear{{Burrows} et~al.,}{{Burrows}
  et~al.}{2005}]{Burrows2005a}
{Burrows} D.~N.,  et~al., 2005, \mn@doi [\ssr] {10.1007/s11214-005-5097-2},
  \href {https://ui.adsabs.harvard.edu/abs/2005SSRv..120..165B} {120, 165}

\bibitem[\protect\citeauthoryear{{Byler}, {Dalcanton}, {Conroy}  \&
  {Johnson}}{{Byler} et~al.}{2017}]{Byler2017a}
{Byler} N.,  {Dalcanton} J.~J.,  {Conroy} C.,   {Johnson} B.~D.,  2017, \mn@doi
  [\apj] {10.3847/1538-4357/aa6c66}, \href
  {https://ui.adsabs.harvard.edu/abs/2017ApJ...840...44B} {840, 44}

\bibitem[\protect\citeauthoryear{{Calzetti}, {Armus}, {Bohlin}, {Kinney},
  {Koornneef}  \& {Storchi-Bergmann}}{{Calzetti} et~al.}{2000}]{Calzetti2000a}
{Calzetti} D.,  {Armus} L.,  {Bohlin} R.~C.,  {Kinney} A.~L.,  {Koornneef} J.,
   {Storchi-Bergmann} T.,  2000, \mn@doi [\apj] {10.1086/308692}, \href
  {http://adsabs.harvard.edu/abs/2000ApJ...533..682C} {533, 682}

\bibitem[\protect\citeauthoryear{{Cao} et~al.,}{{Cao} et~al.}{2013}]{Cao2013}
{Cao} Y.,  et~al., 2013, \mn@doi [\apjl] {10.1088/2041-8205/775/1/L7}, \href
  {https://ui.adsabs.harvard.edu/abs/2013ApJ...775L...7C} {775, L7}

\bibitem[\protect\citeauthoryear{{Cardelli}, {Clayton}  \& {Mathis}}{{Cardelli}
  et~al.}{1989}]{Cardelli1989}
{Cardelli} J.~A.,  {Clayton} G.~C.,   {Mathis} J.~S.,  1989, \mn@doi [\apj]
  {10.1086/167900}, \href
  {https://ui.adsabs.harvard.edu/abs/1989ApJ...345..245C} {345, 245}

\bibitem[\protect\citeauthoryear{{Carrick}, {Turnbull}, {Lavaux}  \&
  {Hudson}}{{Carrick} et~al.}{2015}]{Carrick2015}
{Carrick} J.,  {Turnbull} S.~J.,  {Lavaux} G.,   {Hudson} M.~J.,  2015, \mn@doi
  [\mnras] {10.1093/mnras/stv547}, \href
  {https://ui.adsabs.harvard.edu/abs/2015MNRAS.450..317C} {450, 317}

\bibitem[\protect\citeauthoryear{{Chabrier}}{{Chabrier}}{2003}]{Chabrier2003a}
{Chabrier} G.,  2003, \mn@doi [\pasp] {10.1086/376392}, \href
  {http://adsabs.harvard.edu/abs/2003PASP..115..763C} {115, 763}

\bibitem[\protect\citeauthoryear{{Chambers} et~al.,}{{Chambers}
  et~al.}{2016}]{Chambers2016a}
{Chambers} K.~C.,  et~al., 2016, arXiv e-prints, \href
  {https://ui.adsabs.harvard.edu/abs/2016arXiv161205560C} {p. arXiv:1612.05560}

\bibitem[\protect\citeauthoryear{{Chatzopoulos}, {Wheeler}  \&
  {Vinko}}{{Chatzopoulos} et~al.}{2012}]{Chatzopoulos12}
{Chatzopoulos} E.,  {Wheeler} J.~C.,   {Vinko} J.,  2012, \mn@doi [\apj]
  {10.1088/0004-637X/746/2/121}, \href
  {https://ui.adsabs.harvard.edu/abs/2012ApJ...746..121C} {746, 121}

\bibitem[\protect\citeauthoryear{{Chatzopoulos}, {Wheeler}, {Vinko}, {Horvath}
  \& {Nagy}}{{Chatzopoulos} et~al.}{2013}]{Chatzopoulos2013}
{Chatzopoulos} E.,  {Wheeler} J.~C.,  {Vinko} J.,  {Horvath} Z.~L.,   {Nagy}
  A.,  2013, \mn@doi [\apj] {10.1088/0004-637X/773/1/76}, \href
  {https://ui.adsabs.harvard.edu/abs/2013ApJ...773...76C} {773, 76}

\bibitem[\protect\citeauthoryear{{Chevalier}}{{Chevalier}}{1982}]{Chevalier1982}
{Chevalier} R.~A.,  1982, \mn@doi [\apj] {10.1086/160126}, \href
  {https://ui.adsabs.harvard.edu/abs/1982ApJ...258..790C} {258, 790}

\bibitem[\protect\citeauthoryear{{Colgate} \& {McKee}}{{Colgate} \&
  {McKee}}{1969}]{Colgate1969}
{Colgate} S.~A.,  {McKee} C.,  1969, \mn@doi [\apj] {10.1086/150102}, \href
  {https://ui.adsabs.harvard.edu/abs/1969ApJ...157..623C} {157, 623}

\bibitem[\protect\citeauthoryear{{Conroy}, {Gunn}  \& {White}}{{Conroy}
  et~al.}{2009}]{Conroy2009a}
{Conroy} C.,  {Gunn} J.~E.,   {White} M.,  2009, \mn@doi [\apj]
  {10.1088/0004-637X/699/1/486}, \href
  {https://ui.adsabs.harvard.edu/abs/2009ApJ...699..486C} {699, 486}

\bibitem[\protect\citeauthoryear{{Coughlin} et~al.,}{{Coughlin}
  et~al.}{2023}]{Coughlin2023}
{Coughlin} M.~W.,  et~al., 2023, \mn@doi [\apjs] {10.3847/1538-4365/acdee1},
  \href {https://ui.adsabs.harvard.edu/abs/2023ApJS..267...31C} {267, 31}

\bibitem[\protect\citeauthoryear{Craig et~al.,}{Craig et~al.}{2017}]{ccdproc}
Craig M.,  et~al., 2017, astropy/ccdproc: v1.3.0.post1,
  \mn@doi{10.5281/zenodo.1069648}, \url
  {https://doi.org/10.5281/zenodo.1069648}

\bibitem[\protect\citeauthoryear{{Curti}, {Cresci}, {Mannucci}, {Marconi},
  {Maiolino}  \& {Esposito}}{{Curti} et~al.}{2017}]{Curti2017a}
{Curti} M.,  {Cresci} G.,  {Mannucci} F.,  {Marconi} A.,  {Maiolino} R.,
  {Esposito} S.,  2017, \mn@doi [\mnras] {10.1093/mnras/stw2766}, \href
  {https://ui.adsabs.harvard.edu/abs/2017MNRAS.465.1384C} {465, 1384}

\bibitem[\protect\citeauthoryear{{Das} et~al.,}{{Das}
  et~al.}{2024}]{Das2024_zaw}
{Das} K.~K.,  et~al., 2024, \mn@doi [\apjl] {10.3847/2041-8213/ad527a}, \href
  {https://ui.adsabs.harvard.edu/abs/2024ApJ...969L..11D} {969, L11}

\bibitem[\protect\citeauthoryear{{Dekany} et~al.,}{{Dekany}
  et~al.}{2020}]{Dekany2020}
{Dekany} R.,  et~al., 2020, \mn@doi [\pasp] {10.1088/1538-3873/ab4ca2}, \href
  {https://ui.adsabs.harvard.edu/abs/2020PASP..132c8001D} {132, 038001}

\bibitem[\protect\citeauthoryear{{Dessart}}{{Dessart}}{2024}]{Dessart2024}
{Dessart} L.,  2024, in Supernova Remnants III: An Odyssey in Space after
  Stellar Death. p.~37

\bibitem[\protect\citeauthoryear{{Dessart}, {Hillier}, {Audit}, {Livne}  \&
  {Waldman}}{{Dessart} et~al.}{2016}]{Dessart2016}
{Dessart} L.,  {Hillier} D.~J.,  {Audit} E.,  {Livne} E.,   {Waldman} R.,
  2016, \mn@doi [\mnras] {10.1093/mnras/stw336}, \href
  {https://ui.adsabs.harvard.edu/abs/2016MNRAS.458.2094D} {458, 2094}

\bibitem[\protect\citeauthoryear{{Dessart}, {Hillier}  \&
  {Kuncarayakti}}{{Dessart} et~al.}{2022}]{Dessart2022}
{Dessart} L.,  {Hillier} D.~J.,   {Kuncarayakti} H.,  2022, \mn@doi [\aap]
  {10.1051/0004-6361/202142436}, \href
  {https://ui.adsabs.harvard.edu/abs/2022A&A...658A.130D} {658, A130}

\bibitem[\protect\citeauthoryear{{Dong} et~al.,}{{Dong}
  et~al.}{2024}]{Dong2024}
{Dong} Y.,  et~al., 2024, \mn@doi [\apj] {10.3847/1538-4357/ad8de6}, \href
  {https://ui.adsabs.harvard.edu/abs/2024ApJ...977..254D} {977, 254}

\bibitem[\protect\citeauthoryear{{Drout} et~al.,}{{Drout}
  et~al.}{2014}]{Drout2014}
{Drout} M.~R.,  et~al., 2014, \mn@doi [\apj] {10.1088/0004-637X/794/1/23},
  \href {https://ui.adsabs.harvard.edu/abs/2014ApJ...794...23D} {794, 23}

\bibitem[\protect\citeauthoryear{{Ertl}, {Woosley}, {Sukhbold}  \&
  {Janka}}{{Ertl} et~al.}{2020}]{Ertl2020}
{Ertl} T.,  {Woosley} S.~E.,  {Sukhbold} T.,   {Janka} H.~T.,  2020, \mn@doi
  [\apj] {10.3847/1538-4357/ab6458}, \href
  {https://ui.adsabs.harvard.edu/abs/2020ApJ...890...51E} {890, 51}

\bibitem[\protect\citeauthoryear{{Evans} et~al.,}{{Evans}
  et~al.}{2007}]{Evans2007a}
{Evans} P.~A.,  et~al., 2007, \mn@doi [\aap] {10.1051/0004-6361:20077530},
  \href {https://ui.adsabs.harvard.edu/abs/2007A&A...469..379E} {469, 379}

\bibitem[\protect\citeauthoryear{{Evans} et~al.,}{{Evans}
  et~al.}{2009}]{Evans2009a}
{Evans} P.~A.,  et~al., 2009, \mn@doi [\mnras]
  {10.1111/j.1365-2966.2009.14913.x}, \href
  {https://ui.adsabs.harvard.edu/abs/2009MNRAS.397.1177E} {397, 1177}

\bibitem[\protect\citeauthoryear{{Falk}, {Lattimer}  \& {Margolis}}{{Falk}
  et~al.}{1977}]{falk1977}
{Falk} S.~W.,  {Lattimer} J.~M.,   {Margolis} S.~H.,  1977, \mn@doi [\nat]
  {10.1038/270700a0}, \href
  {https://ui.adsabs.harvard.edu/abs/1977Natur.270..700F} {270, 700}

\bibitem[\protect\citeauthoryear{{Fassia} et~al.,}{{Fassia}
  et~al.}{2001}]{Fassia2001}
{Fassia} A.,  et~al., 2001, \mn@doi [\mnras]
  {10.1046/j.1365-8711.2001.04282.x}, \href
  {https://ui.adsabs.harvard.edu/abs/2001MNRAS.325..907F} {325, 907}

\bibitem[\protect\citeauthoryear{{Flewelling} et~al.,}{{Flewelling}
  et~al.}{2020}]{Flewelling2020}
{Flewelling} H.~A.,  et~al., 2020, \mn@doi [\apjs] {10.3847/1538-4365/abb82d},
  \href {https://ui.adsabs.harvard.edu/abs/2020ApJS..251....7F} {251, 7}

\bibitem[\protect\citeauthoryear{{Foley}, {Smith}, {Ganeshalingam}, {Li},
  {Chornock}  \& {Filippenko}}{{Foley} et~al.}{2007}]{2007ApJ...657L.105F}
{Foley} R.~J.,  {Smith} N.,  {Ganeshalingam} M.,  {Li} W.,  {Chornock} R.,
  {Filippenko} A.~V.,  2007, \mn@doi [\apjl] {10.1086/513145}, \href
  {https://ui.adsabs.harvard.edu/abs/2007ApJ...657L.105F} {657, L105}

\bibitem[\protect\citeauthoryear{{Foreman-Mackey}, {Sick}  \&
  {Johnson}}{{Foreman-Mackey} et~al.}{2014}]{ForemanMackey2014a}
{Foreman-Mackey} D.,  {Sick} J.,   {Johnson} B.,  2014, {Python-Fsps: Python
  Bindings To Fsps (V0.1.1)}, \mn@doi{10.5281/zenodo.12157}

\bibitem[\protect\citeauthoryear{{Fransson} \& {Chevalier}}{{Fransson} \&
  {Chevalier}}{1989}]{1989ApJ...343..323F}
{Fransson} C.,  {Chevalier} R.~A.,  1989, \mn@doi [\apj] {10.1086/167707},
  \href {https://ui.adsabs.harvard.edu/abs/1989ApJ...343..323F} {343, 323}

\bibitem[\protect\citeauthoryear{{Fraser}}{{Fraser}}{2020}]{Fraser2020}
{Fraser} M.,  2020, \mn@doi [Royal Society Open Science] {10.1098/rsos.200467},
  \href {https://ui.adsabs.harvard.edu/abs/2020RSOS....700467F} {7, 200467}

\bibitem[\protect\citeauthoryear{{Fremling}}{{Fremling}}{2023}]{2023TNSTR2892....1F}
{Fremling} C.,  2023, Transient Name Server Discovery Report, \href
  {https://ui.adsabs.harvard.edu/abs/2023TNSTR2892....1F} {2023-2892, 1}

\bibitem[\protect\citeauthoryear{{Fremling} et~al.,}{{Fremling}
  et~al.}{2014}]{Fremling2014A&A...565A.114F}
{Fremling} C.,  et~al., 2014, \mn@doi [\aap] {10.1051/0004-6361/201423884},
  \href {https://ui.adsabs.harvard.edu/abs/2014A&A...565A.114F} {565, A114}

\bibitem[\protect\citeauthoryear{{Fremling} et~al.,}{{Fremling}
  et~al.}{2016}]{Fleming2016}
{Fremling} C.,  et~al., 2016, \mn@doi [\aap] {10.1051/0004-6361/201628275},
  \href {https://ui.adsabs.harvard.edu/abs/2016A&A...593A..68F} {593, A68}

\bibitem[\protect\citeauthoryear{{Gal-Yam} et~al.,}{{Gal-Yam}
  et~al.}{2014}]{GalYam2014NatureFlash}
{Gal-Yam} A.,  et~al., 2014, \mn@doi [\nat] {10.1038/nature13304}, \href
  {https://ui.adsabs.harvard.edu/abs/2014Natur.509..471G} {509, 471}

\bibitem[\protect\citeauthoryear{{Gal-Yam} et~al.,}{{Gal-Yam}
  et~al.}{2022}]{GalYamNature2022Icn}
{Gal-Yam} A.,  et~al., 2022, \mn@doi [\nat] {10.1038/s41586-021-04155-1}, \href
  {https://ui.adsabs.harvard.edu/abs/2022Natur.601..201G} {601, 201}

\bibitem[\protect\citeauthoryear{{Gangopadhyay}}{{Gangopadhyay}}{2024}]{Anjasha2024review}
{Gangopadhyay} A.,  2024, \mn@doi [arXiv e-prints] {10.48550/arXiv.2411.04107},
  \href {https://ui.adsabs.harvard.edu/abs/2024arXiv241104107G} {p.
  arXiv:2411.04107}

\bibitem[\protect\citeauthoryear{{Gangopadhyay} et~al.,}{{Gangopadhyay}
  et~al.}{2020}]{Gangopadhyay2020}
{Gangopadhyay} A.,  et~al., 2020, \mn@doi [\apj] {10.3847/1538-4357/ab6328},
  \href {https://ui.adsabs.harvard.edu/abs/2020ApJ...889..170G} {889, 170}

\bibitem[\protect\citeauthoryear{{Gangopadhyay} et~al.,}{{Gangopadhyay}
  et~al.}{2022}]{Gangopadhyay2022}
{Gangopadhyay} A.,  et~al., 2022, \mn@doi [\apj] {10.3847/1538-4357/ac6187},
  \href {https://ui.adsabs.harvard.edu/abs/2022ApJ...930..127G} {930, 127}

\bibitem[\protect\citeauthoryear{{Gangopadhyay} et~al.,}{{Gangopadhyay}
  et~al.}{2025}]{2025Gangopadhyay}
{Gangopadhyay} A.,  et~al., 2025, \mn@doi [\mnras] {10.1093/mnras/staf187},
  \href {https://ui.adsabs.harvard.edu/abs/2025MNRAS.537.2898G} {537, 2898}

\bibitem[\protect\citeauthoryear{{Gehrels} et~al.,}{{Gehrels}
  et~al.}{2004}]{gehrels2004}
{Gehrels} N.,  et~al., 2004, \mn@doi [\apj] {10.1086/422091}, \href
  {https://ui.adsabs.harvard.edu/abs/2004ApJ...611.1005G} {611, 1005}

\bibitem[\protect\citeauthoryear{{Graham} et~al.,}{{Graham}
  et~al.}{2019}]{Graham2019}
{Graham} M.~J.,  et~al., 2019, \mn@doi [\pasp] {10.1088/1538-3873/ab006c},
  \href {https://ui.adsabs.harvard.edu/abs/2019PASP..131g8001G} {131, 078001}

\bibitem[\protect\citeauthoryear{{HI4PI Collaboration} et~al.,}{{HI4PI
  Collaboration} et~al.}{2016}]{HI4PI2016a}
{HI4PI Collaboration} et~al., 2016, \mn@doi [\aap]
  {10.1051/0004-6361/201629178}, \href
  {https://ui.adsabs.harvard.edu/abs/2016A&A...594A.116H} {594, A116}

\bibitem[\protect\citeauthoryear{Hamuy}{Hamuy}{2003}]{Hamuy_2003}
Hamuy M.,  2003, \mn@doi [The Astrophysical Journal] {10.1086/344689}, 582, 905

\bibitem[\protect\citeauthoryear{{Ho} et~al.,}{{Ho} et~al.}{2023}]{Ho2023}
{Ho} A. Y.~Q.,  et~al., 2023, \mn@doi [\apj] {10.3847/1538-4357/acc533}, \href
  {https://ui.adsabs.harvard.edu/abs/2023ApJ...949..120H} {949, 120}

\bibitem[\protect\citeauthoryear{{Hosseinzadeh} et~al.,}{{Hosseinzadeh}
  et~al.}{2017}]{Hosseinzadeh2017}
{Hosseinzadeh} G.,  et~al., 2017, \mn@doi [\apj] {10.3847/1538-4357/836/2/158},
  \href {https://ui.adsabs.harvard.edu/abs/2017ApJ...836..158H} {836, 158}

\bibitem[\protect\citeauthoryear{Hunter}{Hunter}{2007}]{4160265}
Hunter J.~D.,  2007, \mn@doi [Computing in Science & Engineering]
  {10.1109/MCSE.2007.55}, 9, 90

\bibitem[\protect\citeauthoryear{{Irani} et~al.,}{{Irani}
  et~al.}{2024}]{Irani2024}
{Irani} I.,  et~al., 2024, \mn@doi [\apj] {10.3847/1538-4357/ad3de8}, \href
  {https://ui.adsabs.harvard.edu/abs/2024ApJ...970...96I} {970, 96}

\bibitem[\protect\citeauthoryear{{Johnson}, {Leja}, {Conroy}  \&
  {Speagle}}{{Johnson} et~al.}{2021}]{Johnson2021a}
{Johnson} B.~D.,  {Leja} J.,  {Conroy} C.,   {Speagle} J.~S.,  2021, \mn@doi
  [\apjs] {10.3847/1538-4365/abef67}, \href
  {https://ui.adsabs.harvard.edu/abs/2021ApJS..254...22J} {254, 22}

\bibitem[\protect\citeauthoryear{{Karamehmetoglu} et~al.,}{{Karamehmetoglu}
  et~al.}{2017}]{Karamehmetoglu2017}
{Karamehmetoglu} E.,  et~al., 2017, \mn@doi [\aap]
  {10.1051/0004-6361/201629619}, \href
  {https://ui.adsabs.harvard.edu/abs/2017A&A...602A..93K} {602, A93}

\bibitem[\protect\citeauthoryear{{Karamehmetoglu} et~al.,}{{Karamehmetoglu}
  et~al.}{2021}]{Karamehmetoglu2021}
{Karamehmetoglu} E.,  et~al., 2021, \mn@doi [\aap]
  {10.1051/0004-6361/201936308}, \href
  {https://ui.adsabs.harvard.edu/abs/2021A&A...649A.163K} {649, A163}

\bibitem[\protect\citeauthoryear{{Kawabata} et~al.,}{{Kawabata}
  et~al.}{2008}]{2008SPIE.7014E..4LK}
{Kawabata} K.~S.,  et~al., 2008, in {McLean} I.~S.,  {Casali} M.~M.,  eds,
  Society of Photo-Optical Instrumentation Engineers (SPIE) Conference Series
  Vol. 7014, Ground-based and Airborne Instrumentation for Astronomy II. p.
  70144L, \mn@doi{10.1117/12.788569}

\bibitem[\protect\citeauthoryear{{Kennicutt}}{{Kennicutt}}{1998}]{Kennicutt1998a}
{Kennicutt} Robert~C. J.,  1998, \mn@doi [\araa]
  {10.1146/annurev.astro.36.1.189}, \href
  {https://ui.adsabs.harvard.edu/abs/1998ARA&A..36..189K} {36, 189}

\bibitem[\protect\citeauthoryear{{Khazov} et~al.,}{{Khazov}
  et~al.}{2016}]{Khazov2016}
{Khazov} D.,  et~al., 2016, \mn@doi [\apj] {10.3847/0004-637X/818/1/3}, \href
  {https://ui.adsabs.harvard.edu/abs/2016ApJ...818....3K} {818, 3}

\bibitem[\protect\citeauthoryear{{Kilpatrick} \& {Foley}}{{Kilpatrick} \&
  {Foley}}{2018}]{Kilpatrick2018}
{Kilpatrick} C.~D.,  {Foley} R.~J.,  2018, \mn@doi [\mnras]
  {10.1093/mnras/sty2435}, \href
  {https://ui.adsabs.harvard.edu/abs/2018MNRAS.481.2536K} {481, 2536}

\bibitem[\protect\citeauthoryear{{Kim} et~al.,}{{Kim} et~al.}{2022}]{Kim2022}
{Kim} Y.~L.,  et~al., 2022, \mn@doi [\pasp] {10.1088/1538-3873/ac50a0}, \href
  {https://ui.adsabs.harvard.edu/abs/2022PASP..134b4505K} {134, 024505}

\bibitem[\protect\citeauthoryear{{Kool} et~al.,}{{Kool}
  et~al.}{2021}]{Kool2021}
{Kool} E.~C.,  et~al., 2021, \mn@doi [\aap] {10.1051/0004-6361/202039137},
  \href {https://ui.adsabs.harvard.edu/abs/2021A&A...652A.136K} {652, A136}

\bibitem[\protect\citeauthoryear{{Kumar} et~al.,}{{Kumar}
  et~al.}{2022}]{2022growth}
{Kumar} H.,  et~al., 2022, \mn@doi [\aj] {10.3847/1538-3881/ac7bea}, \href
  {https://ui.adsabs.harvard.edu/abs/2022AJ....164...90K} {164, 90}

\bibitem[\protect\citeauthoryear{{Lang}}{{Lang}}{2014}]{Lang2014a}
{Lang} D.,  2014, \mn@doi [\aj] {10.1088/0004-6256/147/5/108}, \href
  {https://ui.adsabs.harvard.edu/\#abs/2014AJ....147..108L} {147, 108}

\bibitem[\protect\citeauthoryear{{Langer}}{{Langer}}{2012}]{Langer2012}
{Langer} N.,  2012, \mn@doi [\araa] {10.1146/annurev-astro-081811-125534},
  \href {https://ui.adsabs.harvard.edu/abs/2012ARA&A..50..107L} {50, 107}

\bibitem[\protect\citeauthoryear{{Lusk} \& {Baron}}{{Lusk} \&
  {Baron}}{2017}]{2017PASP..129d4202L}
{Lusk} J.~A.,  {Baron} E.,  2017, \mn@doi [\pasp] {10.1088/1538-3873/aa5e49},
  \href {https://ui.adsabs.harvard.edu/abs/2017PASP..129d4202L} {129, 044202}

\bibitem[\protect\citeauthoryear{{Madau} \& {Dickinson}}{{Madau} \&
  {Dickinson}}{2014}]{Madau2014a}
{Madau} P.,  {Dickinson} M.,  2014, \mn@doi [\araa]
  {10.1146/annurev-astro-081811-125615}, \href
  {https://ui.adsabs.harvard.edu/abs/2014ARA&A..52..415M} {52, 415}

\bibitem[\protect\citeauthoryear{{Maeda} \& {Moriya}}{{Maeda} \&
  {Moriya}}{2022}]{MaedaMoriya2022}
{Maeda} K.,  {Moriya} T.~J.,  2022, \mn@doi [\apj] {10.3847/1538-4357/ac4672},
  \href {https://ui.adsabs.harvard.edu/abs/2022ApJ...927...25M} {927, 25}

\bibitem[\protect\citeauthoryear{{Mainzer} et~al.,}{{Mainzer}
  et~al.}{2014}]{Mainzer2014a}
{Mainzer} A.,  et~al., 2014, \mn@doi [\apj] {10.1088/0004-637X/792/1/30}, \href
  {https://ui.adsabs.harvard.edu/abs/2014ApJ...792...30M} {792, 30}

\bibitem[\protect\citeauthoryear{Mandigo-Stoba, Fremling  \&
  Kasliwal}{Mandigo-Stoba et~al.}{2021}]{dbsp_drp:arxiv}
Mandigo-Stoba M.~S.,  Fremling C.,   Kasliwal M.~M.,  2021, DBSP\_DRP: A Python
  package for automated spectroscopic data reduction of DBSP data (\mn@eprint
  {arXiv} {2107.12339})

\bibitem[\protect\citeauthoryear{Mandigo-Stoba, Fremling  \&
  Kasliwal}{Mandigo-Stoba et~al.}{2022a}]{dbsp_drp:zenodo}
Mandigo-Stoba M.~S.,  Fremling C.,   Kasliwal M.~M.,  2022a, {DBSP\_DRP: A
  Python package for automated spectroscopic data reduction of DBSP data},
  \mn@doi{10.5281/zenodo.6241526}, \url
  {https://doi.org/10.5281/zenodo.6241526}

\bibitem[\protect\citeauthoryear{Mandigo-Stoba, Fremling  \&
  Kasliwal}{Mandigo-Stoba et~al.}{2022b}]{dbsp_drp:joss}
Mandigo-Stoba M.~S.,  Fremling C.,   Kasliwal M.~M.,  2022b, \mn@doi [Journal
  of Open Source Software] {10.21105/joss.03612}, 7, 3612

\bibitem[\protect\citeauthoryear{{Martins} \& {Hillier}}{{Martins} \&
  {Hillier}}{2012}]{Martins2012}
{Martins} F.,  {Hillier} D.~J.,  2012, \mn@doi [\aap]
  {10.1051/0004-6361/201219788}, \href
  {https://ui.adsabs.harvard.edu/abs/2012A&A...545A..95M} {545, A95}

\bibitem[\protect\citeauthoryear{{Masci} et~al.,}{{Masci}
  et~al.}{2019}]{Masci2019}
{Masci} F.~J.,  et~al., 2019, \mn@doi [\pasp] {10.1088/1538-3873/aae8ac}, \href
  {https://ui.adsabs.harvard.edu/abs/2019PASP..131a8003M} {131, 018003}

\bibitem[\protect\citeauthoryear{{Matsubayashi} et~al.,}{{Matsubayashi}
  et~al.}{2019}]{2019PASJ...71..102M}
{Matsubayashi} K.,  et~al., 2019, \mn@doi [\pasj] {10.1093/pasj/psz087}, \href
  {https://ui.adsabs.harvard.edu/abs/2019PASJ...71..102M} {71, 102}

\bibitem[\protect\citeauthoryear{{Matzner} \& {McKee}}{{Matzner} \&
  {McKee}}{1999}]{Matzner1999}
{Matzner} C.~D.,  {McKee} C.~F.,  1999, \mn@doi [\apj] {10.1086/306571}, \href
  {https://ui.adsabs.harvard.edu/abs/1999ApJ...510..379M} {510, 379}

\bibitem[\protect\citeauthoryear{{McCully} \& {Tewes}}{{McCully} \&
  {Tewes}}{2019}]{asctroscrappy}
{McCully} C.,  {Tewes} M.,  2019, {Astro-SCRAPPY: Speedy Cosmic Ray
  Annihilation Package in Python}, Astrophysics Source Code Library, record
  ascl:1907.032

\bibitem[\protect\citeauthoryear{{Meisner}, {Lang}  \& {Schlegel}}{{Meisner}
  et~al.}{2017}]{Meisner2017a}
{Meisner} A.~M.,  {Lang} D.,   {Schlegel} D.~J.,  2017, \mn@doi [\aj]
  {10.3847/1538-3881/153/1/38}, \href
  {https://ui.adsabs.harvard.edu/\#abs/2017AJ....153...38M} {153, 38}

\bibitem[\protect\citeauthoryear{{Moriya}}{{Moriya}}{2013}]{Moriya2013}
{Moriya} T.,  2013, PhD thesis, University of Tokyo, Department of Astronomy

\bibitem[\protect\citeauthoryear{{Moriya}, {Yoon}, {Gr{\"a}fener}  \&
  {Blinnikov}}{{Moriya} et~al.}{2017}]{Moriya2017}
{Moriya} T.~J.,  {Yoon} S.-C.,  {Gr{\"a}fener} G.,   {Blinnikov} S.~I.,  2017,
  \mn@doi [\mnras] {10.1093/mnrasl/slx056}, \href
  {https://ui.adsabs.harvard.edu/abs/2017MNRAS.469L.108M} {469, L108}

\bibitem[\protect\citeauthoryear{{Moriya}, {Mueller}, {Blinnikov}, {Ushakova},
  {Sorokina}, {Tauris}  \& {Heger}}{{Moriya} et~al.}{2025}]{Moriya2025}
{Moriya} T.~J.,  {Mueller} B.,  {Blinnikov} S.~I.,  {Ushakova} M.,  {Sorokina}
  E.~I.,  {Tauris} T.~M.,   {Heger} A.,  2025, \mn@doi [arXiv e-prints]
  {10.48550/arXiv.2507.05506}, \href
  {https://ui.adsabs.harvard.edu/abs/2025arXiv250705506M} {p. arXiv:2507.05506}

\bibitem[\protect\citeauthoryear{{Nagao} et~al.,}{{Nagao}
  et~al.}{2023}]{Nagao2023}
{Nagao} T.,  et~al., 2023, \mn@doi [\aap] {10.1051/0004-6361/202346084}, \href
  {https://ui.adsabs.harvard.edu/abs/2023A&A...673A..27N} {673, A27}

\bibitem[\protect\citeauthoryear{{Nicholl}}{{Nicholl}}{2018}]{nicholl2018_superbol}
{Nicholl} M.,  2018, \mn@doi [Research Notes of the American Astronomical
  Society] {10.3847/2515-5172/aaf799}, \href
  {https://ui.adsabs.harvard.edu/abs/2018RNAAS...2..230N} {2, 230}

\bibitem[\protect\citeauthoryear{{Oke}}{{Oke}}{1990}]{Oke1990}
{Oke} J.~B.,  1990, \mn@doi [\aj] {10.1086/115444}, \href
  {https://ui.adsabs.harvard.edu/abs/1990AJ.....99.1621O} {99, 1621}

\bibitem[\protect\citeauthoryear{{Oke} \& {Gunn}}{{Oke} \&
  {Gunn}}{1982}]{OkeandGunn}
{Oke} J.~B.,  {Gunn} J.~E.,  1982, \mn@doi [\pasp] {10.1086/131027}, \href
  {https://ui.adsabs.harvard.edu/abs/1982PASP...94..586O} {94, 586}

\bibitem[\protect\citeauthoryear{{Oke} et~al.,}{{Oke} et~al.}{1995}]{Oke1995}
{Oke} J.~B.,  et~al., 1995, \mn@doi [\pasp] {10.1086/133562}, \href
  {https://ui.adsabs.harvard.edu/abs/1995PASP..107..375O} {107, 375}

\bibitem[\protect\citeauthoryear{{Osterbrock} \& {Ferland}}{{Osterbrock} \&
  {Ferland}}{2006}]{Osterbrock2006a}
{Osterbrock} D.~E.,  {Ferland} G.~J.,  2006, {Astrophysics of gaseous nebulae
  and active galactic nuclei}

\bibitem[\protect\citeauthoryear{{Pastorello} et~al.,}{{Pastorello}
  et~al.}{2007}]{2007Natur.447..829P}
{Pastorello} A.,  et~al., 2007, \mn@doi [\nat] {10.1038/nature05825}, \href
  {https://ui.adsabs.harvard.edu/abs/2007Natur.447..829P} {447, 829}

\bibitem[\protect\citeauthoryear{{Pastorello} et~al.,}{{Pastorello}
  et~al.}{2015a}]{PastorelloSN2010al}
{Pastorello} A.,  et~al., 2015a, \mn@doi [\mnras] {10.1093/mnras/stu2745},
  \href {https://ui.adsabs.harvard.edu/abs/2015MNRAS.449.1921P} {449, 1921}

\bibitem[\protect\citeauthoryear{{Pastorello} et~al.,}{{Pastorello}
  et~al.}{2015b}]{Pastorello2015-OGLE}
{Pastorello} A.,  et~al., 2015b, \mn@doi [\mnras] {10.1093/mnras/stu2621},
  \href {https://ui.adsabs.harvard.edu/abs/2015MNRAS.449.1941P} {449, 1941}

\bibitem[\protect\citeauthoryear{{Pastorello} et~al.,}{{Pastorello}
  et~al.}{2015c}]{Pastorello2015-15ed}
{Pastorello} A.,  et~al., 2015c, \mn@doi [\mnras] {10.1093/mnras/stv1812},
  \href {https://ui.adsabs.harvard.edu/abs/2015MNRAS.453.3649P} {453, 3649}

\bibitem[\protect\citeauthoryear{{Pellegrino} et~al.,}{{Pellegrino}
  et~al.}{2022a}]{Pellegrino2022Ibn}
{Pellegrino} C.,  et~al., 2022a, \mn@doi [\apj] {10.3847/1538-4357/ac3e63},
  \href {https://ui.adsabs.harvard.edu/abs/2022ApJ...926..125P} {926, 125}

\bibitem[\protect\citeauthoryear{{Pellegrino} et~al.,}{{Pellegrino}
  et~al.}{2022b}]{Pellegrino2022Icn}
{Pellegrino} C.,  et~al., 2022b, \mn@doi [\apj] {10.3847/1538-4357/ac8ff6},
  \href {https://ui.adsabs.harvard.edu/abs/2022ApJ...938...73P} {938, 73}

\bibitem[\protect\citeauthoryear{{Perley} et~al.,}{{Perley}
  et~al.}{2019}]{Perley2019}
{Perley} D.~A.,  et~al., 2019, \mn@doi [\mnras] {10.1093/mnras/sty3420}, \href
  {https://ui.adsabs.harvard.edu/abs/2019MNRAS.484.1031P} {484, 1031}

\bibitem[\protect\citeauthoryear{{Perley} et~al.,}{{Perley}
  et~al.}{2020}]{Perley2020}
{Perley} D.~A.,  et~al., 2020, \mn@doi [\apj] {10.3847/1538-4357/abbd98}, \href
  {https://ui.adsabs.harvard.edu/abs/2020ApJ...904...35P} {904, 35}

\bibitem[\protect\citeauthoryear{{Perley} et~al.,}{{Perley}
  et~al.}{2022}]{PerleyIcn}
{Perley} D.~A.,  et~al., 2022, \mn@doi [\apj] {10.3847/1538-4357/ac478e}, \href
  {https://ui.adsabs.harvard.edu/abs/2022ApJ...927..180P} {927, 180}

\bibitem[\protect\citeauthoryear{{Piascik}, {Steele}, {Bates}, {Mottram},
  {Smith}, {Barnsley}  \& {Bolton}}{{Piascik} et~al.}{2014}]{Piascik2014}
{Piascik} A.~S.,  {Steele} I.~A.,  {Bates} S.~D.,  {Mottram} C.~J.,  {Smith}
  R.~J.,  {Barnsley} R.~M.,   {Bolton} B.,  2014, in {Ramsay} S.~K.,  {McLean}
  I.~S.,   {Takami} H.,  eds,  Society of Photo-Optical Instrumentation
  Engineers (SPIE) Conference Series Vol. 9147, Ground-based and Airborne
  Instrumentation for Astronomy V. p. 91478H, \mn@doi{10.1117/12.2055117}

\bibitem[\protect\citeauthoryear{{Poznanski}, {Prochaska}  \&
  {Bloom}}{{Poznanski} et~al.}{2012}]{2012Poznanski}
{Poznanski} D.,  {Prochaska} J.~X.,   {Bloom} J.~S.,  2012, \mn@doi [\mnras]
  {10.1111/j.1365-2966.2012.21796.x}, \href
  {https://ui.adsabs.harvard.edu/abs/2012MNRAS.426.1465P} {426, 1465}

\bibitem[\protect\citeauthoryear{{Prabhu} \& {Anupama}}{{Prabhu} \&
  {Anupama}}{2010}]{2010ASInC...1..193P}
{Prabhu} T.~P.,  {Anupama} G.~C.,  2010, in Astronomical Society of India
  Conference Series.

\bibitem[\protect\citeauthoryear{{Prochaska} et~al.,}{{Prochaska}
  et~al.}{2019}]{Prochaska2019}
{Prochaska} J.~X.,  et~al., 2019, {pypeit/PypeIt: Releasing for DOI},
  \mn@doi{10.5281/zenodo.3506873}

\bibitem[\protect\citeauthoryear{Prochaska et~al.,}{Prochaska
  et~al.}{2020}]{pypeit:joss_pub}
Prochaska J.~X.,  et~al., 2020, \mn@doi [Journal of Open Source Software]
  {10.21105/joss.02308}, 5, 2308

\bibitem[\protect\citeauthoryear{Prochaska et~al.,}{Prochaska
  et~al.}{2021}]{pypeit:zenodov_v1_6}
Prochaska J.~X.,  et~al., 2021, pypeit/PypeIt: Version 1.6.0,
  \mn@doi{10.5281/zenodo.5548381}, \url
  {https://doi.org/10.5281/zenodo.5548381}

\bibitem[\protect\citeauthoryear{{Pursiainen} et~al.,}{{Pursiainen}
  et~al.}{2023}]{Pursiainen2023}
{Pursiainen} M.,  et~al., 2023, \mn@doi [\apjl] {10.3847/2041-8213/ad103d},
  \href {https://ui.adsabs.harvard.edu/abs/2023ApJ...959L..10P} {959, L10}

\bibitem[\protect\citeauthoryear{{Qin} \& {Zabludoff}}{{Qin} \&
  {Zabludoff}}{2024}]{QinZabludoff2024}
{Qin} Y.-J.,  {Zabludoff} A.,  2024, \mn@doi [\mnras] {10.1093/mnras/stae1921},
  \href {https://ui.adsabs.harvard.edu/abs/2024MNRAS.533.3517Q} {533, 3517}

\bibitem[\protect\citeauthoryear{{Riess} et~al.,}{{Riess}
  et~al.}{2011}]{2011ApJ...730..119R}
{Riess} A.~G.,  et~al., 2011, \mn@doi [\apj] {10.1088/0004-637X/730/2/119},
  \href {https://ui.adsabs.harvard.edu/abs/2011ApJ...730..119R} {730, 119}

\bibitem[\protect\citeauthoryear{{Rigault} et~al.,}{{Rigault}
  et~al.}{2019}]{Rigault2019}
{Rigault} M.,  et~al., 2019, \mn@doi [\aap] {10.1051/0004-6361/201935344},
  \href {https://ui.adsabs.harvard.edu/abs/2019A&A...627A.115R} {627, A115}

\bibitem[\protect\citeauthoryear{{Roberson}, {Fremling}  \&
  {Kasliwal}}{{Roberson} et~al.}{2022}]{Roberson2022}
{Roberson} M.,  {Fremling} C.,   {Kasliwal} M.,  2022, \mn@doi [The Journal of
  Open Source Software] {10.21105/joss.03612}, \href
  {https://ui.adsabs.harvard.edu/abs/2022JOSS....7.3612R} {7, 3612}

\bibitem[\protect\citeauthoryear{{Roming} et~al.,}{{Roming}
  et~al.}{2005}]{2005roming}
{Roming} P.~W.~A.,  et~al., 2005, \mn@doi [Space Science Reviews]
  {10.1007/s11214-005-5095-4}, \href
  {http://adsabs.harvard.edu/abs/2005SSRv..120...95R} {120, 95}

\bibitem[\protect\citeauthoryear{{Sarin}, {H{\"u}bner}, {Omand}, {Setzer}  \&
  et al.}{{Sarin} et~al.}{2024}]{sarin_redback}
{Sarin} N.,  {H{\"u}bner} M.,  {Omand} C. M.~B.,  {Setzer} C.~N.,   et al.
  2024, \mn@doi [\mnras] {10.1093/mnras/stae1238}, \href
  {https://ui.adsabs.harvard.edu/abs/2024MNRAS.531.1203S} {531, 1203}

\bibitem[\protect\citeauthoryear{{Schlafly} \& {Finkbeiner}}{{Schlafly} \&
  {Finkbeiner}}{2011}]{milkyway_reddening}
{Schlafly} E.~F.,  {Finkbeiner} D.~P.,  2011, \mn@doi [\apj]
  {10.1088/0004-637X/737/2/103}, \href
  {https://ui.adsabs.harvard.edu/abs/2011ApJ...737..103S} {737, 103}

\bibitem[\protect\citeauthoryear{{Schulze} et~al.,}{{Schulze}
  et~al.}{2021}]{Schulze2021a}
{Schulze} S.,  et~al., 2021, \mn@doi [\apjs] {10.3847/1538-4365/abff5e}, \href
  {https://ui.adsabs.harvard.edu/abs/2021ApJS..255...29S} {255, 29}

\bibitem[\protect\citeauthoryear{{Schulze} et~al.,}{{Schulze}
  et~al.}{2024}]{SteveIen}
{Schulze} S.,  et~al., 2024, \mn@doi [arXiv e-prints]
  {10.48550/arXiv.2409.02054}, \href
  {https://ui.adsabs.harvard.edu/abs/2024arXiv240902054S} {p. arXiv:2409.02054}

\bibitem[\protect\citeauthoryear{{Shivvers} et~al.,}{{Shivvers}
  et~al.}{2016}]{2016MNRAS.461.3057S}
{Shivvers} I.,  et~al., 2016, \mn@doi [\mnras] {10.1093/mnras/stw1528}, \href
  {https://ui.adsabs.harvard.edu/abs/2016MNRAS.461.3057S} {461, 3057}

\bibitem[\protect\citeauthoryear{{Skrutskie} et~al.,}{{Skrutskie}
  et~al.}{2006}]{Skrutskie2006a}
{Skrutskie} M.~F.,  et~al., 2006, \mn@doi [\aj] {10.1086/498708}, \href
  {https://ui.adsabs.harvard.edu/abs/2006AJ....131.1163S} {131, 1163}

\bibitem[\protect\citeauthoryear{{Smith}}{{Smith}}{2017}]{Smith2017}
{Smith} N.,  2017, in {Alsabti} A.~W.,  {Murdin} P.,  eds, , Handbook of
  Supernovae.
p.~403, \mn@doi{10.1007/978-3-319-21846-5_38}

\bibitem[\protect\citeauthoryear{{Smith}, {Li}, {Silverman}, {Ganeshalingam}
  \& {Filippenko}}{{Smith} et~al.}{2011}]{Smith2011}
{Smith} N.,  {Li} W.,  {Silverman} J.~M.,  {Ganeshalingam} M.,   {Filippenko}
  A.~V.,  2011, \mn@doi [\mnras] {10.1111/j.1365-2966.2011.18763.x}, \href
  {https://ui.adsabs.harvard.edu/abs/2011MNRAS.415..773S} {415, 773}

\bibitem[\protect\citeauthoryear{{Sollerman}, {Chu}, {Dahiwale}  \&
  {Fremling}}{{Sollerman} et~al.}{2023}]{2023TNSCR3181....1S}
{Sollerman} J.,  {Chu} M.,  {Dahiwale} A.,   {Fremling} C.,  2023, Transient
  Name Server Classification Report, \href
  {https://ui.adsabs.harvard.edu/abs/2023TNSCR3181....1S} {2023-3181, 1}

\bibitem[\protect\citeauthoryear{{Sorokina}, {Blinnikov}, {Nomoto}, {Quimby}
  \& {Tolstov}}{{Sorokina} et~al.}{2016}]{Sorokina2016}
{Sorokina} E.,  {Blinnikov} S.,  {Nomoto} K.,  {Quimby} R.,   {Tolstov} A.,
  2016, \mn@doi [\apj] {10.3847/0004-637X/829/1/17}, \href
  {https://ui.adsabs.harvard.edu/abs/2016ApJ...829...17S} {829, 17}

\bibitem[\protect\citeauthoryear{{Speagle}}{{Speagle}}{2020}]{Speagle2020}
{Speagle} J.~S.,  2020, \mn@doi [\mnras] {10.1093/mnras/staa278}, \href
  {https://ui.adsabs.harvard.edu/abs/2020MNRAS.493.3132S} {493, 3132}

\bibitem[\protect\citeauthoryear{{Steele} et~al.,}{{Steele}
  et~al.}{2004}]{Steele2004}
{Steele} I.~A.,  et~al., 2004, in {Oschmann} Jr. J.~M.,  ed.,  Society of
  Photo-Optical Instrumentation Engineers (SPIE) Conference Series Vol. 5489,
  Ground-based Telescopes. pp 679--692, \mn@doi{10.1117/12.551456}

\bibitem[\protect\citeauthoryear{Sun, Maund, Hirai, Crowther  \&
  Podsiadlowski}{Sun et~al.}{2019}]{Sun2019}
Sun N.-C.,  Maund J.~R.,  Hirai R.,  Crowther P.~A.,   Podsiadlowski P.,  2019,
  \mn@doi [Monthly Notices of the Royal Astronomical Society]
  {10.1093/mnras/stz3431}, 491, 6000–6019

\bibitem[\protect\citeauthoryear{{Suzuki}, {Moriya}  \& {Takiwaki}}{{Suzuki}
  et~al.}{2019}]{Suzuki2019}
{Suzuki} A.,  {Moriya} T.~J.,   {Takiwaki} T.,  2019, \mn@doi [\apj]
  {10.3847/1538-4357/ab5a83}, \href
  {https://ui.adsabs.harvard.edu/abs/2019ApJ...887..249S} {887, 249}

\bibitem[\protect\citeauthoryear{{Tauris}, {Sanyal}, {Yoon}  \&
  {Langer}}{{Tauris} et~al.}{2013}]{Tauris2013}
{Tauris} T.~M.,  {Sanyal} D.,  {Yoon} S.~C.,   {Langer} N.,  2013, \mn@doi
  [\aap] {10.1051/0004-6361/201321662}, \href
  {https://ui.adsabs.harvard.edu/abs/2013A&A...558A..39T} {558, A39}

\bibitem[\protect\citeauthoryear{{Teja} et~al.,}{{Teja}
  et~al.}{2023}]{2023teja}
{Teja} R.~S.,  et~al., 2023, \mn@doi [\apjl] {10.3847/2041-8213/acef20}, \href
  {https://ui.adsabs.harvard.edu/abs/2023ApJ...954L..12T} {954, L12}

\bibitem[\protect\citeauthoryear{{Tody}}{{Tody}}{1986}]{Tody1986}
{Tody} D.,  1986, in {Crawford} D.~L.,  ed.,  Society of Photo-Optical
  Instrumentation Engineers (SPIE) Conference Series Vol. 627, Instrumentation
  in astronomy VI. p.~733, \mn@doi{10.1117/12.968154}

\bibitem[\protect\citeauthoryear{{Valenti} et~al.,}{{Valenti}
  et~al.}{2008}]{Valenti2008}
{Valenti} S.,  et~al., 2008, \mn@doi [\mnras]
  {10.1111/j.1365-2966.2007.12647.x}, \href
  {https://ui.adsabs.harvard.edu/abs/2008MNRAS.383.1485V} {383, 1485}

\bibitem[\protect\citeauthoryear{{Van Dyk} et~al.,}{{Van Dyk}
  et~al.}{2018}]{VanDyk2018}
{Van Dyk} S.~D.,  et~al., 2018, \mn@doi [\apj] {10.3847/1538-4357/aac32c},
  \href {https://ui.adsabs.harvard.edu/abs/2018ApJ...860...90V} {860, 90}

\bibitem[\protect\citeauthoryear{{Virtanen} et~al.,}{{Virtanen}
  et~al.}{2020}]{2020zndo...4406806V}
{Virtanen} P.,  et~al., 2020, {scipy/scipy: SciPy 1.6.0},
  \mn@doi{10.5281/zenodo.4406806}

\bibitem[\protect\citeauthoryear{{Wang} et~al.,}{{Wang}
  et~al.}{2024}]{Wang2024}
{Wang} Z.~Y.,  et~al., 2024, \mn@doi [\aap] {10.1051/0004-6361/202451131},
  \href {https://ui.adsabs.harvard.edu/abs/2024A&A...691A.156W} {691, A156}

\bibitem[\protect\citeauthoryear{{Woosley}}{{Woosley}}{2019}]{Woosley2019}
{Woosley} S.~E.,  2019, \mn@doi [\apj] {10.3847/1538-4357/ab1b41}, \href
  {https://ui.adsabs.harvard.edu/abs/2019ApJ...878...49W} {878, 49}

\bibitem[\protect\citeauthoryear{{Woosley} \& {Weaver}}{{Woosley} \&
  {Weaver}}{1995}]{Woosley1995}
{Woosley} S.~E.,  {Weaver} T.~A.,  1995, \mn@doi [\apjs] {10.1086/192237},
  \href {https://ui.adsabs.harvard.edu/abs/1995ApJS..101..181W} {101, 181}

\bibitem[\protect\citeauthoryear{{Wright} et~al.,}{{Wright}
  et~al.}{2010}]{Wright2010a}
{Wright} E.~L.,  et~al., 2010, \mn@doi [\aj] {10.1088/0004-6256/140/6/1868},
  \href {http://adsabs.harvard.edu/abs/2010AJ....140.1868W} {140, 1868}

\bibitem[\protect\citeauthoryear{{Wright} et~al.,}{{Wright}
  et~al.}{2016}]{Wright2016a}
{Wright} A.~H.,  et~al., 2016, \mn@doi [\mnras] {10.1093/mnras/stw832}, \href
  {http://adsabs.harvard.edu/abs/2016MNRAS.460..765W} {460, 765}

\bibitem[\protect\citeauthoryear{{Wu} \& {Fuller}}{{Wu} \&
  {Fuller}}{2022}]{Wu2022}
{Wu} S.~C.,  {Fuller} J.,  2022, \mn@doi [\apjl] {10.3847/2041-8213/ac9b3d},
  \href {https://ui.adsabs.harvard.edu/abs/2022ApJ...940L..27W} {940, L27}

\bibitem[\protect\citeauthoryear{{Yaron} \& {Gal-Yam}}{{Yaron} \&
  {Gal-Yam}}{2012}]{wiserep2012}
{Yaron} O.,  {Gal-Yam} A.,  2012, \mn@doi [\pasp] {10.1086/666656}, \href
  {https://ui.adsabs.harvard.edu/abs/2012PASP..124..668Y} {124, 668}

\bibitem[\protect\citeauthoryear{{van Baal}, {Jerkstrand}, {Wongwathanarat}  \&
  {Janka}}{{van Baal} et~al.}{2023}]{VanBaal2023}
{van Baal} B. F.~A.,  {Jerkstrand} A.,  {Wongwathanarat} A.,   {Janka} H.-T.,
  2023, \mn@doi [\mnras] {10.1093/mnras/stad1488}, \href
  {https://ui.adsabs.harvard.edu/abs/2023MNRAS.523..954V} {523, 954}

\bibitem[\protect\citeauthoryear{{van Dokkum}}{{van Dokkum}}{2001}]{lacosmic}
{van Dokkum} P.~G.,  2001, \mn@doi [\pasp] {10.1086/323894}, \href
  {https://ui.adsabs.harvard.edu/abs/2001PASP..113.1420V} {113, 1420}

\bibitem[\protect\citeauthoryear{van~der Walt, Colbert  \& Varoquaux}{van~der
  Walt et~al.}{2011}]{5725236}
van~der Walt S.,  Colbert S.~C.,   Varoquaux G.,  2011, \mn@doi [Computing in
  Science & Engineering] {10.1109/MCSE.2011.37}, 13, 22

\bibitem[\protect\citeauthoryear{van~der Walt, Crellin-Quick  \& Bloom}{van~der
  Walt et~al.}{2019}]{vanderWalt2019}
van~der Walt S.~J.,  Crellin-Quick A.,   Bloom J.~S.,  2019, \mn@doi [Journal
  of Open Source Software] {10.21105/joss.01247}, 4, 1247

\makeatother
\end{thebibliography}
